\documentclass[a4paper,fleqn,usenatbib]{mnras}

\usepackage{newtxtext,newtxmath}
\usepackage[T1]{fontenc}

\usepackage[english]{babel}
\usepackage{graphicx}
\usepackage{fleqn}
\usepackage{amssymb}
\usepackage{amsmath}
\usepackage{booktabs}
\usepackage{hyperref}
\usepackage{comment}
\usepackage{footmisc}
\usepackage{enumitem}
\setlist[enumerate]{
  labelsep=8pt,
  labelindent=0.\parindent,
 itemindent=0pt,
  leftmargin=*,
}

\usepackage{color, colortbl}

\newcommand*{\mycdot}{\kern-.2em\cdot\kern-.2em}
\renewcommand{\S}{Section}
\newcommand{\F}{Fig.}
\newcommand{\eq}{equation}

\newcommand{\msun}{\mathrm{M}_\odot}
\newcommand{\rsun}{\mathrm{R}_\odot}

\newcommand{\au}{\,\textsc{au}}

\newcommand{\yr}{\mathrm{yr}}
\newcommand{\Myr}{\mathrm{Myr}}
\newcommand{\Gyr}{\mathrm{Gyr}}

\renewcommand{\fint}{f_\mathrm{int}}
\newcommand{\fnon}{f_\mathrm{no}}
\newcommand{\fdyn}{f_\mathrm{dyninst}}

\newcommand{\nmc}{N_\mathrm{MC}}
\newcommand{\ns}{N_\star}
\newcommand{\nsys}{N_\mathrm{sys}}
\newcommand{\nlevel}{N_\mathrm{levels}}

\newcommand{\figsize}{0.47}

\pubyear{2020}

\usepackage{listings}

\usepackage{color}

\allowdisplaybreaks

\begin{document}
%\onecolumn
\title[MSC census]{A census of main-sequence interactions in the {\it Multiple Star Catalog}}

\author[Hamers]{Adrian S. Hamers$^{1}$\thanks{E-mail: hamers@mpa-garching.mpg.de} \\
$^{1}$Max-Planck-Institut f\"{u}r Astrophysik, Karl-Schwarzschild-Str. 1, 85741 Garching, Germany}
\date{Accepted 2020 April 21. Received 2020 March 30; in original form 2020 February 20.}

\label{firstpage}
\pagerange{\pageref{firstpage}--\pageref{lastpage}}
\maketitle

\begin{abstract}  % 231 words
Statistics of hierarchical systems containing three or more stars are continuously improving. The {\it Multiple Star Catalogue} (MSC) is currently the most comprehensive catalogue of multiple-star systems and contains component masses, orbital periods, and additional information. The systems in the MSC are interesting for several reasons, including the long-term dynamical evolution of few-body systems. Although the secular evolution of triples and quadruples has been explored before, a systematic study of the systems in the MSC including also quintuples and sextuples has not been carried out. Here, we explore the main-sequence (MS) evolution of stars from the MSC based on approximately $2\times10^5$ secular dynamical integrations. We estimate statistical probabilities for strong interactions during the MS such as tidal evolution and mass transfer, and the onset of dynamical instability. Depending on the assumed model for the unknown orbital elements, we find that the fraction of noninteracting systems is largest for triples ($\sim 0.9$), and decreases to $\sim 0.6$-$0.8$ for sextuples. The fraction of strong interactions increases from $\sim 0.1$ to $\sim 0.2$ from triples to sextuples, and the fraction of dynamically unstable systems increases from $\sim 0.001$ to $\sim 0.1$-$0.2$. The larger fractions of strong interactions and dynamical instability in systems with increasing multiplicity can be attributed to increasingly complex secular evolution in these systems. Our results indicate that a significant fraction of high-multiplicity systems interact or become dynamically unstable already during the MS, with an increasing importance as the number of stars increases.
\end{abstract}


\section{Introduction}
\label{sect:introduction}
Hierarchical multiple-star systems with $\ns=3$ or more stars are interesting for a number of reasons. Their complex orbital architecture encodes information that can be used to constrain single and binary star formation (e.g., \citealt{1973bmss.book.....B,2011ApJ...734...55P}). Also, they give rise to a rich variety of dynamical evolution, since (hierarchical) orbits in systems with more than two bodies are usually not static. So far, the implications of dynamical evolution in multiple-star systems have been addressed in most detail for triples. In hierarchical triples, Lidov-Kozai (LK) oscillations (\citealt{1962P&SS....9..719L,1962AJ.....67..591K}; see \citealt{2016ARA&A..54..441N} for a review) can drive high eccentricities in the inner orbit on potentially long timescales. High eccentricities can couple with strong tidal evolution, producing short-period binaries (e.g., \citealt{1979A&A....77..145M,2001ApJ...562.1012E,2006Ap&SS.304...75E,2007ApJ...669.1298F,2013MNRAS.430.2262H,2014ApJ...793..137N,2016ComAC...3....6T,2017MNRAS.467.3066A,2018MNRAS.479.4749B,2019MNRAS.488.2480R}), although it has recently been cast into doubt whether this effect is responsible for producing an enhanced population of short-period binaries \citep{2018ApJ...854...44M,2019AJ....158..222T,2020MNRAS.491.5158T}. In addition, LK cycles in triples can couple with stellar evolution, giving rise to strong interactions during or after the main-sequence (MS; e.g., \citealt{2013MNRAS.430.2262H,2016ComAC...3....6T,2016MNRAS.460.3494S,2017ApJ...841...77A,2018A&A...610A..22T,2019ApJ...882...24H}). 

The long-term dynamical evolution becomes more complex as the number of stars is increased. For quadruples, which occur in two long-term stable configurations (`2+2': two binaries orbiting each other, and `3+1': a triple orbited by a fourth body). The secular dynamical evolution of quadruples has been studied by a number of authors \citep{2013MNRAS.435..943P,2015MNRAS.449.4221H,2016MNRAS.461.3964V,2017MNRAS.470.1657H,2018MNRAS.476.4234F,2018MNRAS.474.3547G,2019MNRAS.483.4060L,2019MNRAS.486.4781F}, who have generally found that the efficiency to attain high eccentricities in quadruples is higher compared to equivalent triples, i.e., if two stars would be replaced by a single star. Also, when combined with stellar evolution, quadruples can give rise to a wide range of outcomes \citep{2018MNRAS.478..620H}.

However, even more complex hierarchical systems with five or more bodies are known to exist, and the statistics of high-multiplicity systems are steadily improving (e.g., \citealt{2018AJ....156...48T,2018AJ....156..194T,2019AJ....157...91T,2019AJ....158..222T}). The {\it Multiple Star Catalogue} (MSC; \citealt{1997A&AS..124...75T,2018ApJS..235....6T}) in particular contains information of hierarchical systems with $\ns=3$ up to and including $\ns=7$ stars. The MSC is not based on a volume-limited sample of stars, and is therefore distorted by observational selection effects. However, it has the advantage of being the most comprehensive catalogue of multiple-star systems to date, including information on the component masses, orbital periods, and additional information. In particular, it currently encompasses the largest database of systems with $\ns>4$ stars. 

Although the tools to efficiently study the long-term dynamical evolution of $\ns>4$ systems exist \citep{2016MNRAS.459.2827H,2018MNRAS.476.4139H}, there is no study to date that explores these systems. In this work, we address this omission and consider the dynamical evolution of multiple systems with $3 \leq \ns \leq 6$ from the MSC\footnote{Since the number of hierarchical $\ns=7$ systems in the MSC with all periods and masses known or estimated is only 2, we ignore these systems in our statistical considerations.}. Based on secular dynamical integrations of approximately $2\times10^5$ systems, we study the probability of interactions during the (shortest) MS lifetime of the stars, including strong interactions inducing tidal evolution and possibly mass transfer, and the onset of dynamical instability of the system. Our focus is on the probability and delay-time of such interactions, and their dependence on $\ns$. These quantities give insight into the importance and efficiency of the decay of $\ns\geq3$ systems after their formation, and during their MS lifetime. A detailed study of the evolution and outcomes following such interactions, as well as the post-MS evolution, is left for future work. 

This paper is structured as follows. In \S~\ref{sect:meth}, we describe our methodology of extracting data from the MSC and sampling the unknown orbital parameters. In \S~\ref{sect:results}, we show a number of examples, and give our main results of the interaction fractions, as well as focus on the orbital distributions. We give a discussion in \S~\ref{sect:discussion}, and conclude in \S~\ref{sect:conclusions}.

\section{Methodology}
\label{sect:meth}
\subsection{Initial conditions}
\label{sect:meth:IC}

We extract data of systems with $\ns=3$ to $\ns=6$ stars from the MSC\footnote{The MSC database can be downloaded at \href{http://www.ctio.noao.edu/~atokovin/stars/}{http://www.ctio.noao.edu/~atokovin/stars/}. For this work, we retrieved the database on February 7 2020 (the database was last updated on July 25 2019).} \citep{1997A&AS..124...75T,2018ApJS..235....6T}. We restrict to hierarchical systems and ignore `trapezium'-type systems (indicated in the MSC with the symbol `t'), for which the separations between resolved components are comparable and it cannot be established if the system is hierarchical. Also, we ignore systems in which not all orbital periods $P$ and component masses $m_i$ are known or estimated\footnote{Long orbital periods in the MSC are estimated from projected separations and are therefore estimates of the true orbital period; the difference can be a factor of up to $\sim 3$ \citep{2014AJ....147...86T}.}, but do not otherwise impose a cut on these quantities. The typical primary star in the MSC has a mass between 0.5 and 3 $\msun$. In the MSC, the spectral types of all components are only known in a subset of systems. To maximise the number of available systems, we do not reject systems with known giant stars. Therefore, the `MS lifetime' on which we base the duration of our integrations (see \S~\ref{sect:meth:num} below) can be an overestimate of the true remaining lifetime of the components in the system. Also, stellar evolution is beyond the scope of this paper and is ignored in our integrations.

\begin{figure}
\center
\includegraphics[scale = 0.42, trim = 0mm 0mm 0mm 10mm]{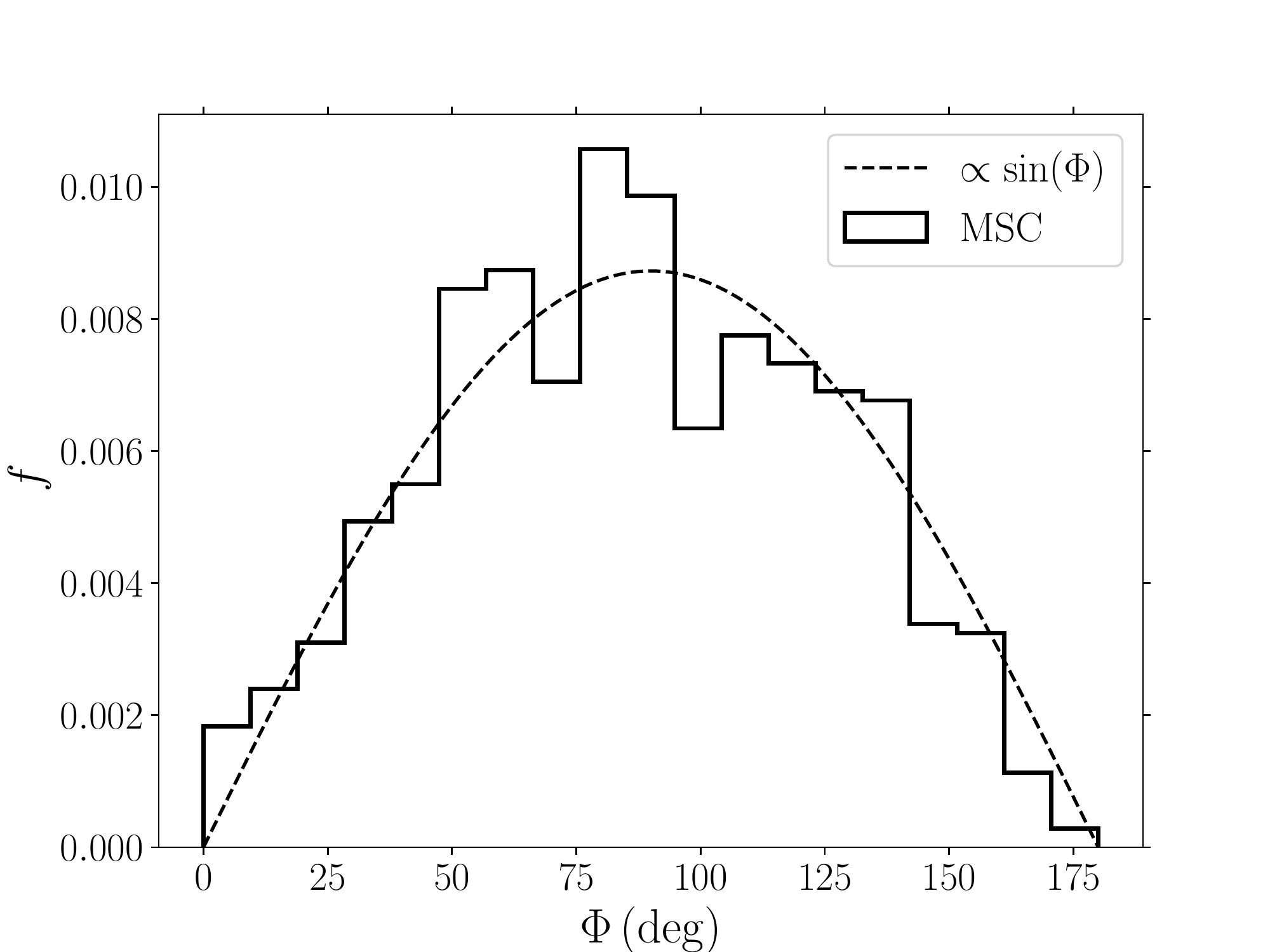}
\caption{ Normalised distribution of the mutual inclination $\Phi$ in the MSC (solid black lines), based on data in the \texttt{orb} file where orbital orientation data is available. The black dashed line shows a curve $\propto \sin (\Phi)$, which corresponds to an isotropic distribution. }
\label{fig:MSC_mut_incls}
\end{figure}

With these assumptions, we thus reduce the MSC to 1841 hierarchical systems, subdivided into 1504 triples, 281 quadruples, 48 quintuples, and 8 sextuples. For each orbit in the system, we compute the semimajor axis from the orbital period and total enclosed orbital mass using Kepler's law. In addition, we use the \texttt{orb} file from the MSC to extract the orbital eccentricity $e$ where available (evidently, $e$ is not known for all orbits in all systems). The \texttt{orb} file also contains information on the orbital orientations (inclinations $i$, arguments of periapsis $\omega$, and longitudes of the ascending node $\Omega$). However, in many cases, the orbital elements are derived from visual orbits which do not distinguish between the two orbital nodes, leaving a $180^\circ$ ambiguity in the longitude of the ascending node, $\Omega$ (e.g., \citealt{2017ApJ...844..103T}). Therefore, we choose not to use the orbital orientation data from the MSC, but instead sample from assumed distributions (see below). Nevertheless, we remark that the available orientation data in the MSC yield a distribution of the mutual inclination which is consistent with isotropic orientations (see \F~\ref{fig:MSC_mut_incls}), and the latter case is included in our models (see below). 

In some cases, multiple entries exist in the MSC for the orbital properties of the same orbit corresponding to different observations. In those cases, we adopt the mean eccentricity of these multiple entries if they are not zero\footnote{Zero values in the MSC indicate unknown quantities.}. 

For each selected system from the MSC (with at least some missing orbital elements), we generate $\nmc$ additional realisations in which we sample the unknown orbital parameters from assumed distributions. Since the number of systems in the MSC with a given number of stars decreases as $\ns$ increases, we set $\nmc$ depending on $\ns$ in order to obtain reasonable statistics also for systems with large $\ns$. Of course, since we sample the unknown orbital elements only, increasing $\nmc$ does not increase the statistical information on the masses and orbital periods, i.e., the statistics are still bound by the limited number of known systems with large $\ns$. We set $\nmc = 40$ for systems with $\ns=3$ or $\ns=4$; for $\ns=5$, we set $\nmc = 200$; for $\ns=6$, we set $\nmc = 400$. Note that from these systems, we reject those that do not satisfy our initial requirements (see below).

\begin{table}
\begin{tabular}{cp{3.5cm}p{3cm}}
\toprule
Model & Eccentricities $e$ & Mutual inclinations $\Phi$ \\
\midrule
A & Flat & Isotropic \\
B & Flat & Low $\Phi$ for $a<50\,\au$; isotropic for $a\geq50\,\au$ \\
C & Sine function for $P<100\,\yr$; thermal for $P\geq100\,\yr$ & Isotropic \\
D & Sine function for $P<100\,\yr$; thermal for $P\geq100\,\yr$ & Low $\Phi$ for $a<50\,\au$; isotropic for $a\geq50\,\au$ \\
\bottomrule
\end{tabular}
\caption{ Summary of the four different models for the unknown orbital parameters in our Monte Carlo simulations. }
\label{table:models}
\end{table}

\begin{figure}
\center
\includegraphics[scale = 0.42, trim = 0mm 0mm 0mm 10mm]{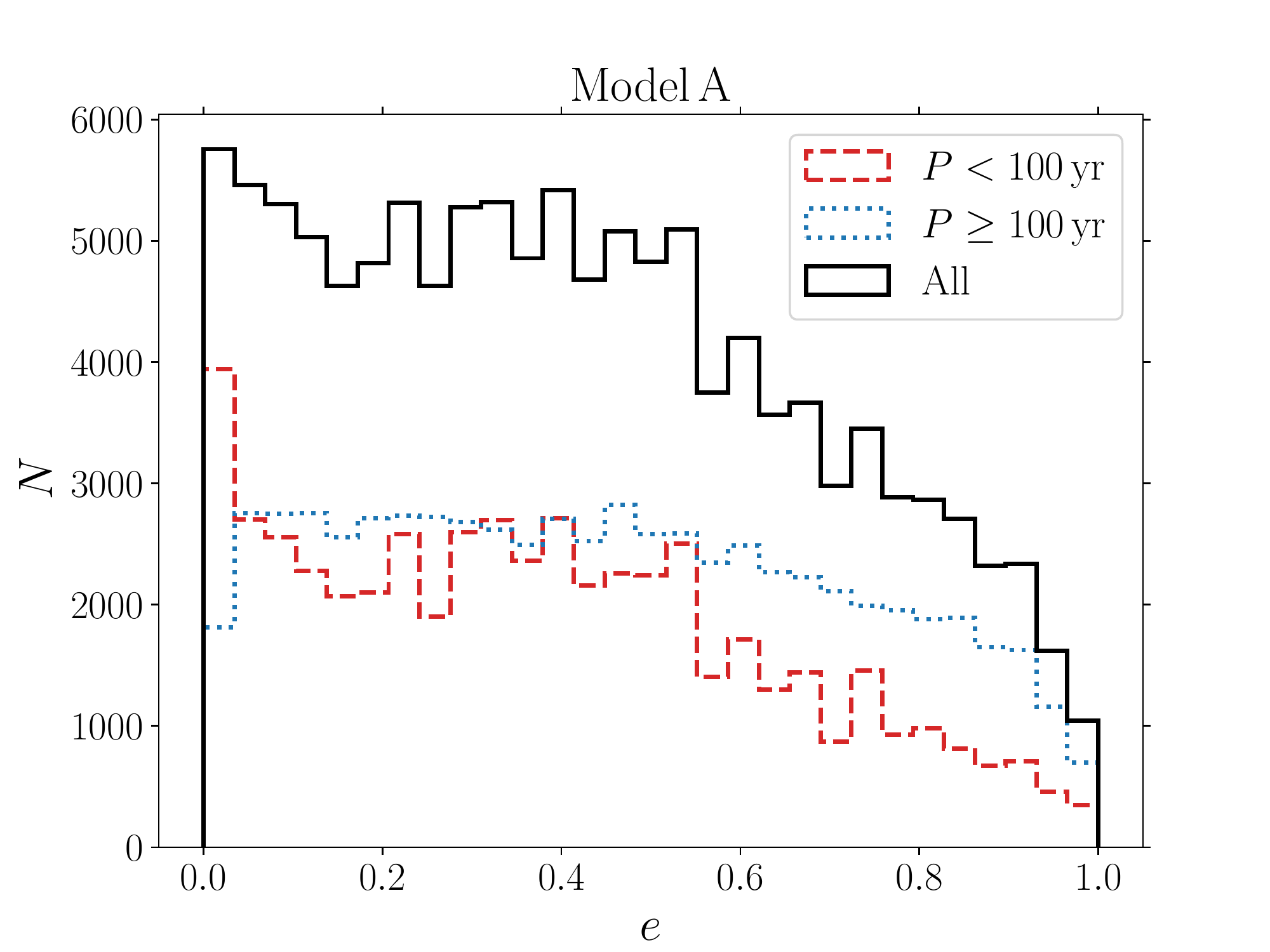}
\includegraphics[scale = 0.42, trim = 0mm 0mm 0mm 10mm]{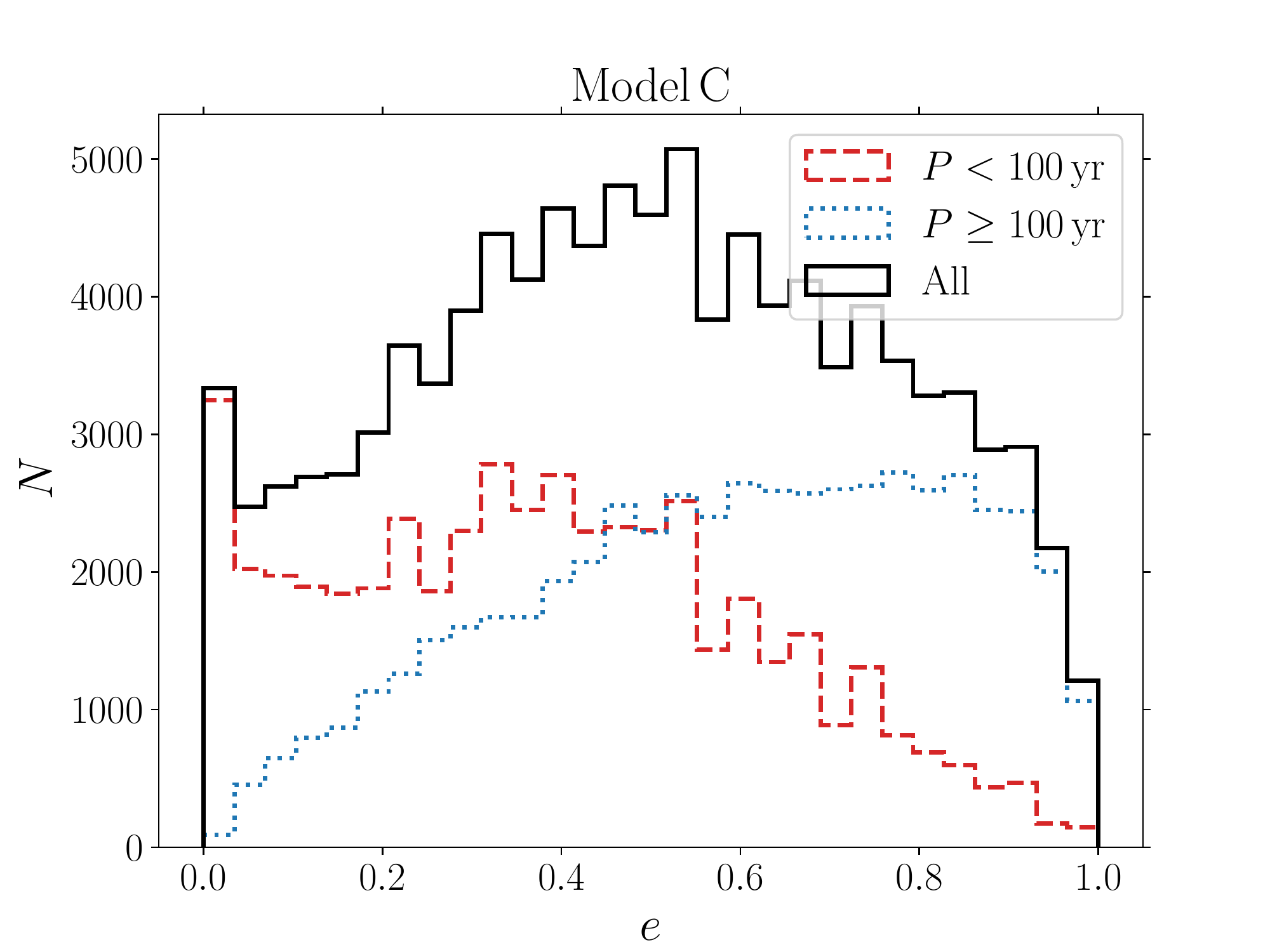}
\caption{ Distributions of orbital eccentricities in our models. Top panel: Model A (the same distribution applies to Model B). Bottom panel: Model C (also applies to Model D). We show the distributions separately for systems with orbital periods $P<100\,\yr$ (red dashed lines) and $P\geq 100\,\yr$ (blue dotted lines); solid black lines show distributions for all orbits. }
\label{fig:IC:e}
\end{figure}

\begin{figure}
\center
\includegraphics[scale = 0.42, trim = 0mm 10mm 0mm 10mm]{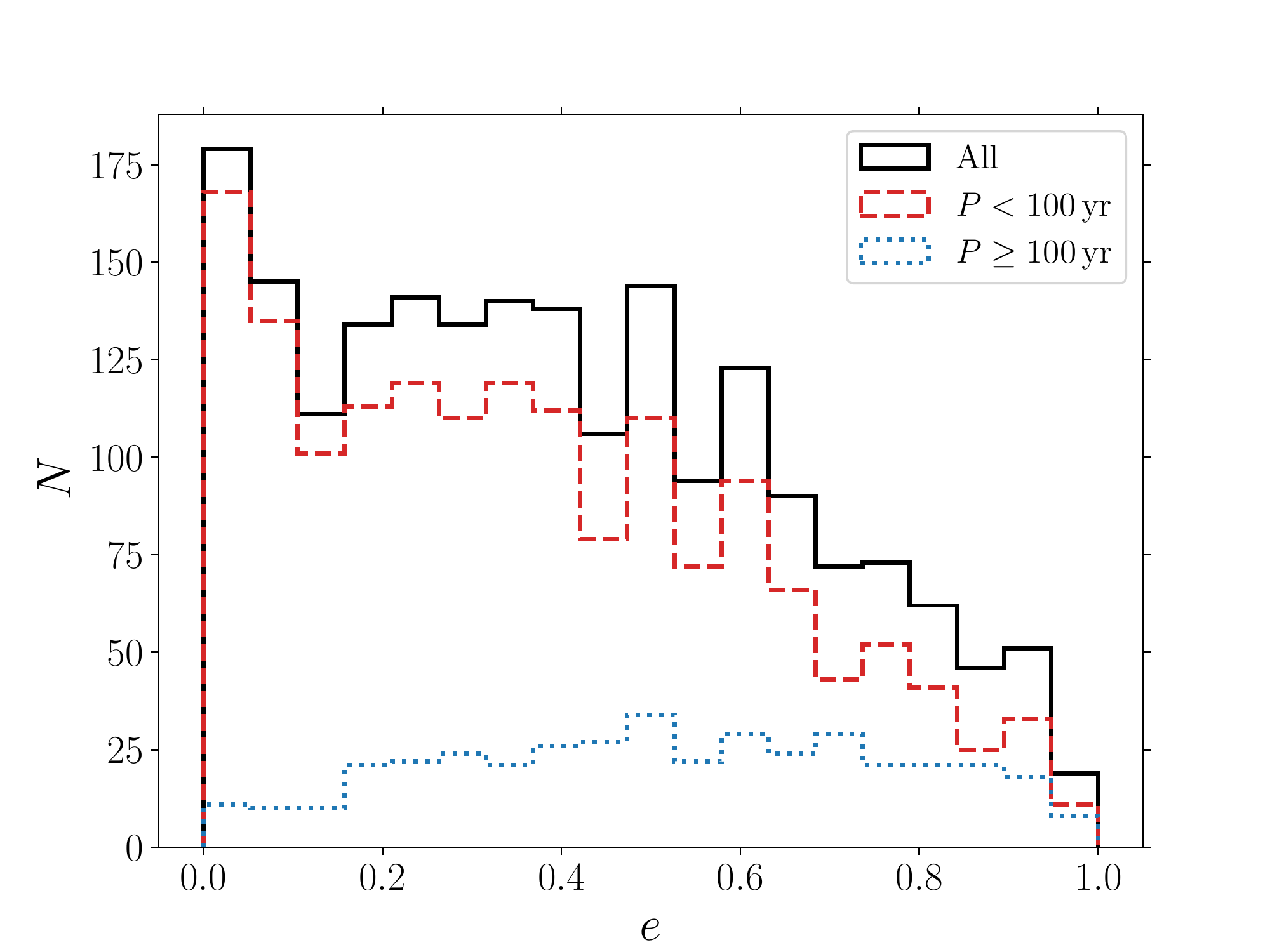}
\caption{Black solid line: distribution of the known eccentricities of all orbits in the MSC of systems with any number of stars and with known masses and orbital periods. The red dashed (blue dotted) lines show the eccentricity distributions for orbits with $P<100\,\yr$ ($P\geq 100 \, \yr$). }
\label{fig:MSC_eccentricities}
\end{figure}

To sample the unknown orbital elements, we adopt four different models, Models A through D. An overview with a brief description of these models is given in Table~\ref{table:models}. 

In models A and B, we assume a flat distribution of the eccentricities ($0.01 < e < 0.99$), subject to dynamical stability constraints (see \S~\ref{sect:meth:num} below for the latter). The resulting eccentricity distributions in models A and B are shown in the top panel of \F~\ref{fig:IC:e}; they are a reasonable first-order approximation to the observed distribution for multiple stars of \citet{2010ApJS..190....1R}, as well as to the eccentricities in the MSC (see \F~\ref{fig:MSC_eccentricities}). 

It should be noted, however, that catalogues such as those from \citet{2010ApJS..190....1R} and the MSC rely on visual orbits for wide binaries, and visual orbits are systematically biased against large eccentricities. In fact, statistical analyses show that wide binaries ($P>100\,\yr$) tend to have on average higher eccentricities than tighter binaries \citep{2016MNRAS.456.2070T}. This is also indicated by the distributions of the available eccentricities in the MSC (see \F~\ref{fig:MSC_eccentricities}), which show typically higher eccentricities for orbits with $P>100\,\yr$. In our Models C and D, we therefore assume a period-dependent eccentricity distribution that approximates the observed distribution of \citet{2016MNRAS.456.2070T}. Specifically, we sample the eccentricity from a sine-distribution $\mathrm{d}N/\mathrm{d} e \propto \sin(\pi e)$ for $P<100\,\yr$, and $\mathrm{d}N/\mathrm{d} e \propto e$ (a `thermal' distribution) for $P\geq100\,\yr$. In both cases, sampled eccentricities are subject to dynamical stability constraints. The resulting distribution is shown in the bottom panel of \F~\ref{fig:IC:e}. As shown, models C and D have typically higher eccentricities compared to models A and B. 

Next, we discuss the orbital orientations ($i,\omega$, and $\Omega$). In models A and C, we assume uniform distributions in the cosines of the inclinations $i$ (ranging between $-1$ and $+1$), and flat distributions in $\omega$ and $\Omega$ (ranging between 0 and $2\pi$). This choice of orbital angles corresponds to isotropic orbital distributions. However, observations suggest that the orbital alignment in triple stars is not isotropic \citep{1993AstL...19..383T,2002A&A...384.1030S}, and, in particular, is correlated with the separation of the tertiary companion \citep{2017ApJ...844..103T}. \citet{2017ApJ...844..103T} found that tight triples with inner semimajor axes less than about 50 $\au$ tend to be aligned in their orbits (with a mean mutual inclination of $\Phi \sim 20^\circ$), whereas triples with wider inner orbits tend to have isotropic relative orbital orientations. 

Although the correlation found by \citet{2017ApJ...844..103T} applies strictly to triple stars, we here (naively) extrapolate it to any hierarchy. To mimic the observational trend of \citet{2017ApJ...844..103T}, in models B and D, we sample the {\it individual} orbital inclination $i$ of any orbit with $a<50\,\au$ from a Gaussian distribution with a mean of $\mu=0^\circ$ and a dispersion of $\sigma=20^\circ$, with individual inclinations restricted to the range $0^\circ< i < 30^\circ$. The angles $\omega$ and $\Omega$ are sampled from flat distributions between 0 and $2\pi$, as in Model A. For orbits with $a<50\,\au$, these choices lead to a smeared-out distribution in the {\it mutual} inclination between orbits, $\Phi$, since, in general for two orbits labeled 1 and 2,
\begin{align}
\cos \Phi = \cos i_1 \cos i_2 + \sin i_1 \sin i_2 \cos(\Omega_1-\Omega_2).
\end{align}
For any orbits with $a\geq50\,\au$, we sample the orbital orientation corresponding to a random orientation, as in Model A for all separations. 

\begin{figure}
\center
\includegraphics[scale = 0.42, trim = 0mm 10mm 0mm 10mm]{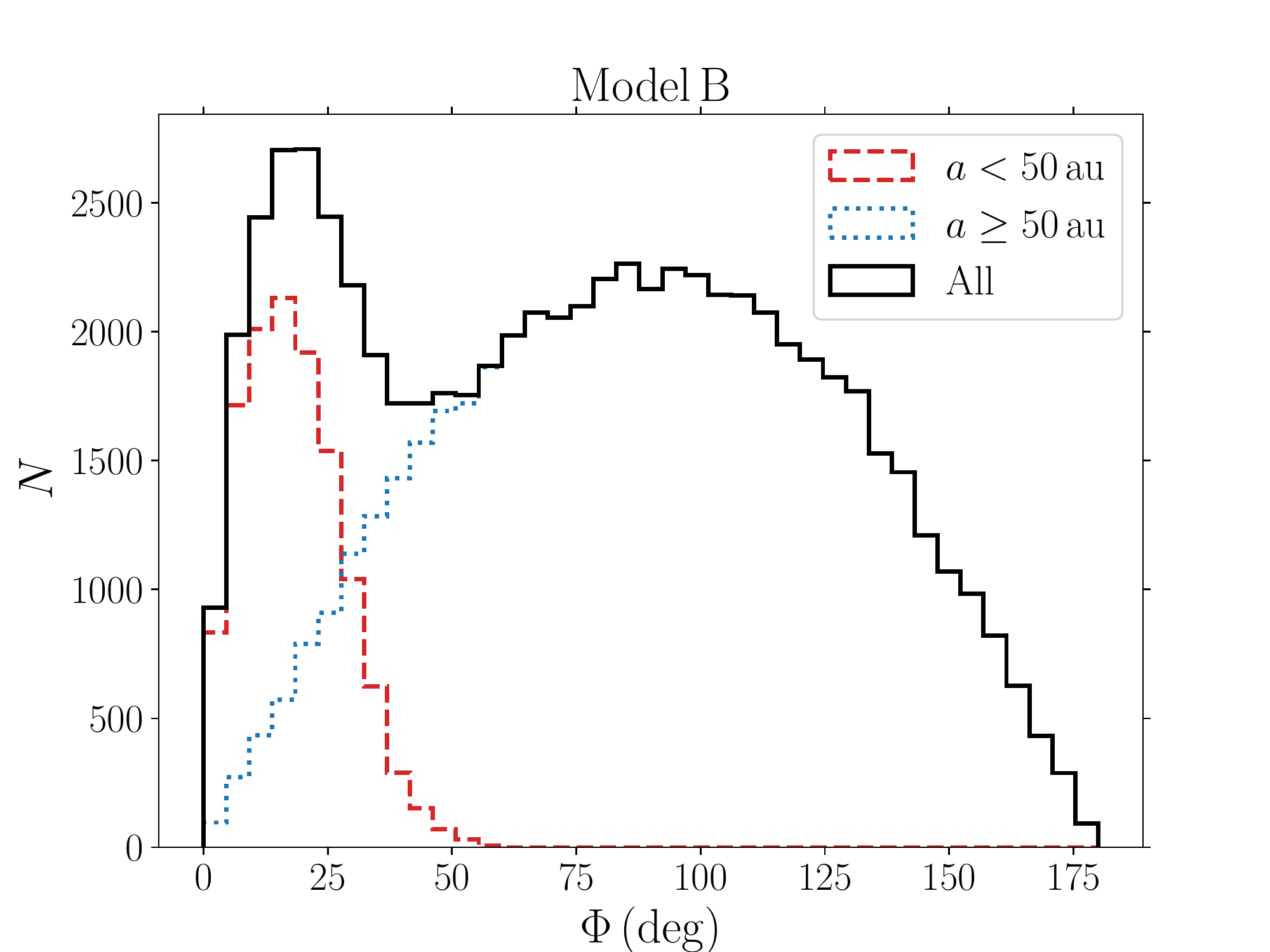}
\caption{Distributions of all initial mutual inclinations in Model B (also applies to Model D; in models A and C, the distribution is $\mathrm{d} N/\mathrm{d} \Phi \propto \sin \Phi$). We distinguish between orbits with $a<50\,\au$ (red dashed line) and $a\geq 50\,\au$ (blue dotted line), as well as show the distribution for all orbits (black solid line). }
\label{fig:IC:orient}
\end{figure}

To illustrate our choices for the orbital orientations, we show in \F~\ref{fig:IC:orient} the distributions for models B and D of all initial mutual inclinations $\Phi$ (evidently, a system with $\ns>3$ has multiple values of $\Phi$). We distinguish between orbits with $a<50\,\au$ (red dashed line) and $a\geq 50\,\au$ (blue dotted line), and also show the distribution for all orbits (black solid line). As required, the distribution of $\Phi$ for $a<50\,\au$ peaks near $\Phi \sim 20^\circ$, and is isotropic ($\mathrm{d} N/\mathrm{d} \Phi \propto \sin \Phi$) for larger separations. The combined distribution shows a large peak around $\Phi=20^\circ$. 

We reject a generated system if it is expected to interact strongly, or be dynamically unstable at the beginning (i.e., before any secular dynamical evolution). We refer to \S~\ref{sect:meth:num} for our definitions of these events.

\subsection{Numerical integrations}
\label{sect:meth:num}
Using the procedure described in \S~\ref{sect:meth:IC}, we obtain $\nsys=51,057$, $51,044$, $44,864$, and $44,850$ realisations of multiple systems in the MSC for models A, B, C, and D, respectively. We subsequently integrate these systems using the secular dynamical code \textsc{SecularMultiple} \citep{2016MNRAS.459.2827H,2018MNRAS.476.4139H} which is freely available\footnote{\href{https://github.com/hamers/secularmultiple}{https://github.com/hamers/secularmultiple}}. \textsc{SecularMultiple} is based on an expansion of the Hamiltonian of the hierarchical system in terms of ratios of orbital separations $x_i$. The Hamiltonian is subsequently averaged, and the orbit-averaged equations of motion are solved numerically. The main advantage of \textsc{SecularMultiple} over `traditional' direct $N$-body codes is that the former captures the long-term secular dynamical evolution (at least approximately) whereas the latter are generally much slower, such that a Monte Carlo study of the scope of this paper would be computationally prohibitively expensive. 

Here, we include Newtonian terms up to and including fifth order (dotriacontupole) in $x_i$ for pairwise interactions, and up to including third order (octupole order) for interactions involving three orbits simultaneously. We also include post-Newtonian (PN) corrections to the first PN order (giving rise to orbital precession) in all orbits, ignoring direct PN interactions between orbits (see, e.g., \citealt{2013ApJ...773..187N,2020arXiv200103654L} for the latter). 

We do not include tidal evolution in the secular equations of motion. Tidal dissipation is still poorly understood, especially at high orbital eccentricity (see, e.g., \citealt{2014ARA&A..52..171O} for a review). Instead, we adopt a simplistic approach in which we consider strong interactions such as tidal evolution, possibly followed by mass transfer, to occur when the periapsis distance of any orbit containing two stars satisfies 
\begin{align}
\label{eq:rpmin}
r_{\mathrm{p},i} < 3 (R_1+R_2),
\end{align}
where $R_1$ and $R_2$ are the radii of the stars in orbit $i$. Here, we assume MS radii for the stars, which we calculate according to\footnote{\label{note1}See \href{https://www.astro.ru.nl/~onnop/education/stev\_utrecht\_notes/}{https://www.astro.ru.nl/~onnop/education/stev\_utrecht\_notes/}.}
\begin{align}
R_i = \left ( \frac{m_i}{\msun} \right )^{0.7} \, \rsun,
\end{align}
with $m_i$ the mass of the star (as reported in the MSC). As discussed above in \S~\ref{sect:meth:IC}, systems that initially satisfy \eq~(\ref{eq:rpmin}) are rejected during the sampling procedure.  We neglect changes of the stellar radii during stellar evolution (see, e.g., \citealt{2013MNRAS.430.2262H,2016ComAC...3....6T,2016MNRAS.460.3494S,2017ApJ...841...77A,2018A&A...610A..22T,2019ApJ...882...24H} for studies of triples including stellar evolution). A comprehensive investigation of the post-MS evolution of systems in the MSC is left for future work. 

During the integrations, we check for dynamical instability of the system. Dynamical instability can occur as a result of high eccentricities induced by secular evolution, in particular in higher-order systems such as quadruples (e.g., \citealt{2017MNRAS.466.4107H,2019MNRAS.482.2262H}). We use the stability criterion of \citet{2001MNRAS.321..398M}, i.e., 
\begin{align}
\frac{a_\mathrm{out}(1-e_\mathrm{out})}{a_\mathrm{in}} > 2.8 \, \left [ (1+q_\mathrm{out}) \frac{1+e_\mathrm{out}}{\sqrt{1-e_\mathrm{out}}} \right ]^{2/5} \, \left (1-0.3\, \frac{\Phi}{\pi} \right )
\end{align}
for stability, and where `in' and `out' refer to the inner and outer orbits, respectively, $q_\mathrm{out} = m_3/(m_1+m_2)$ is the tertiary-to-inner-binary mass ratio, and the mutual inclination $\Phi$ is expressed in radians. This criterion applies strictly only to hierarchical triples, but, for lack of a generalised criterion for higher-multiplicity systems, we extrapolate it to any hierarchical system. Specifically, we apply the criterion to any orbit pair, with the masses appropriately applied (for example, for a 2+2 quadruple, the criterion is applied twice to the two inner orbits, with the `tertiary' mass given by the total mass of the companion binary). We also check for initial dynamical instability using the same procedure, and reject initially dynamically unstable systems (cf. \S~\ref{sect:meth:IC}). 

We determine the integration time of each system by the shortest MS lifetime of its stars, the approximate age of the Galaxy, and, for technical reasons\footnote{In some systems, the LK timescale can be extremely short compared to the shortest MS lifetime or the age of the Galaxy; the simulations in these cases can be very computationally demanding because of the large number of secular oscillations, especially in systems with large $\ns$.}, a maximum number of secular oscillations. Specifically, the integration time of each system, $t_\mathrm{end}$ is set according to
\begin{align}
\label{eq:tend}
t_\mathrm{end} = \underset{i,j}{\mathrm{min}}\left [ t_{\mathrm{MS},i}, t_\mathrm{H}, \alpha t_{\mathrm{LK},j} \right ],
\end{align}
where $t_\mathrm{H} \equiv 10\,\Gyr$, $i$ runs over all bodies in the system, and $j$ runs over all orbits which contain at least one component that itself is an orbit. We approximate the MS timescale for each star $i$ with\footref{note1}
\begin{align}
t_{\mathrm{MS},i} = 10\, \left (\frac{m_i}{\msun} \right )^{-2.8} \, \Gyr,
\end{align}
and we estimate the LK timescale as (e.g., \citealt{1997AJ....113.1915I,2015MNRAS.452.3610A,2016MNRAS.459.2827H})
\begin{align}
t_{\mathrm{LK},j} = \frac{P_j^2}{P_{j,\mathrm{child}}} \frac{M_j}{m_{j,\mathrm{sibling}}} \left (1-e_j^2 \right )^{3/2}.
\end{align}
Here, $P_{j,\mathrm{child}}$ is the orbital period of the corresponding child of orbit $j$, $M_j$ is the total mass of all bodies contained within orbit $j$, and $m_{j,\mathrm{sibling}}$ is the mass of the sibling of the corresponding child in orbit $j$. We set $\alpha=10^4$, i.e., at least on the order of $10^4$ secular oscillations are included in our simulations, unless they exceed the shortest MS lifetime or $t_\mathrm{H}$.

\section{Results}
\label{sect:results}

\begin{figure*}
\center
\includegraphics[scale = 0.43, trim = 0mm 0mm 0mm 10mm]{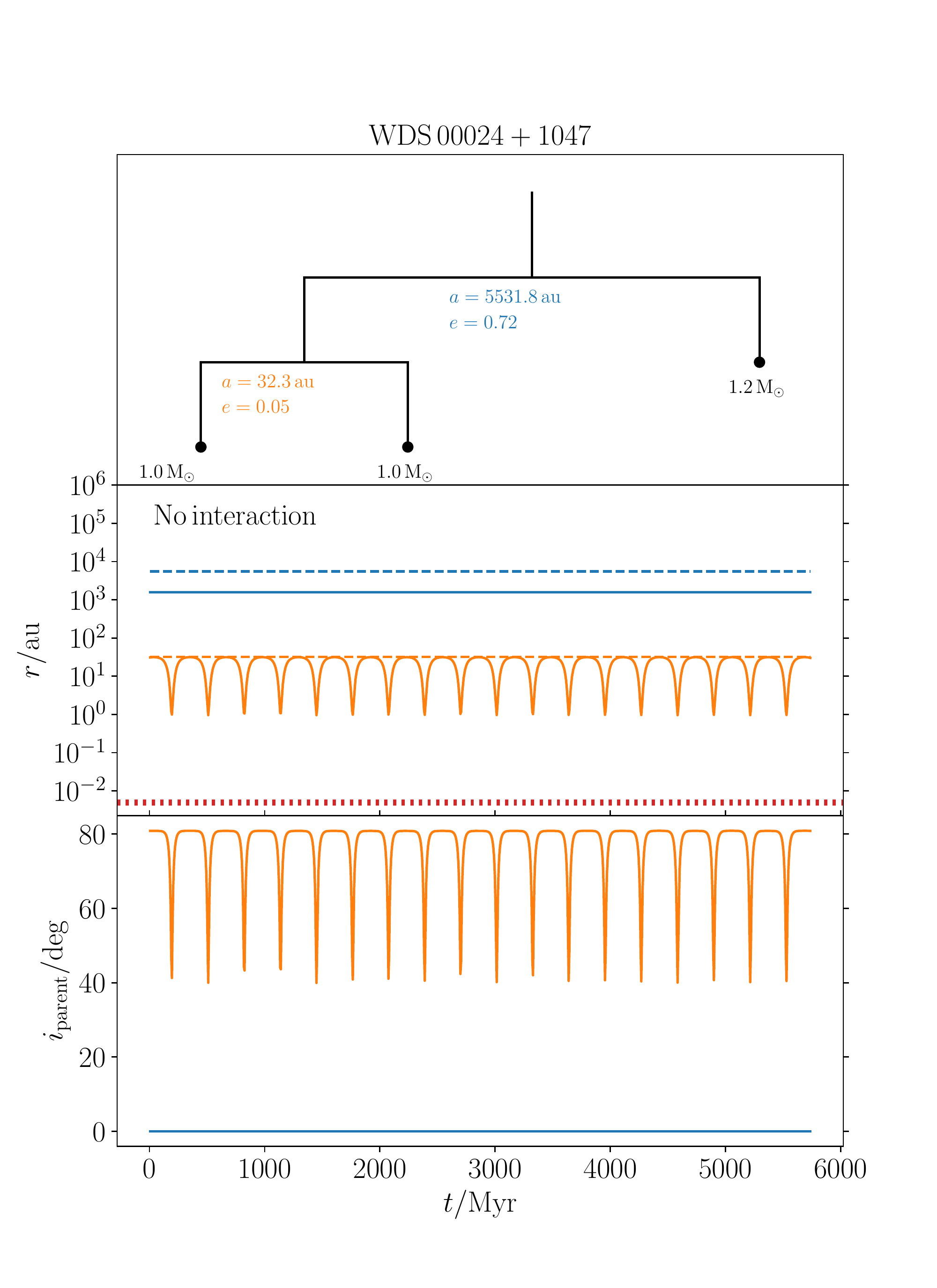}
\includegraphics[scale = 0.43, trim = 0mm 0mm 0mm 10mm]{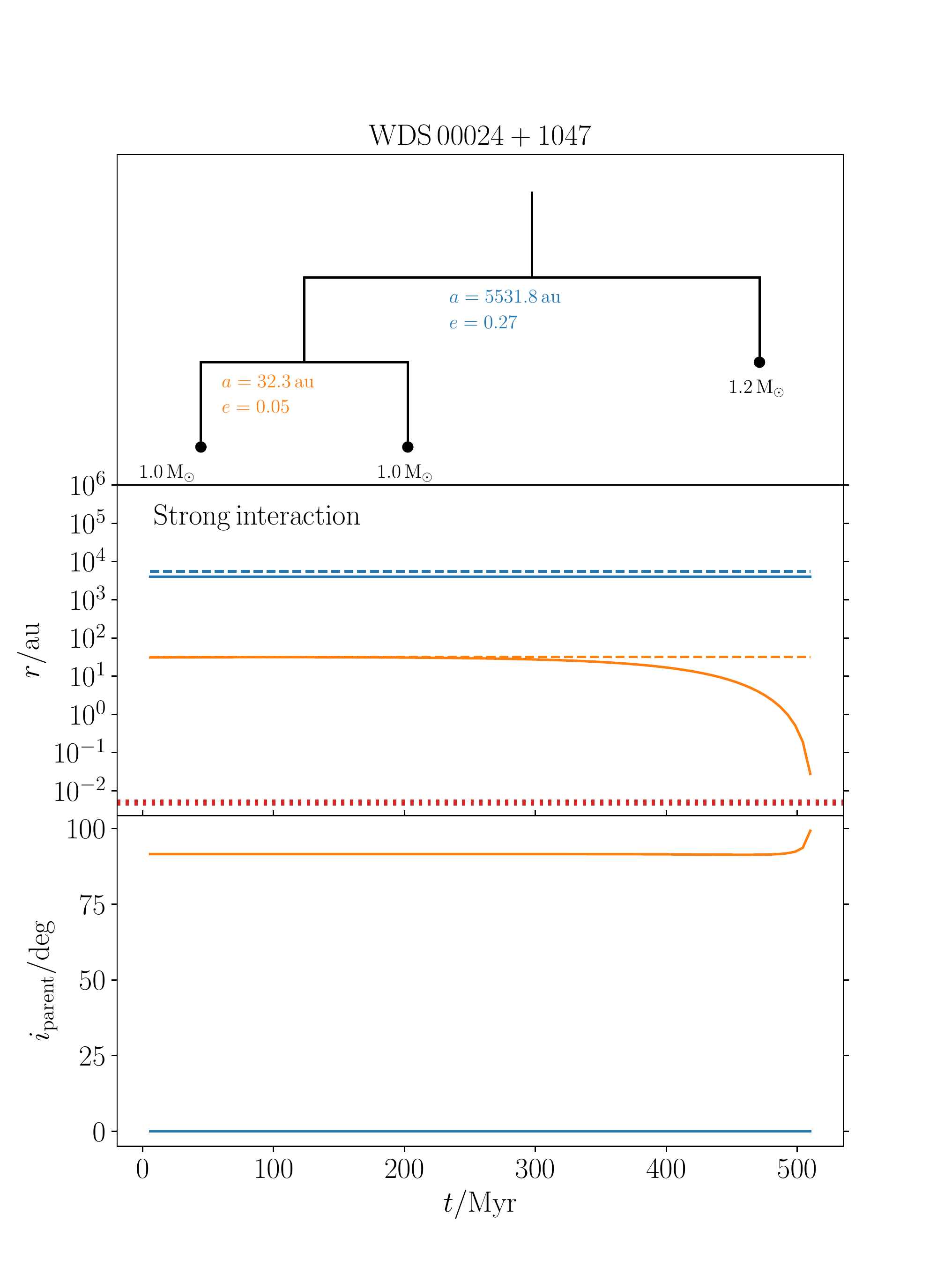}
\includegraphics[scale = 0.43, trim = 0mm 10mm 0mm 15mm]{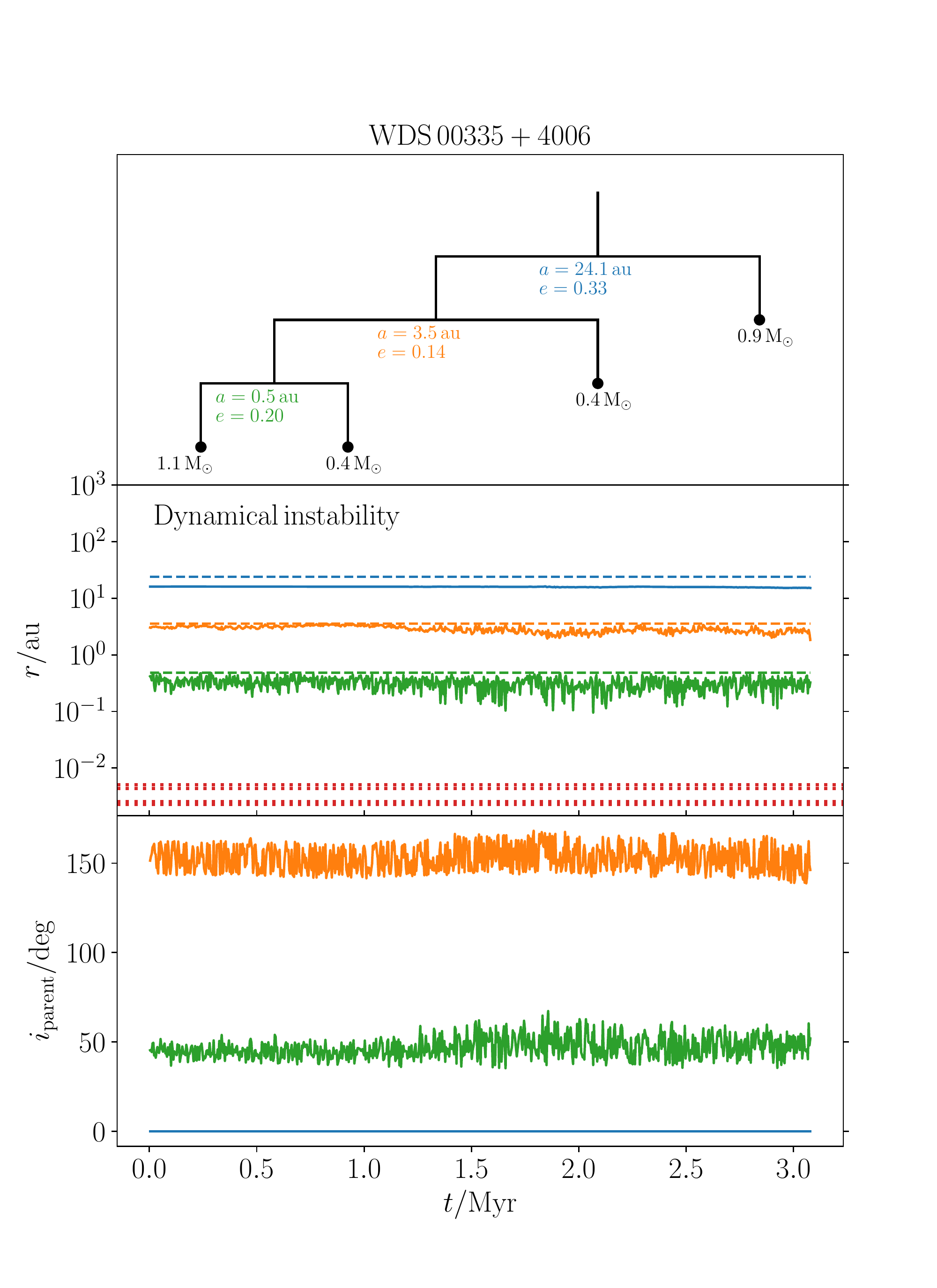}
\includegraphics[scale = 0.43, trim = 0mm 10mm 0mm 15mm]{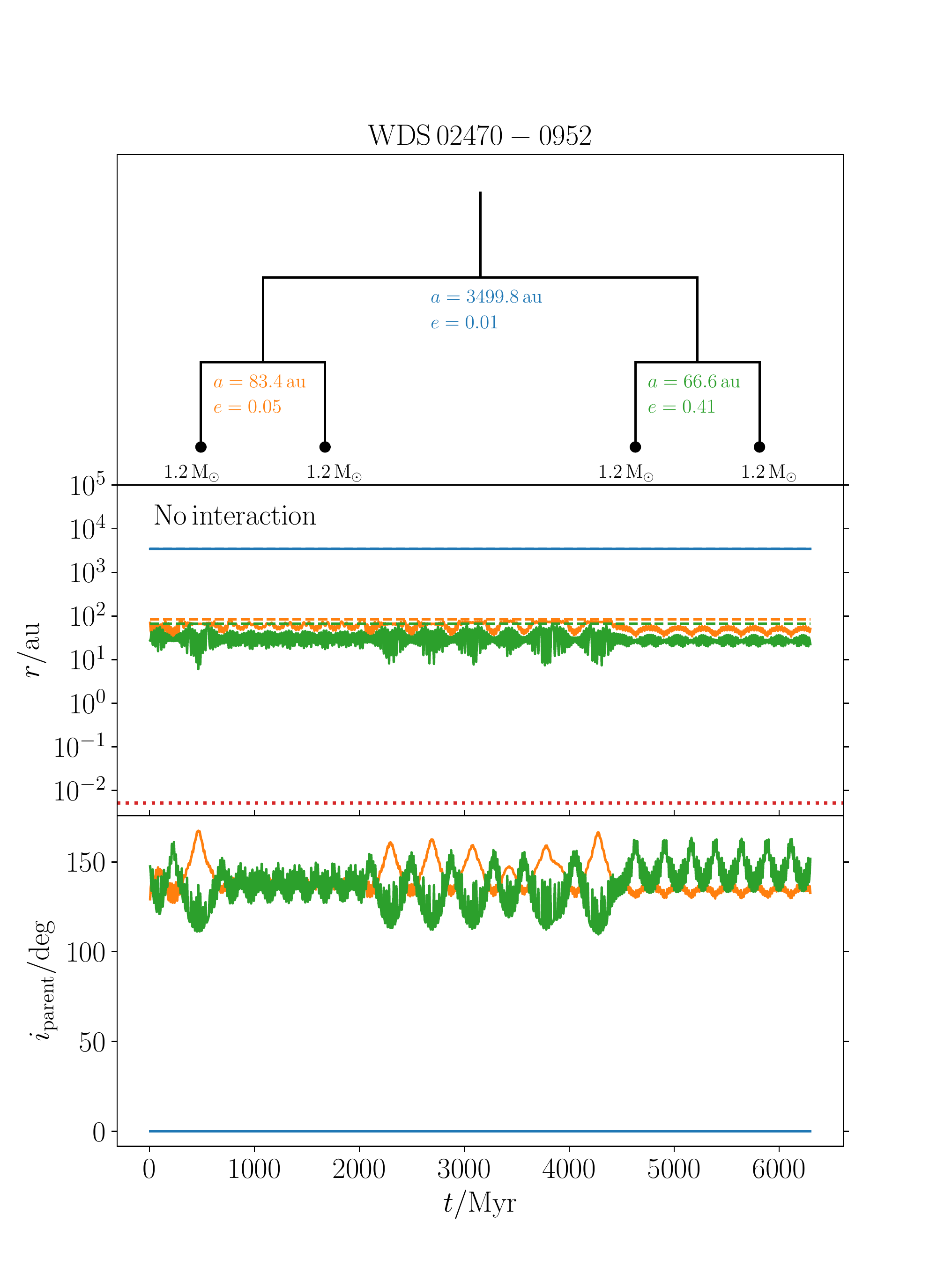}
\caption{ Example evolution of two triples and two quadruples; the WDS name is indicated for each system at the top. Top panels: mobile diagrams. Values of the semimajor and eccentricities are indicated, as well as the masses of the components. Middle panels: semimajor axes (dashed lines) and periapsis distances (solid lines) of all orbits (the colors correspond to the orbits in the mobile diagrams); the horizontal red dotted lines indicate the stellar radii. Bottom panels: inclinations of all orbits relative to their parent orbit, if applicable. The outcome of each system in the simulations is indicated at the top-left part of the middle panels. }
\label{fig:ex1}
\end{figure*}

\begin{figure*}
\center
\includegraphics[scale = 0.43, trim = 0mm 0mm 0mm 10mm]{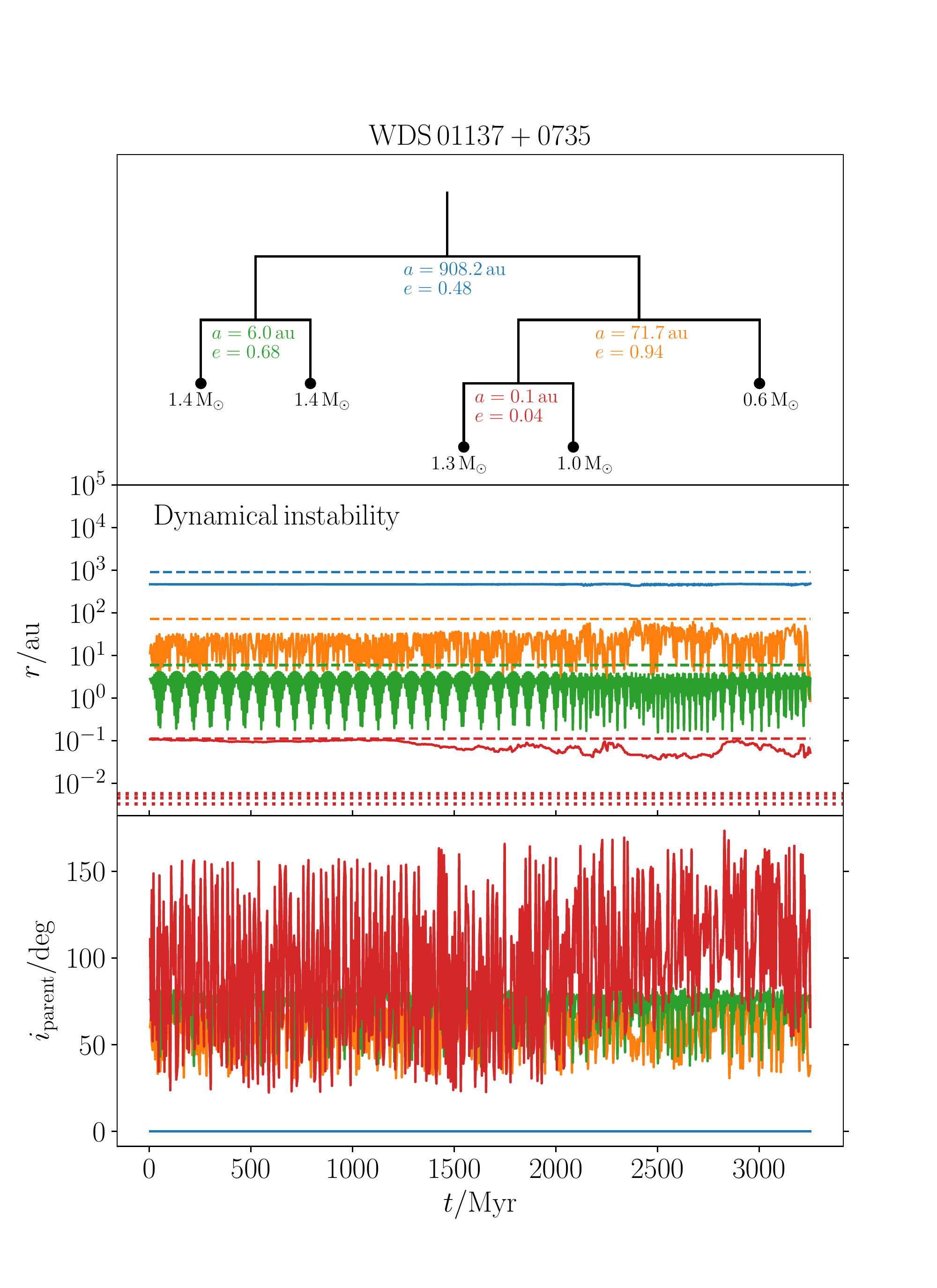}
\includegraphics[scale = 0.43, trim = 0mm 0mm 0mm 10mm]{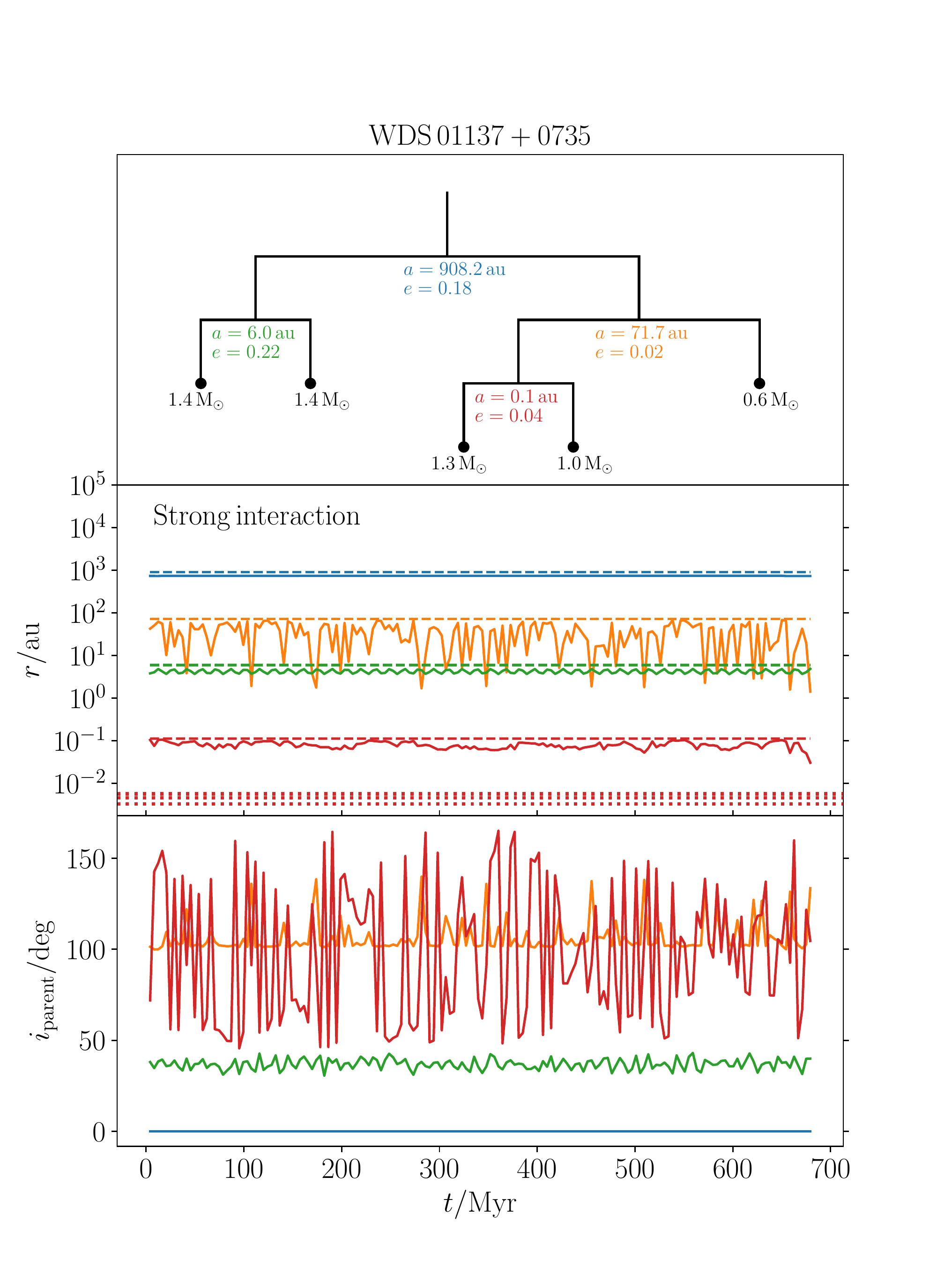}
\includegraphics[scale = 0.43, trim = 0mm 10mm 0mm 15mm]{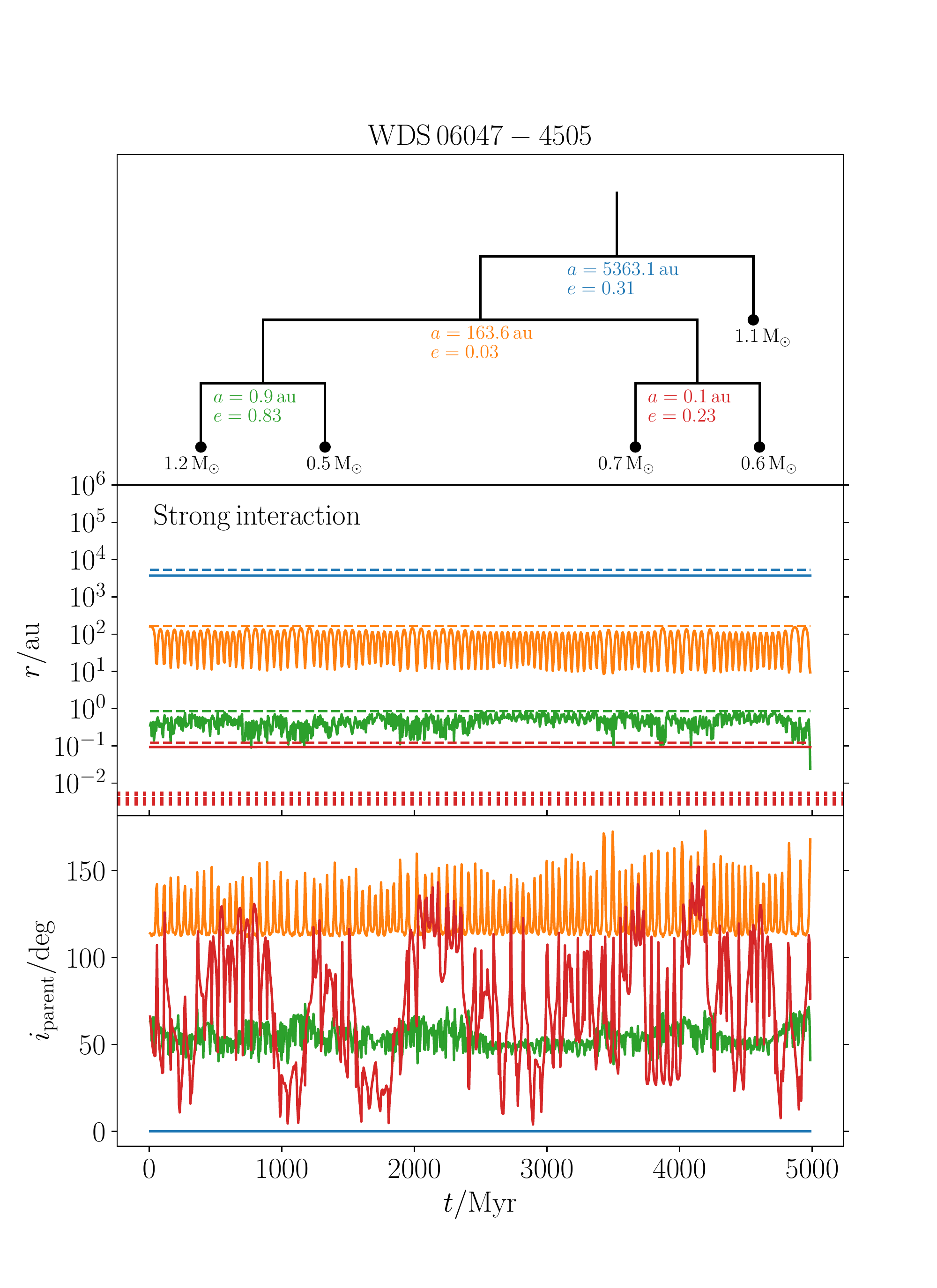}
\includegraphics[scale = 0.43, trim = 0mm 10mm 0mm 15mm]{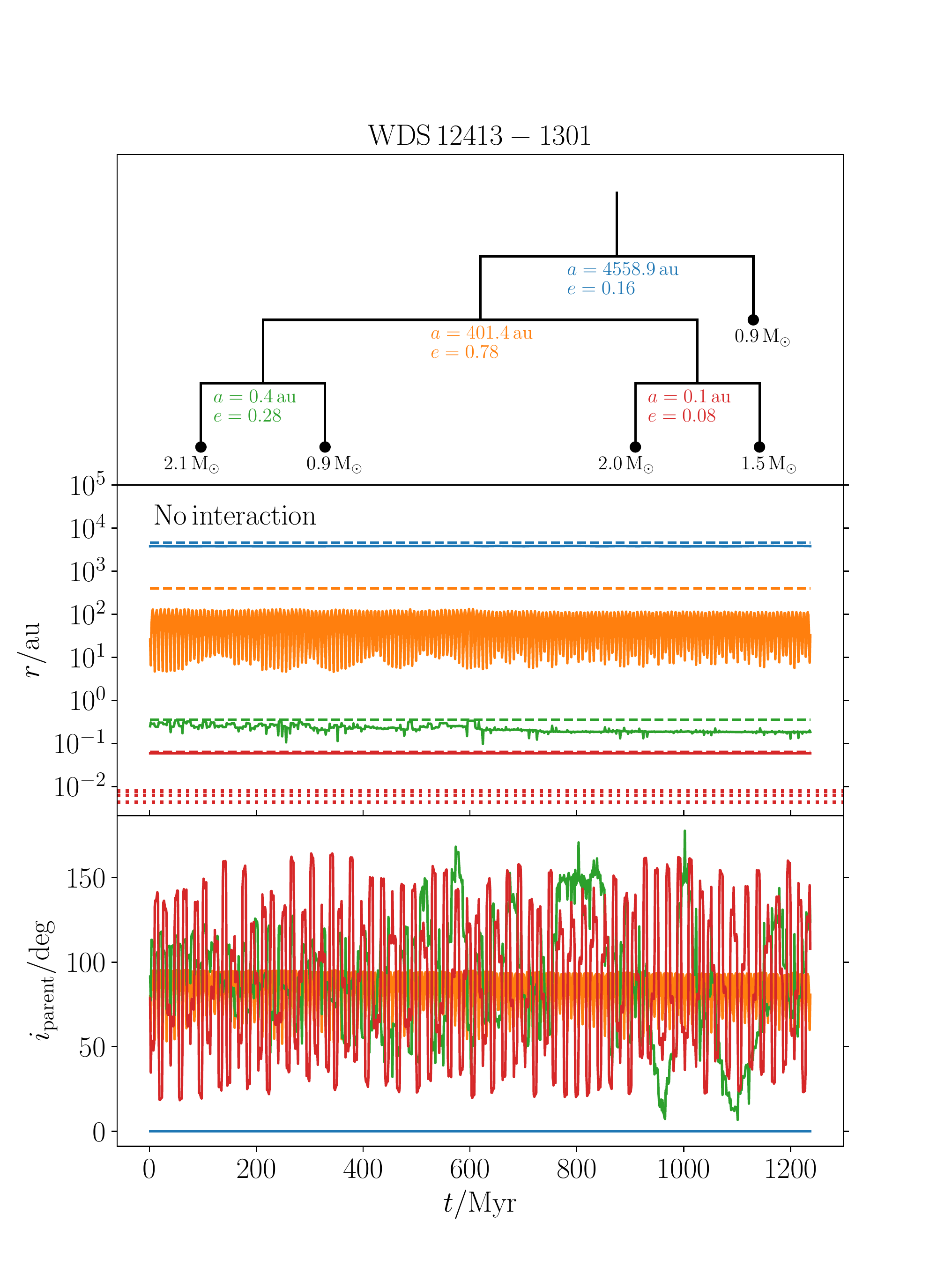}
\caption{ Further examples as in \F~\ref{fig:ex1}, here showing four quintuple systems. }
\label{fig:ex2}
\end{figure*}

\begin{figure*}
\center
\includegraphics[scale = 0.43, trim = 0mm 0mm 0mm 10mm]{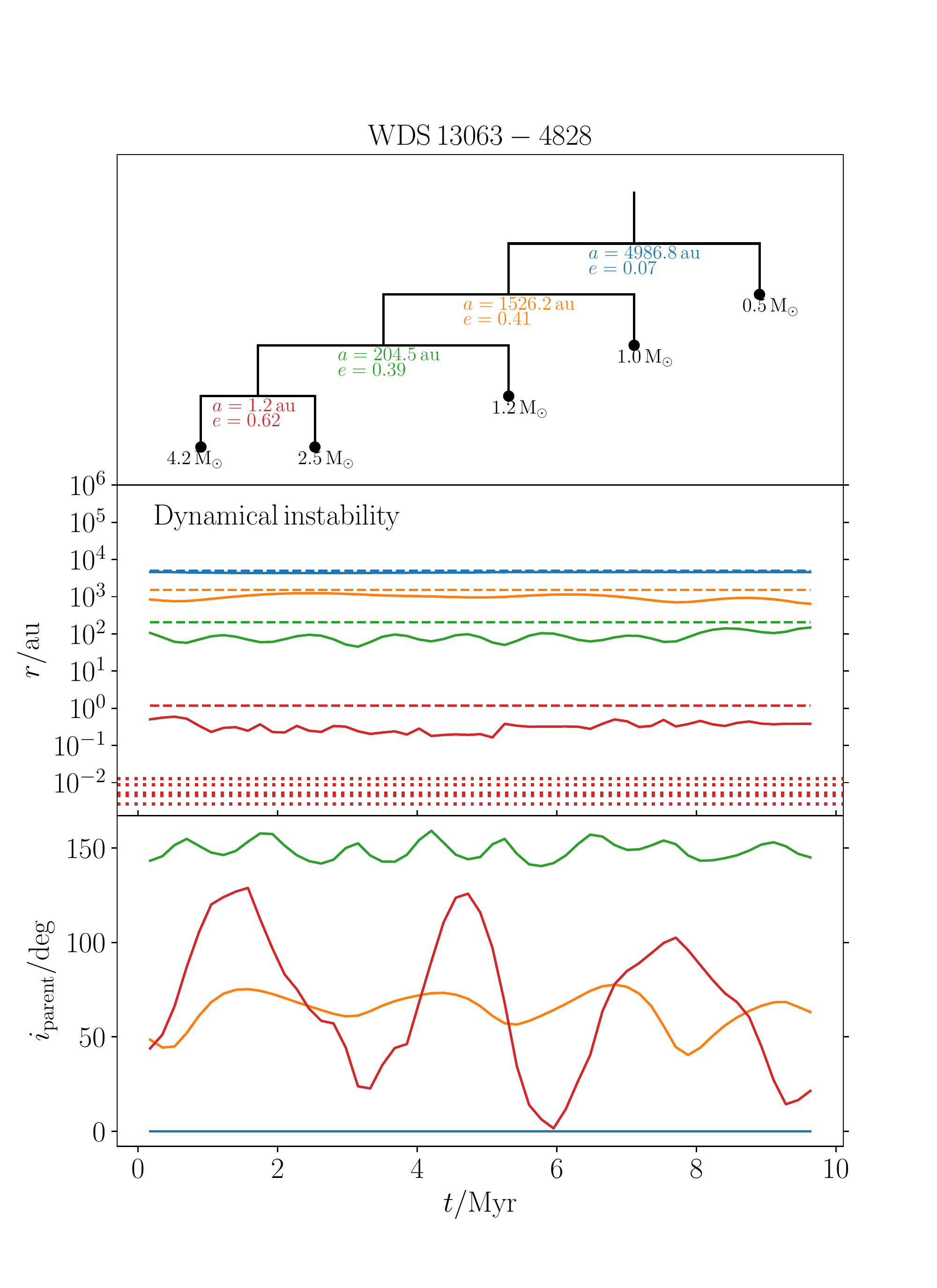}
\includegraphics[scale = 0.43, trim = 0mm 0mm 0mm 10mm]{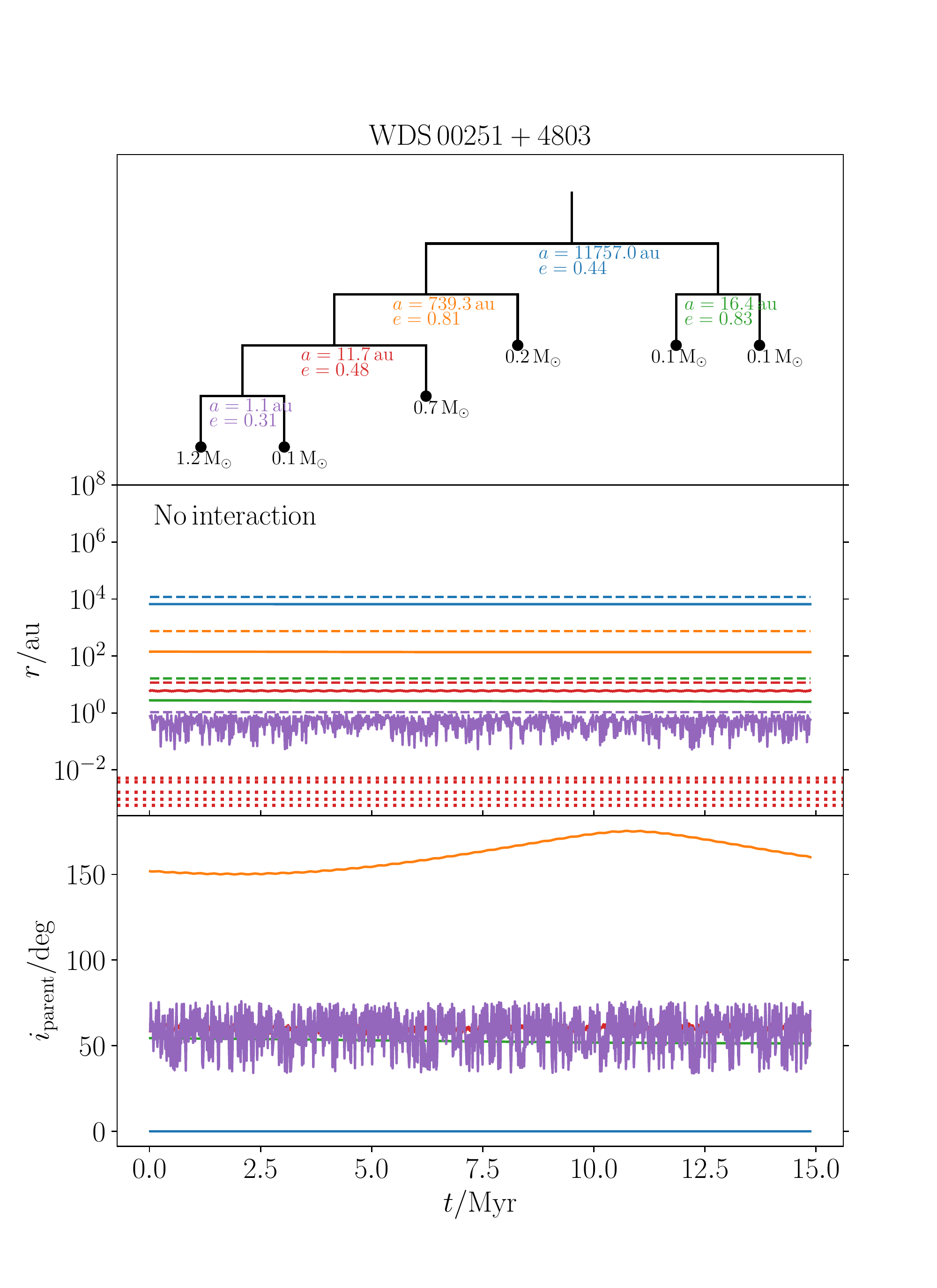}
\includegraphics[scale = 0.43, trim = 0mm 10mm 0mm 15mm]{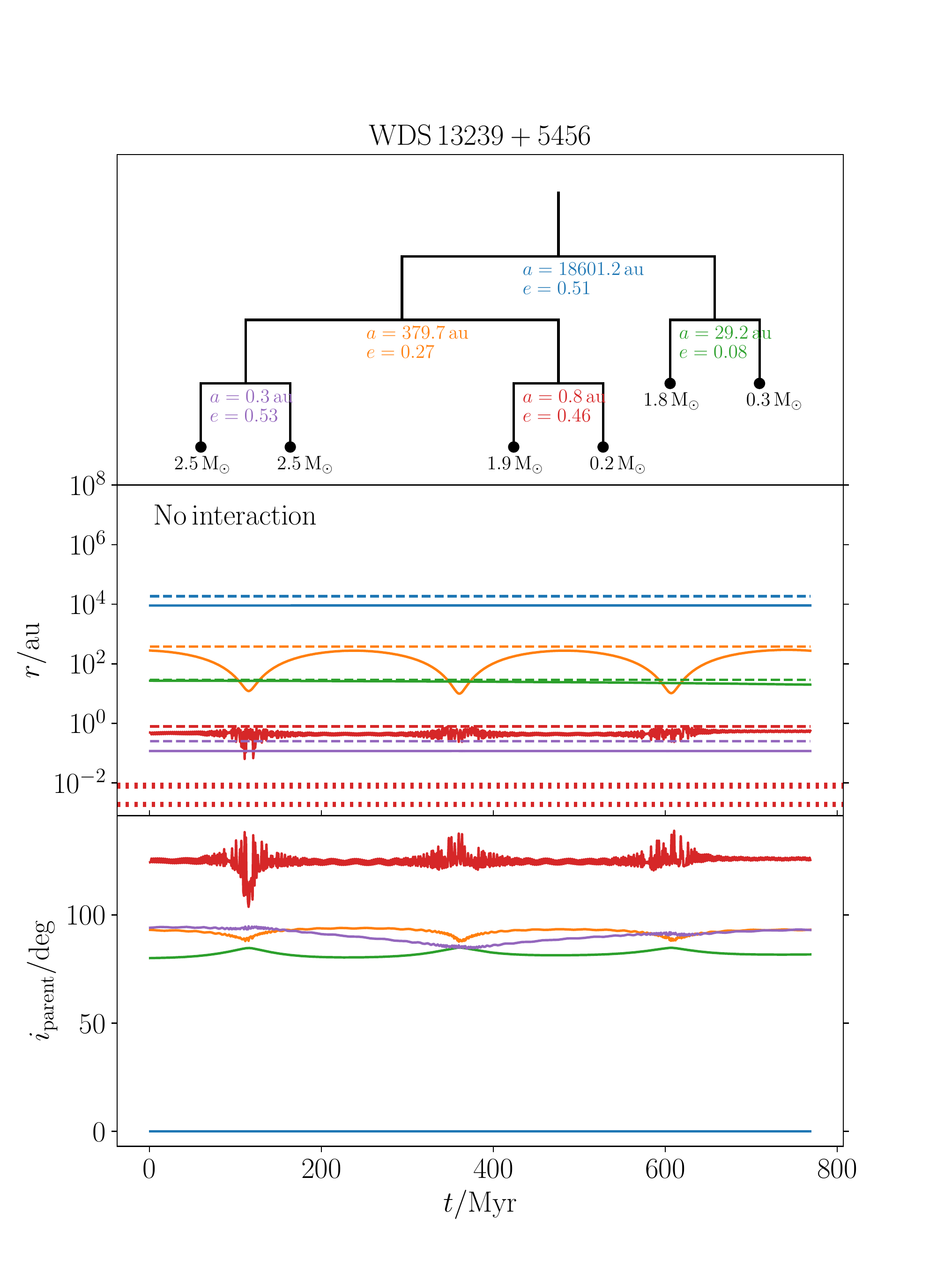}
\includegraphics[scale = 0.43, trim = 0mm 10mm 0mm 15mm]{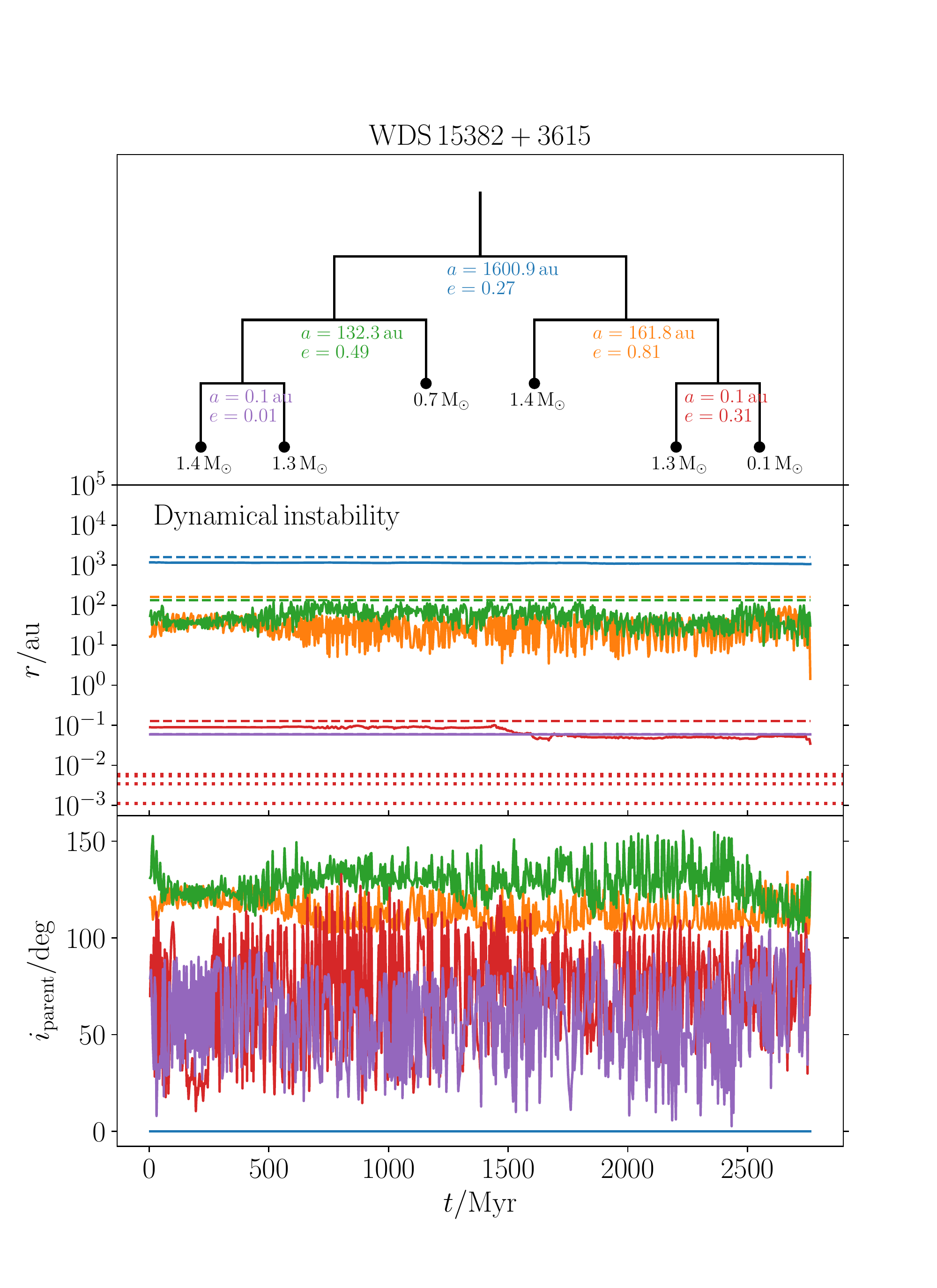}
\caption{ Further examples as in \F~\ref{fig:ex2}, here showing one quintuple, and three sextuple systems. }
\label{fig:ex3}
\end{figure*}

\subsection{Examples}
\label{sect:results:ex}
We first show the time evolution of a few systems to illustrate possible secular dynamical evolution for different types of systems. All the examples discussed in this section were selected from integrations in Model A. \F~\ref{fig:ex1} highlights the evolution of two triple and two quadruple systems. For each system, the top panels indicate the hierarchy and basic orbital properties of the system in terms of a mobile diagram \citep{1968QJRAS...9..388E}. The middle panels show the semimajor axes (dashed lines) and periapsis distances (solid lines) of all orbits (the colors correspond to the orbits in the mobile diagrams); the horizontal red dotted lines indicate the stellar radii. The bottom panels give the inclinations of all orbits relative to their parent orbit, if applicable. The outcome of each system in the simulations (see \S~\ref{sect:meth:num}, and \S~\ref{sect:results:frac} below) is indicated at the top-left part of the middle panels.

The examples of WDS 00024+1047 show prototypical evolution of highly inclined triples, where in one case the eccentricity reached is sufficiently high to trigger a strong interaction after $\sim 500\,\Myr$, whereas in the other case no interaction occurs during the $\sim 6\,\Gyr$ MS lifetime of the system. The example of the 3+1 quadruple WDS 00335+4006 illustrates the onset of dynamical instability, triggered by enhanced eccentricity of the intermediate orbit arising from the secular torque of the outermost orbit. The 2+2 quadruple WDS 02470-0952 shows coupled evolution between the two inner orbits, which is expected based on the similar LK timescales of both inner orbits \citep{2017MNRAS.470.1657H}.

\F~\ref{fig:ex2} illustrates the evolution of four quintuples. In one realisation of the quintuple WDS 01137+0735, a triple orbited by a binary, dynamical instability is triggered when the orbit with $a\approx72\,\au$ becomes highly eccentric. The innermost orbit, with $a=0.1\,\au$, evolves in a complicated, quasi-random fashion. Also, note that the orbit with $a=6\,\au$ does not behave completely regularly on long timescales, which may be due to the 2+2 quadruple-interaction with the orbit with $a\approx 72\,\au$. In another realisation of WDS 01137+0735 with different orbital eccentricities and orientations, the innermost orbit with $a=0.1\,\au$ becomes highly eccentric more rapidly, leading to a strong interaction after $\sim 680 \, \Myr$. WDS 06047-4505 is a 2+2 quadruple orbited by a fifth body. The outer orbit of the quadruple, with $a\approx 164\,\au$, is excited to high eccentricities by the fifth body, which affects the 2+2 quadruple. After $\sim5\,\Gyr$, a strong interaction is triggered in one of the innermost orbits. WDS 12413-1301 is also a 2+2+1 system. The innermost orbits in this system are tight (both have $a<0.5\,\au)$; nonetheless, one of the innermost orbits is excited in eccentricity as the orbit with $a\approx 401\,\au$ is driven to high eccentricity by the outermost fifth body.

Lastly, \F~\ref{fig:ex3} shows one quintuple, and three sextuples. WDS 13063-4828 is a `fully-nested' quintuple, i.e., it has a planetary system-like architecture (note, however, that the orbits can potentially be highly mutually inclined, unlike the Solar system or many exoplanet systems). In the realisation shown in \F~\ref{fig:ex3}, a dynamical instability is triggered relatively early when the orbit with $a\approx 1526\,\au$ is excited in eccentricity. The sextuple WDS 00251+4803 has a similar structure as the previous quintuple WDS 13063-4828, except that the outermost body is now itself a binary. In this particular example, there is no strong secular evolution in all orbits except for the innermost orbit, although a strong interaction is avoided. The sextuple WDS 13239+5456, consisting of a 2+2 quadruple orbited by a binary, shows strongly coupled evolution between the orbit with $a=0.8\,\au$ and with $a\approx 380\,\au$. The latter orbit is driven to high eccentricity by the outermost binary. The orbit with $a=0.3\,\au$ is not strongly affected, likely because secular oscillations are quenched in this tighter orbit by PN precession (e.g., \citealt{2002ApJ...578..775B,2007ApJ...669.1298F,2015MNRAS.447..747L}). In the sextuple WDS 15382+3615, consisting of two triples orbiting each other, a dynamical instability is triggered by excited eccentricity of the orbit with $a\approx 162\,\au$ by secular evolution. The innermost orbits, although compact, show some secular evolution in a complicated way, correlated with the eccentricity of their parent orbits. Interestingly, the `companion triple' to the triple that becomes dynamically unstable in the example shown has similar orbital properties, and evolves in a similar way. In fact, in other realisations of the system (not shown here), we find that dynamical instability can be triggered starting in the other triple system (with the intermediate orbit $a\approx 132\,\au$).

\begin{table}
\begin{tabular}{lccc}
\toprule
\multicolumn{4}{c}{Model A} \\
\midrule
%$\ns$ & $\nsys$ & \multicolumn{3}{c}{$f$}  \\
$\ns$ & $\fnon$ & $\fint$ & $\fdyn$ \\
\midrule
3 & $0.913 \pm 0.005$ & $0.086 \pm 0.001$ & $0.001 \pm 0.000$  \\
4 & $0.781 \pm 0.011$ & $0.194 \pm 0.006$ & $0.025 \pm 0.002$  \\
5 & $0.673 \pm 0.014$ & $0.262 \pm 0.008$ & $0.065 \pm 0.004$  \\
6 & $0.608 \pm 0.024$ & $0.205 \pm 0.014$ & $0.187 \pm 0.013$  \\
\midrule
\midrule
\multicolumn{4}{c}{Model B} \\
\midrule
$\ns$ & $\fnon$ & $\fint$ & $\fdyn$ \\
\midrule
3 & $0.963 \pm 0.005$ & $0.036 \pm 0.001$ & $0.002 \pm 0.000$  \\
4 & $0.866 \pm 0.012$ & $0.111 \pm 0.004$ & $0.023 \pm 0.002$  \\
5 & $0.843 \pm 0.015$ & $0.093 \pm 0.005$ & $0.064 \pm 0.004$  \\
6 & $0.775 \pm 0.027$ & $0.072 \pm 0.008$ & $0.153 \pm 0.012$  \\
\midrule
\midrule
\multicolumn{4}{c}{Model C} \\
\midrule
$\ns$ & $\fnon$ & $\fint$ & $\fdyn$ \\
\midrule
3 & $0.905 \pm 0.005$ & $0.093 \pm 0.002$ & $0.001 \pm 0.000$  \\
4 & $0.784 \pm 0.012$ & $0.189 \pm 0.006$ & $0.027 \pm 0.002$  \\
5 & $0.669 \pm 0.015$ & $0.281 \pm 0.010$ & $0.049 \pm 0.004$  \\
6 & $0.619 \pm 0.029$ & $0.238 \pm 0.018$ & $0.143 \pm 0.014$  \\
\midrule
\multicolumn{4}{c}{Model D} \\
\midrule
$\ns$ & $\fnon$ & $\fint$ & $\fdyn$ \\
\midrule
3 & $0.961 \pm 0.005$ & $0.037 \pm 0.001$ & $0.002 \pm 0.000$  \\
4 & $0.876 \pm 0.013$ & $0.109 \pm 0.005$ & $0.016 \pm 0.002$  \\
5 & $0.866 \pm 0.017$ & $0.083 \pm 0.005$ & $0.051 \pm 0.004$  \\
6 & $0.784 \pm 0.032$ & $0.082 \pm 0.010$ & $0.134 \pm 0.013$  \\
\bottomrule
\end{tabular}
\caption{Outcome fractions of the systems in our Monte Carlo integrations for each of the four models. We distinguish between no interaction during the MS ($\fnon$), a strong interaction ($\fint$), and dynamical instability ($\fdyn$). Poisson errors are given for each fraction. }
\label{table:fractions}
\end{table}

\begin{table}
\begin{tabular}{lccc}
\toprule
\multicolumn{4}{c}{Model A} \\
\midrule
%$\ns$ & $\nsys$ & \multicolumn{3}{c}{$f$}  \\
$\ns$ & $\fnon$ & $\fint$ & $\fdyn$ \\
\midrule
3 & $0.993 \pm 0.005$ & $0.006 \pm 0.000$ & $0.000 \pm 0.000$  \\
4 & $0.929 \pm 0.013$ & $0.063 \pm 0.003$ & $0.008 \pm 0.001$  \\
5 & $0.918 \pm 0.019$ & $0.061 \pm 0.005$ & $0.021 \pm 0.003$  \\
6 & $0.825 \pm 0.032$ & $0.070 \pm 0.009$ & $0.106 \pm 0.011$  \\
\midrule
\midrule
\multicolumn{4}{c}{Model B} \\
\midrule
$\ns$ & $\fnon$ & $\fint$ & $\fdyn$ \\
\midrule
3 & $0.994 \pm 0.005$ & $0.006 \pm 0.000$ & $0.000 \pm 0.000$  \\
4 & $0.950 \pm 0.013$ & $0.045 \pm 0.003$ & $0.005 \pm 0.001$  \\
5 & $0.935 \pm 0.017$ & $0.044 \pm 0.004$ & $0.021 \pm 0.003$  \\
6 & $0.896 \pm 0.031$ & $0.032 \pm 0.006$ & $0.072 \pm 0.009$  \\
\midrule
\midrule
\multicolumn{4}{c}{Model C} \\
\midrule
$\ns$ & $\fnon$ & $\fint$ & $\fdyn$ \\
\midrule
3 & $0.993 \pm 0.005$ & $0.007 \pm 0.000$ & $0.000 \pm 0.000$  \\
4 & $0.933 \pm 0.015$ & $0.058 \pm 0.004$ & $0.009 \pm 0.001$  \\
5 & $0.925 \pm 0.021$ & $0.053 \pm 0.005$ & $0.022 \pm 0.003$  \\
6 & $0.853 \pm 0.040$ & $0.076 \pm 0.012$ & $0.071 \pm 0.011$  \\
\midrule
\multicolumn{4}{c}{Model D} \\
\midrule
$\ns$ & $\fnon$ & $\fint$ & $\fdyn$ \\
\midrule
3 & $0.994 \pm 0.005$ & $0.006 \pm 0.000$ & $0.001 \pm 0.000$  \\
4 & $0.951 \pm 0.014$ & $0.043 \pm 0.003$ & $0.006 \pm 0.001$  \\
5 & $0.944 \pm 0.019$ & $0.035 \pm 0.004$ & $0.022 \pm 0.003$  \\
6 & $0.882 \pm 0.036$ & $0.045 \pm 0.008$ & $0.073 \pm 0.010$  \\
\bottomrule
\end{tabular}
\caption{Similar to Table~\ref{table:fractions}, here removing from the analysis any systems which interact before $1\%$ of $t_\mathrm{end}$ (cf. equation~\ref{eq:tend}). }
\label{table:fractions_long}
\end{table}

\begin{figure*}
\center
\includegraphics[scale = 0.43, trim = 0mm 0mm 0mm 0mm]{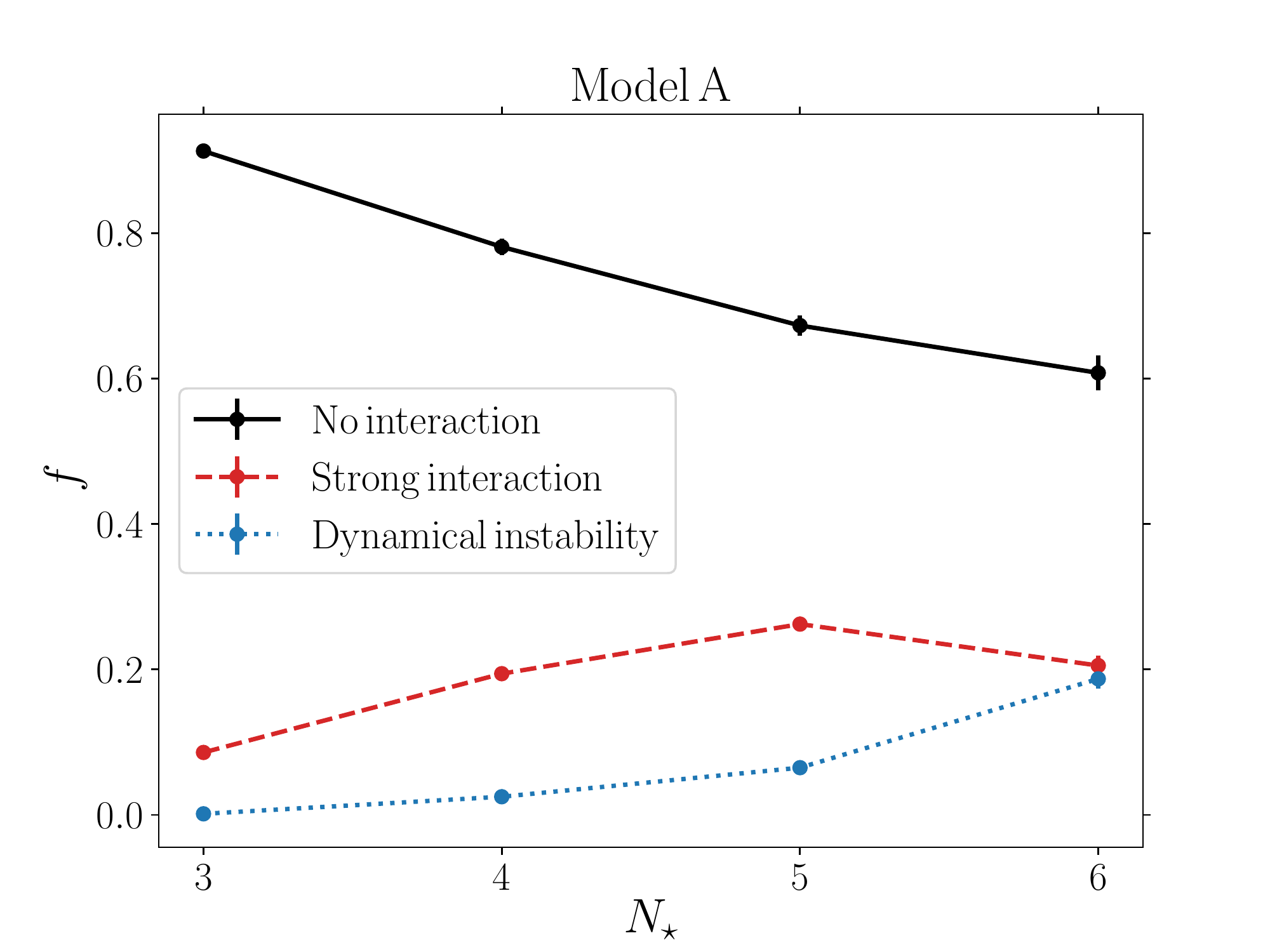}
\includegraphics[scale = 0.43, trim = 0mm 0mm 0mm 0mm]{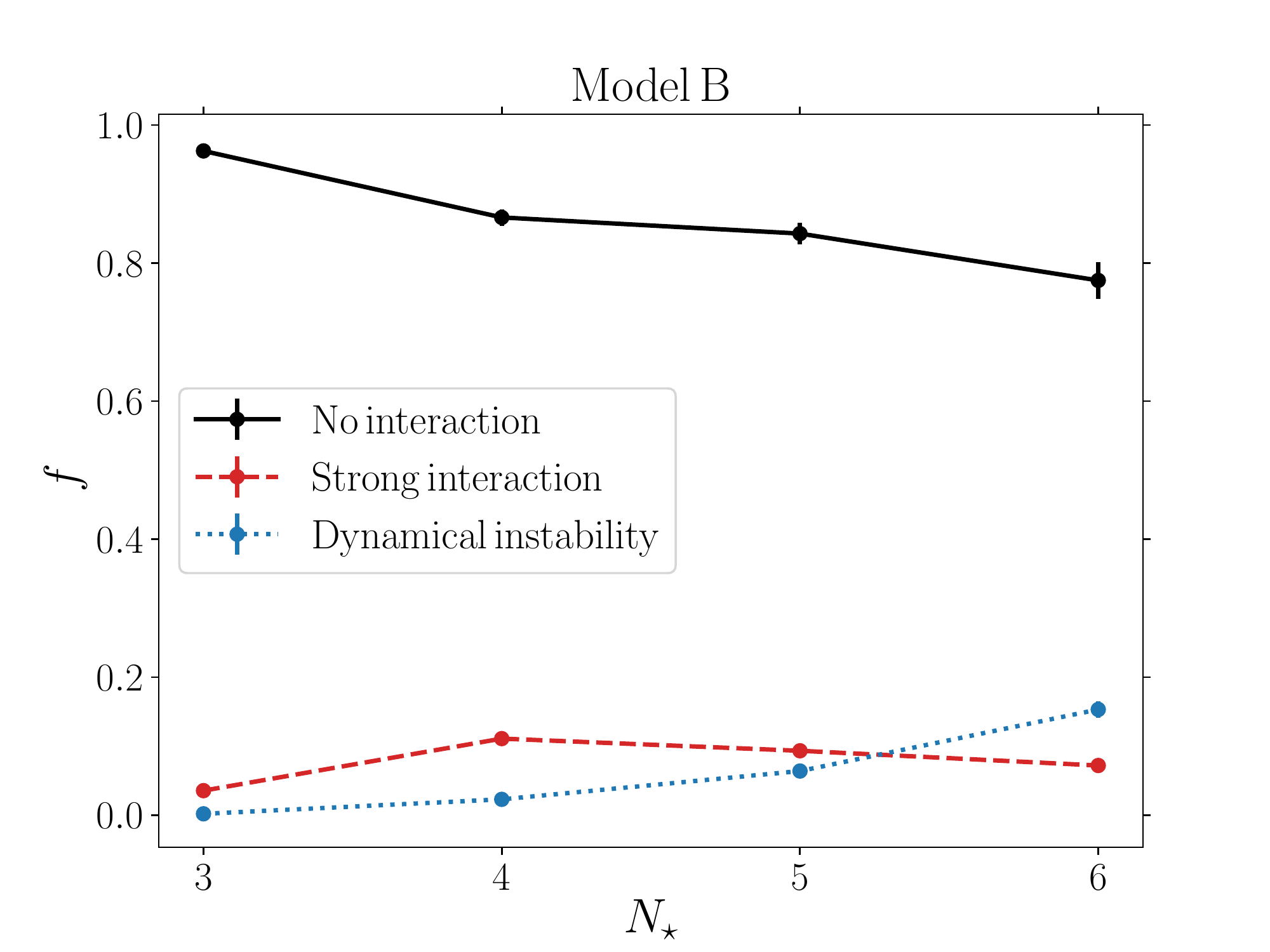}
\includegraphics[scale = 0.43, trim = 0mm 0mm 0mm 0mm]{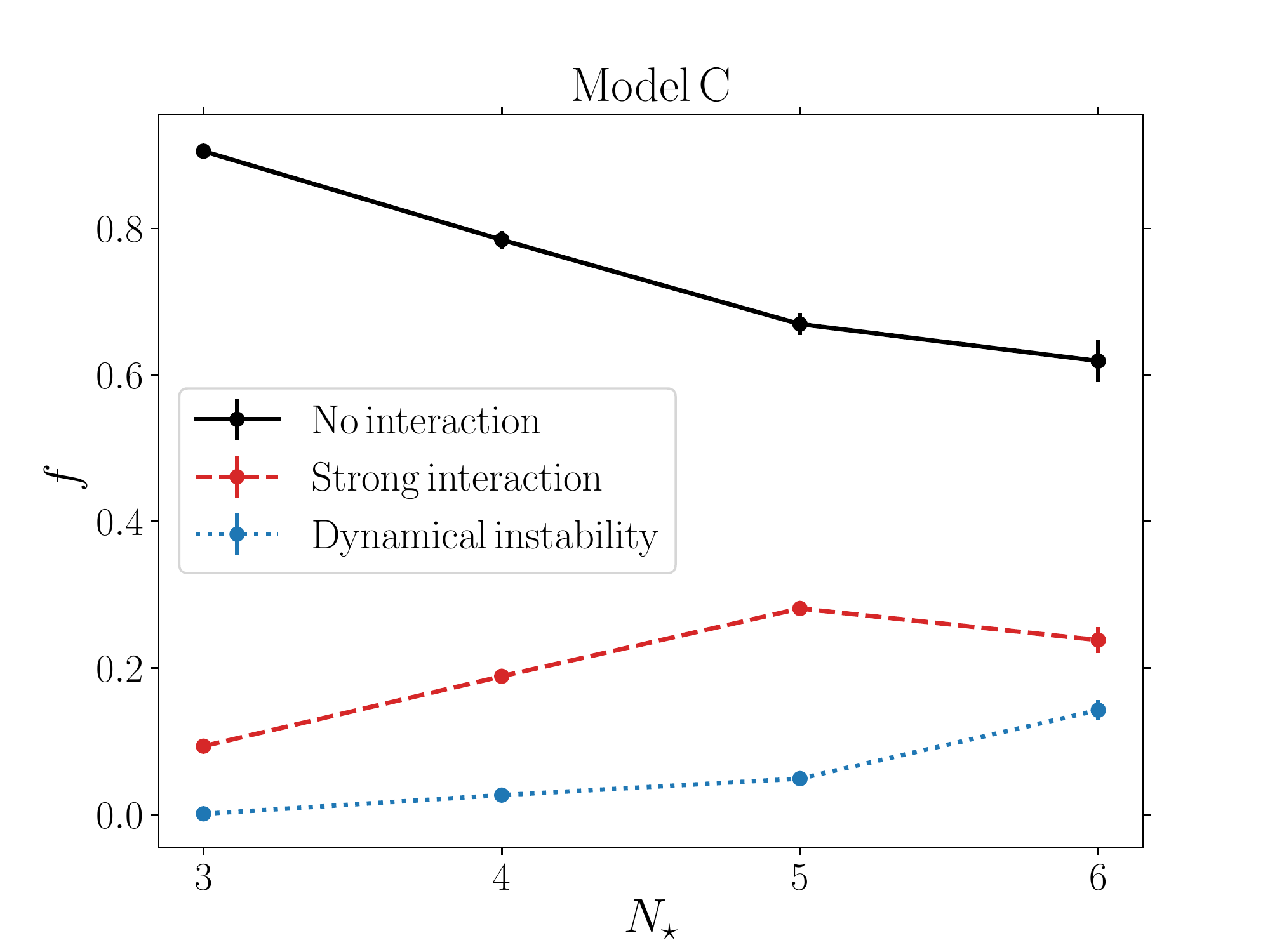}
\includegraphics[scale = 0.43, trim = 0mm 0mm 0mm 0mm]{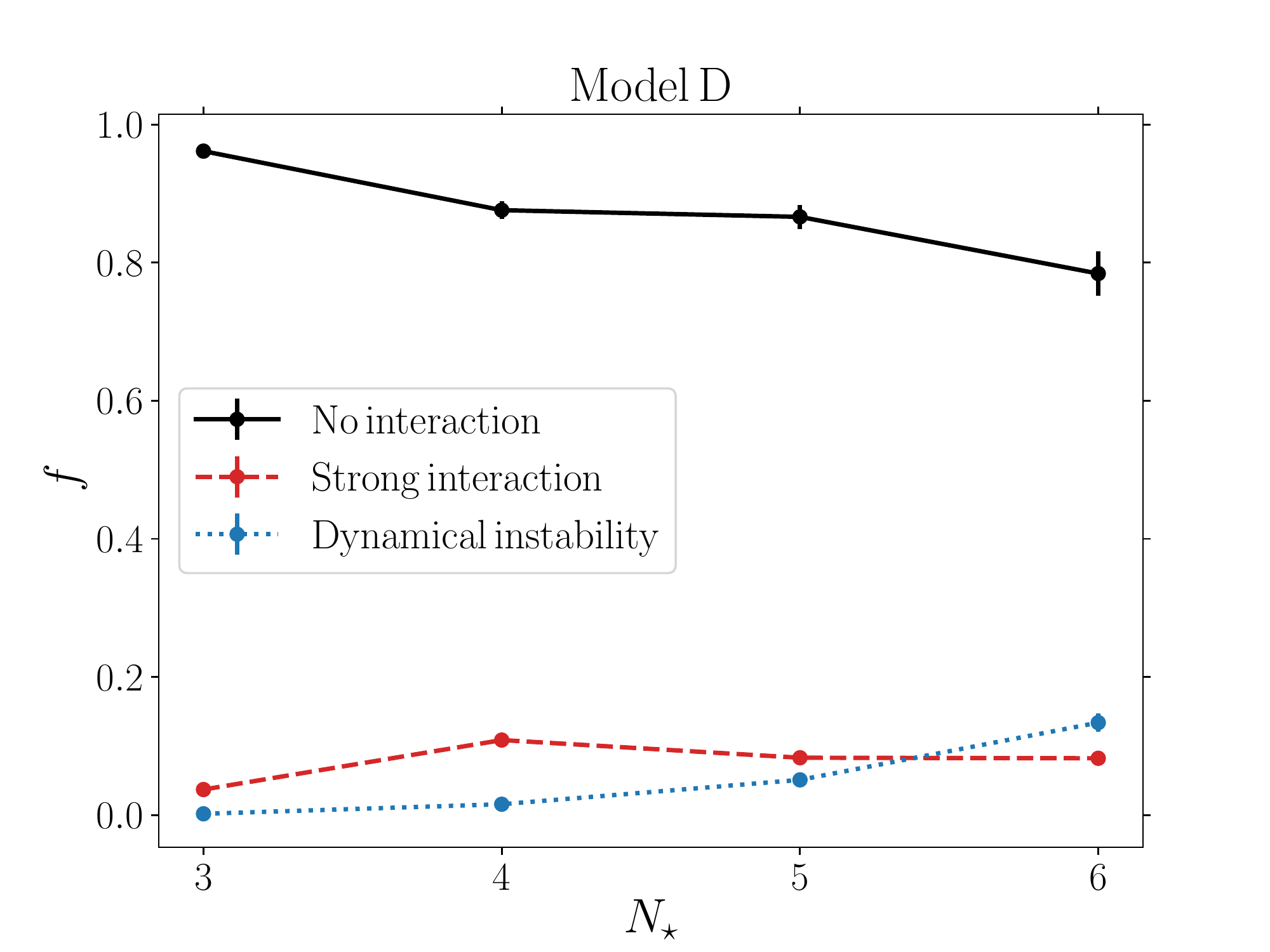}
\caption{ Outcome fractions as a function of the number of stars, $\ns$, for all four models (the same data are shown as in Table~\ref{table:fractions}). Black solid, red dashed, and blue dotted lines correspond to noninteracting, strongly interacting, and dynamical instability systems, respectively. Error bars give the Poisson errors. }
\label{fig:fractions}
\end{figure*}

\subsection{Interaction fractions}
\label{sect:results:frac}
In Table~\ref{table:fractions}, we show the outcome fractions of the systems in our Monte Carlo integrations for all four models. We distinguish between `strong interaction', i.e., an orbit in the system satisfies \eq~(\ref{eq:rpmin}), dynamical instability (see \S~\ref{sect:meth:num}), and none of the above, i.e., no interaction. We also present the same data visually (as a function of $\ns$) in \F~\ref{fig:fractions}.

The majority of systems do not interact during the simulations, although $\fnon$ decreases strongly with increasing $\ns$. For models A and C, $\fnon \sim 0.9$ for $\ns=3$, and it decreases to $\fnon \sim 0.6$ for $\ns=6$. The decrease of $\fnon$ with increasing $\ns$ is associated with significant increases in both $\fint$ and $\fdyn$. An increase in $\fint$ with increasing $\ns$ can be understood from the larger available parameter space for which strong secular evolution can arise in more complex hierarchical systems. This aspect has been explored in quadruples (e.g., \citealt{2013MNRAS.435..943P,2015MNRAS.449.4221H,2017MNRAS.470.1657H,2018MNRAS.474.3547G}), but is demonstrated here for higher-order systems as well. 

The fraction of dynamically unstable systems is nearly zero for triples, but increases strongly with increasing $\ns$. For triples, dynamical instability can only occur for systems that are initially marginally stable, since the onset of instability on the MS is mainly driven by changes of the outer orbit eccentricity, which are typically small (the outer orbit eccentricity is constant at the quadrupole expansion order; changes in outer orbit eccentricity occur starting only at the higher octupole order; see, e.g., \citealt{2000ApJ...535..385F}). However, in 3+1 quadruples (not in 2+2 quadruples, see below), the inner-intermediate pair can efficiently be driven to dynamical instability due to increased eccentricity of the intermediate orbit induced by the fourth body (see, e.g., the example evolution of WDS 00335+4006 in \F~\ref{fig:ex1}). Our results show that this effect becomes even more important in higher-order quintuple and sextuple systems. 

In addition, the interaction fraction $\fint$ {\it decreases} in all models from $\ns=5$ to $\ns=6$, whereas the dynamical instability fraction $\fdyn$ increases significantly. In models B and D, even $\fdyn>\fint$ for $\ns>5$. This highlights the importance of dynamical instability in high-multiplicity systems. 

In Model B, the noninteraction fraction is generally higher compared to Model A. This can be attributed to the typically lower mutual inclinations in Model B (see \F~\ref{fig:IC:orient}), giving rise to weaker secular evolution. Correspondingly, $\fint$ and $\fdyn$ are lower. Nevertheless, $\fint$ and $\fdyn$ still increase significantly with increasing $\ns$, both up to $\sim 0.1$ for $\ns=6$. Models A and C give very similar results, as do models B and D. In other words, the differences in the assumed eccentricity distributions (see \F~\ref{fig:IC:e}) do not lead to major differences in the interaction fractions. The largest differences arise from the assumed inclination distributions, i.e., differences between models A and B, and between models C and D. These results also apply to the orbital distributions (\S~\ref{sect:results:orb}), and the interaction time distributions (\S~\ref{sect:results:time}). 

\begin{figure}
\center
\includegraphics[scale = 0.43, trim = 0mm 0mm 0mm 0mm]{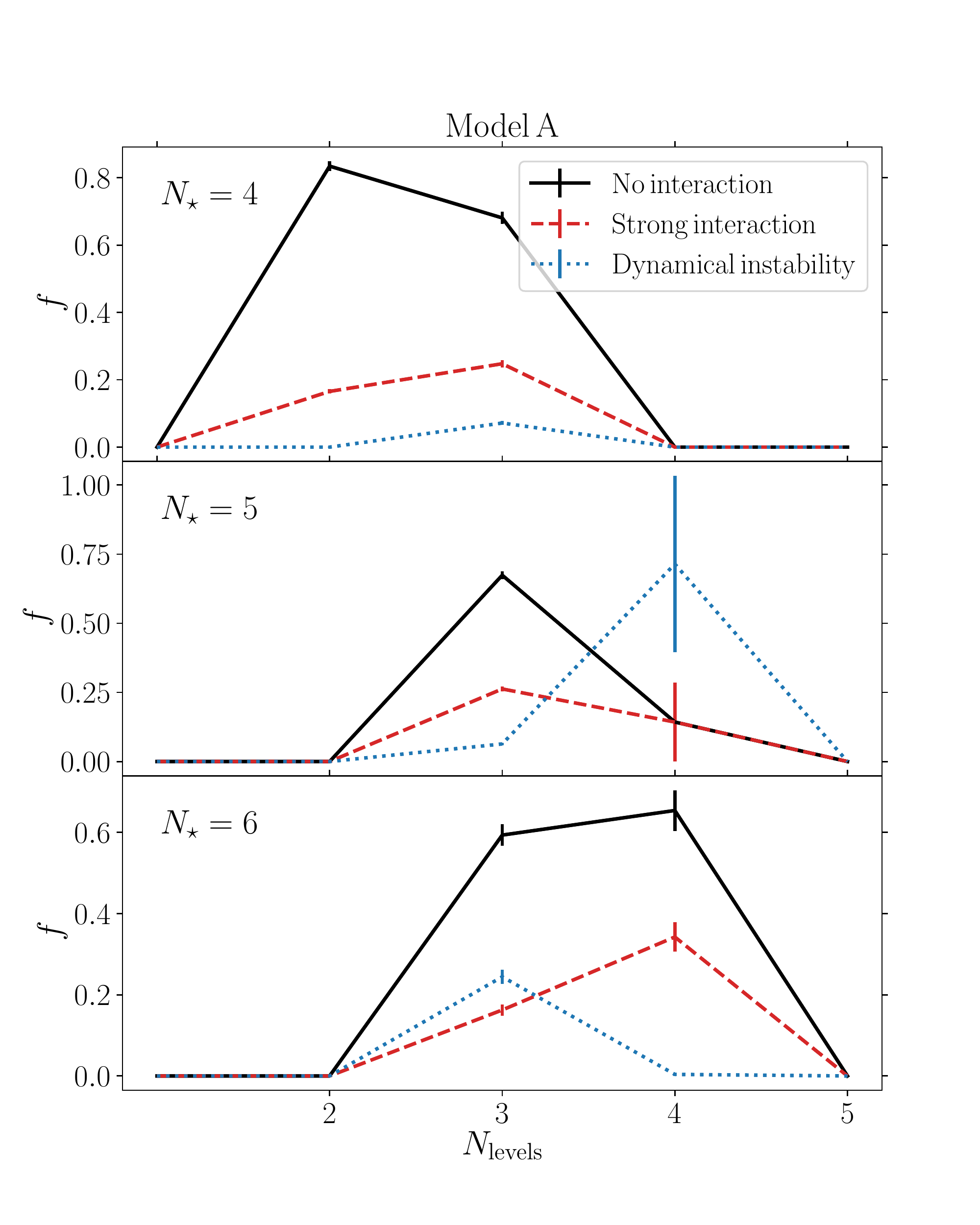}
\caption{ Outcome fractions from Model A for a given number of stars, as a function of $\nlevel$, the number of levels in the system. The fractions are all set to zero if a particular configuration does not apply to a system with given $\ns$. Top, middle and bottom panels correspond to $\ns=4$, 5, and 6, respectively. Note that triples by design only have $\nlevel=2$; for the latter, the fractions can be read off from Table\,\ref{table:fractions}, or \F~\ref{fig:fractions}. Black solid, red dashed, and blue dotted lines correspond to noninteracting, strongly interacting, and dynamical instability systems, respectively. Error bars give the Poisson errors.}
\label{fig:fractions_levels}
\end{figure}

In \F~\ref{fig:fractions_levels}, we break down the outcome fractions with respect to the number of levels in the system (showing results from Model A; other models give qualitatively similar results). A triple has $\nlevel=2$ different levels, whereas higher-order systems generally have $\nlevel \geq 2$. For example, a 2+2 quadruple has $\nlevel=2$, whereas a 3+1 quadruple has $\nlevel=3$. Each panel in \F~\ref{fig:fractions_levels} corresponds to a different $\ns$. For systems with a given $\nlevel$, we plot the fractions of systems that either do not interact, interact strongly, or become dynamically unstable. Generally, for a given $\ns$, we expect systems with larger $\nlevel$ to have a higher probability to become dynamically unstable.

As expected, for quadruples, dynamical stability occurs almost exclusively if $\nlevel=3$, which corresponds to the 3+1 configuration. For quintuples, the dynamical instability fraction is higher for $\nlevel=4$ compared to $\nlevel=3$, although it should be noted that the number of available quintuple systems with $\nlevel=4$ is small (reflected in the large Poisson errors). Interestingly, for sextuples, the dynamical instability fraction is significantly higher for $\nlevel=3$ compared to $\nlevel=4$. However, this result might be biased by the limited number of known sextuple systems in the MSC (see \S~\ref{sect:meth:IC}).

In Table~\ref{table:fractions_long}, we show the interaction fractions, similar to Table~\ref{table:fractions}, but removing from the analysis any systems which interact before $1\%$ of $t_\mathrm{end}$ (cf. equation~\ref{eq:tend}). Since many interactions occur relatively early in the evolution (see \S~\ref{sect:results:time} below), the interaction fractions are markedly lower in this case, although the probability of an interaction at later times is still significant, especially for higher-multiplicity systems. This aspect is discussed further in \S~\ref{sect:discussion:times}. 

\begin{figure*}
\center
\includegraphics[scale = \figsize, trim = 0mm 0mm 0mm 0mm]{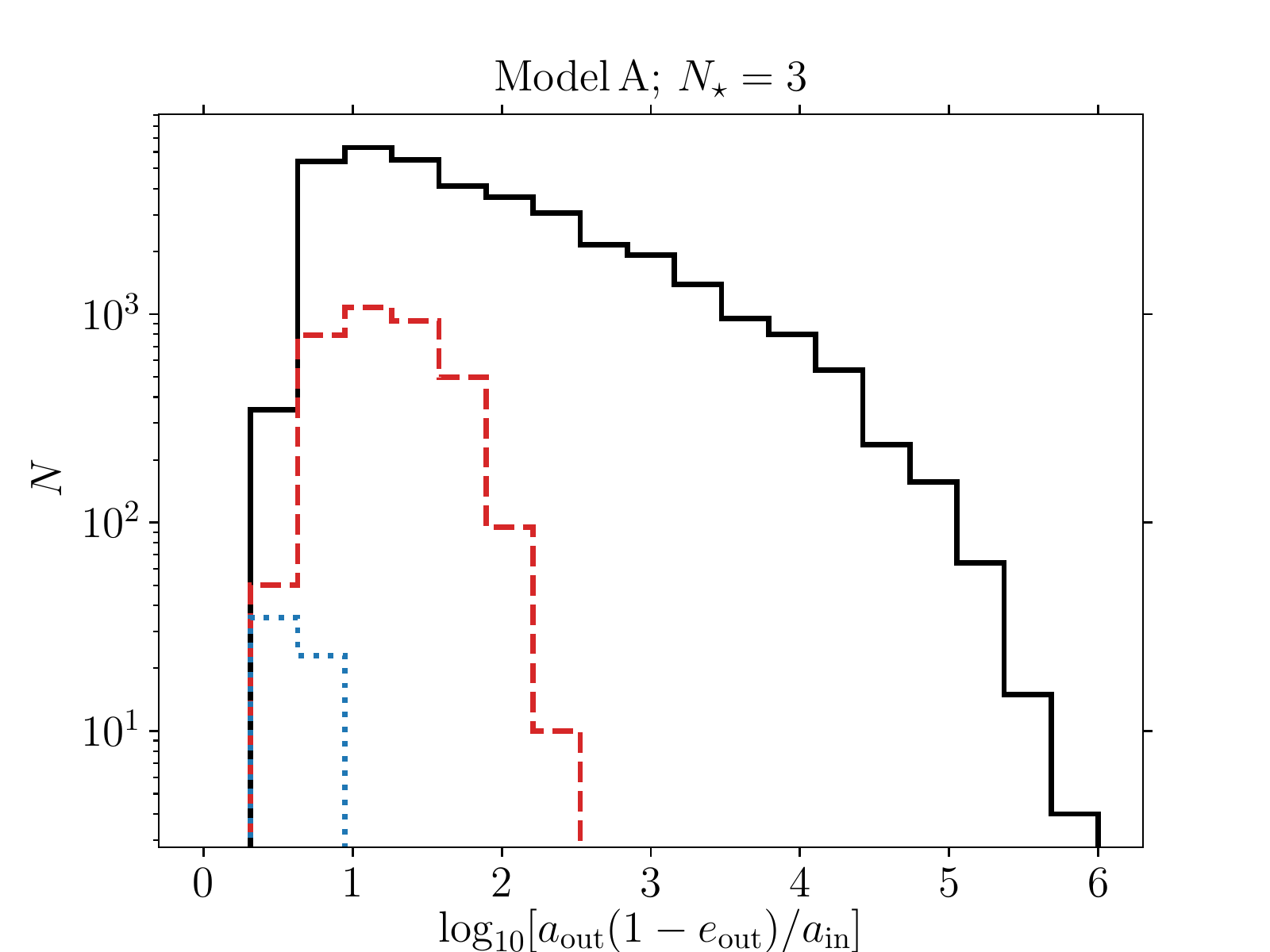}
\includegraphics[scale = \figsize, trim = 0mm 0mm 0mm 0mm]{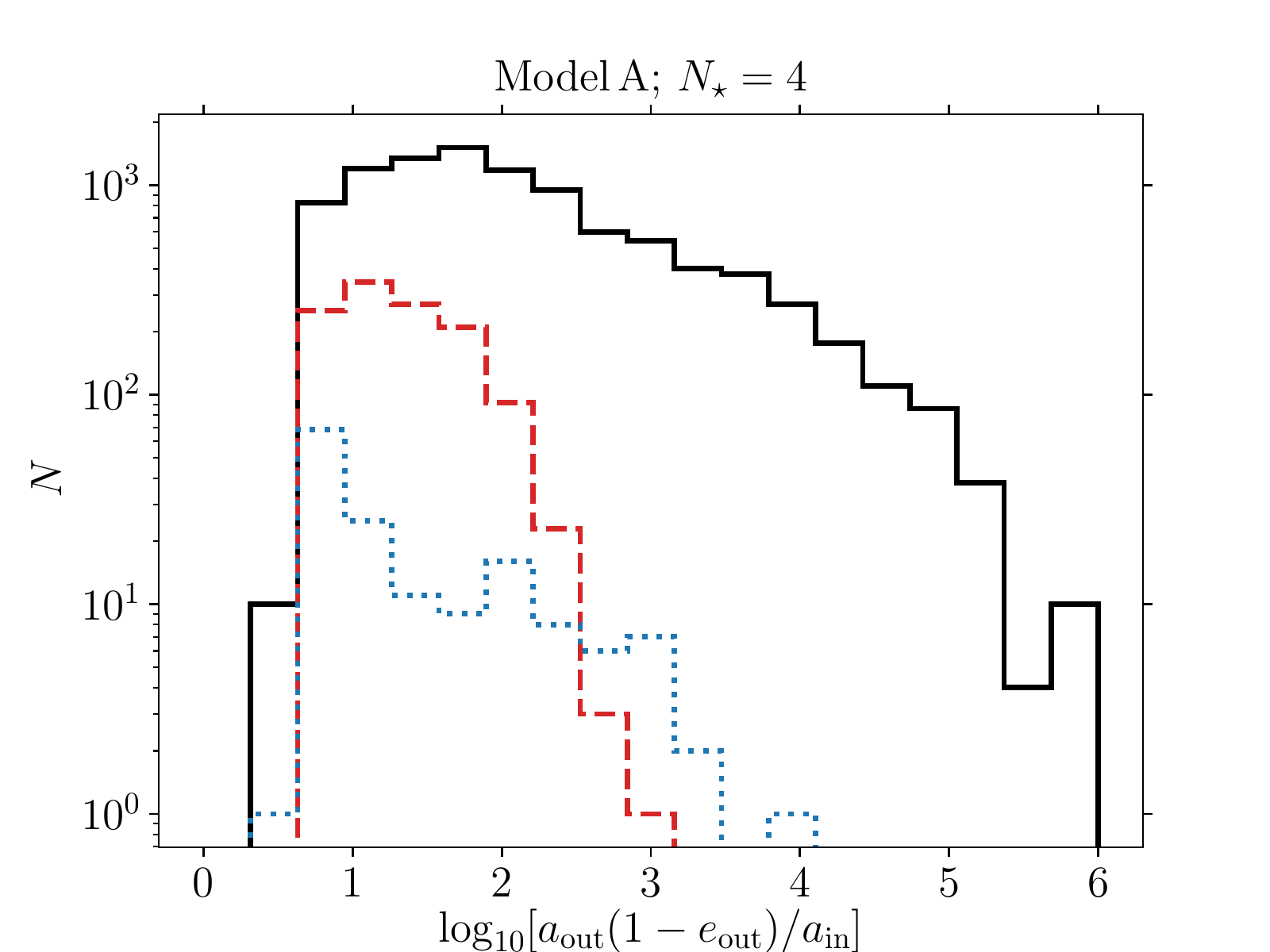}
\includegraphics[scale = \figsize, trim = 0mm 0mm 0mm 0mm]{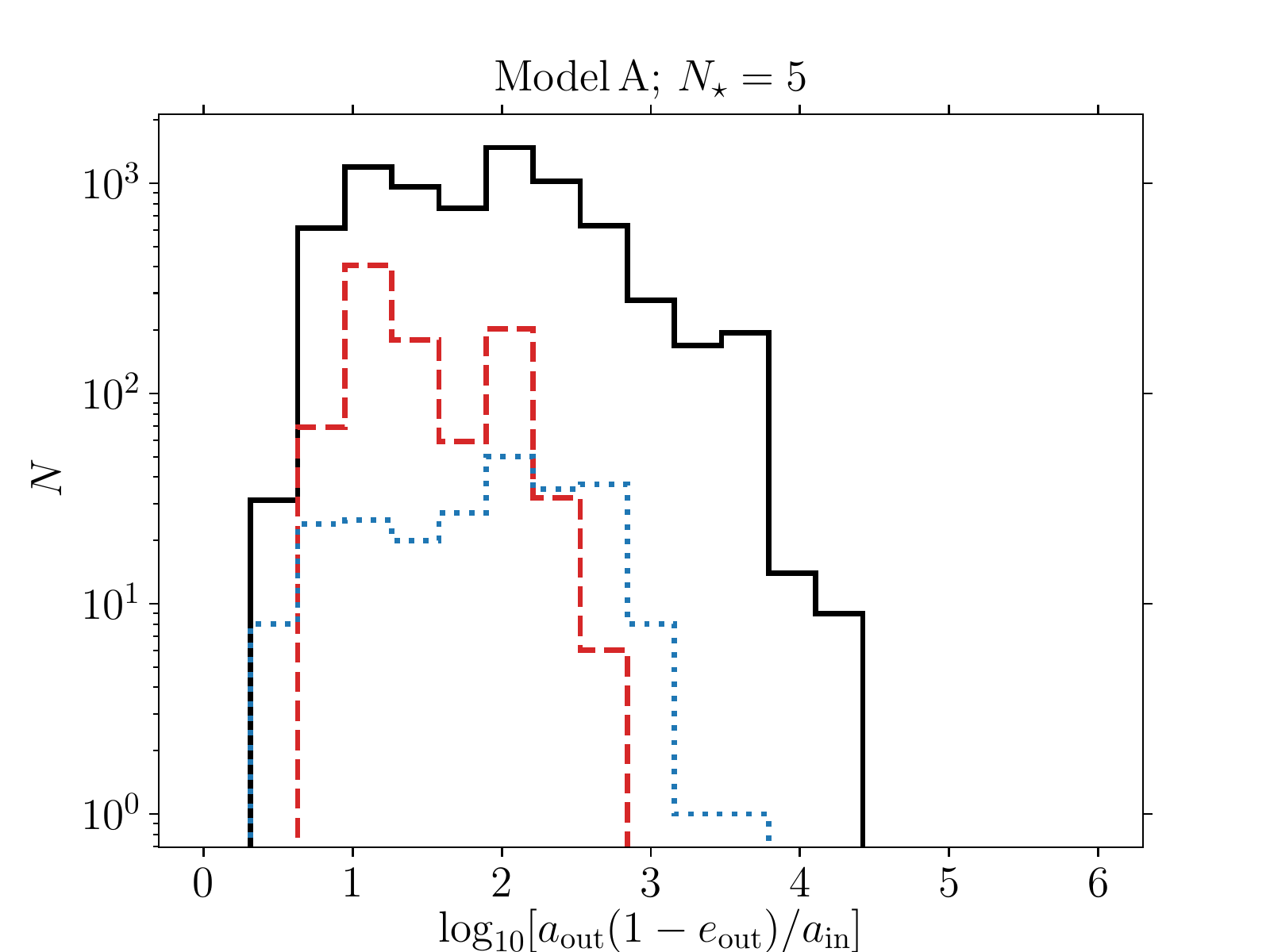}
\includegraphics[scale = \figsize, trim = 0mm 0mm 0mm 0mm]{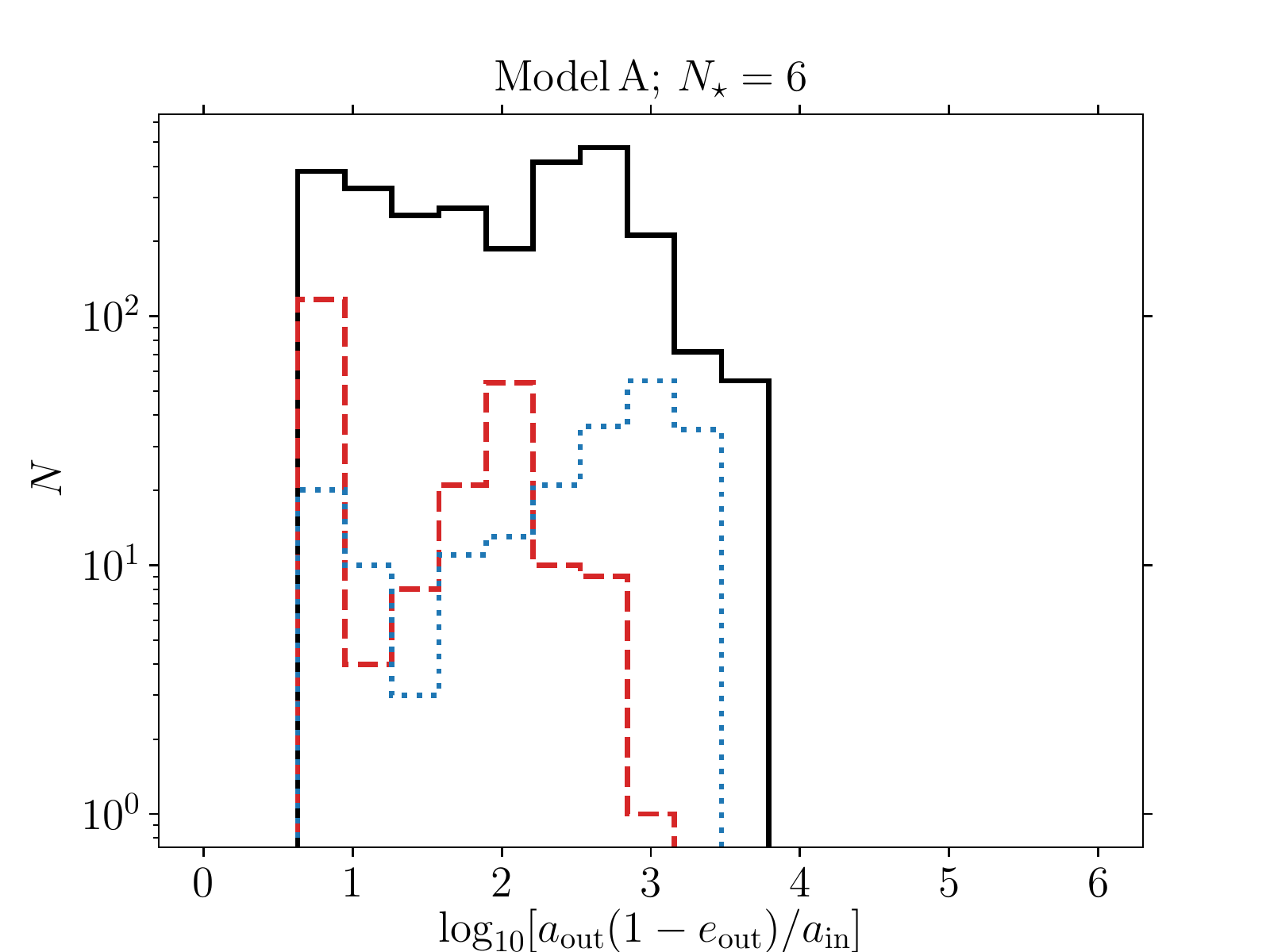}
\includegraphics[scale = \figsize, trim = 0mm 0mm 0mm 0mm]{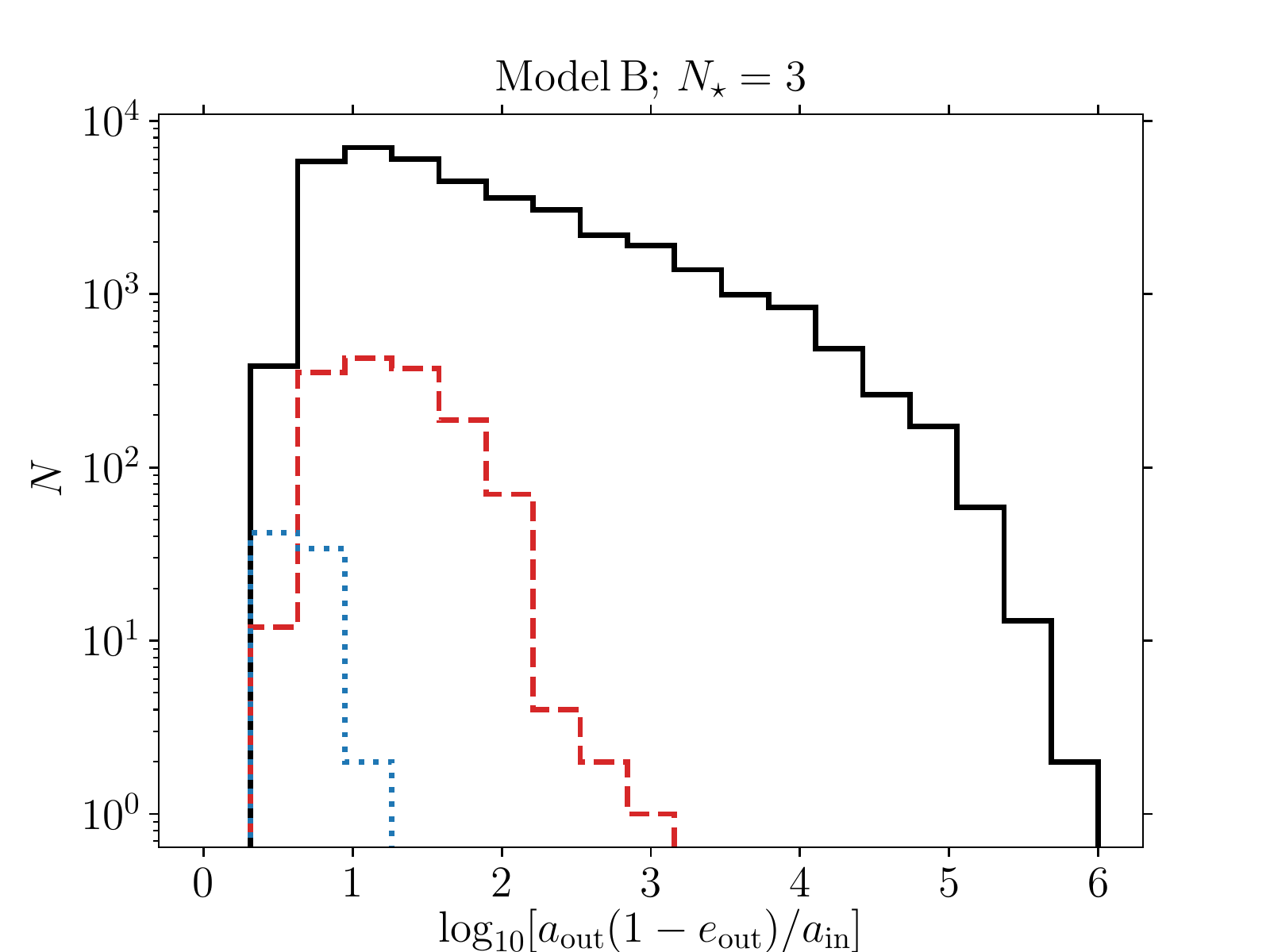}
\includegraphics[scale = \figsize, trim = 0mm 0mm 0mm 0mm]{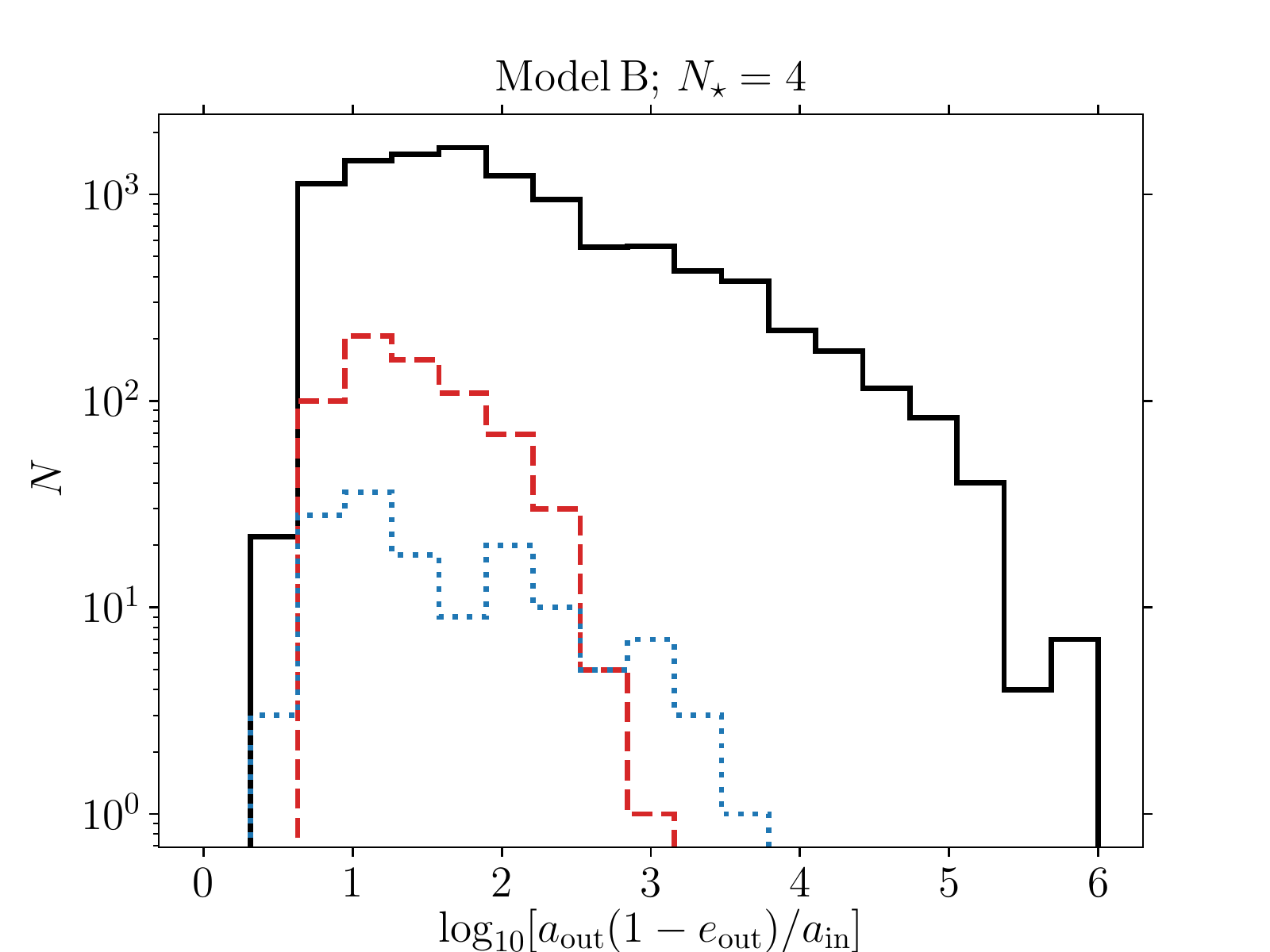}
\includegraphics[scale = \figsize, trim = 0mm 0mm 0mm 0mm]{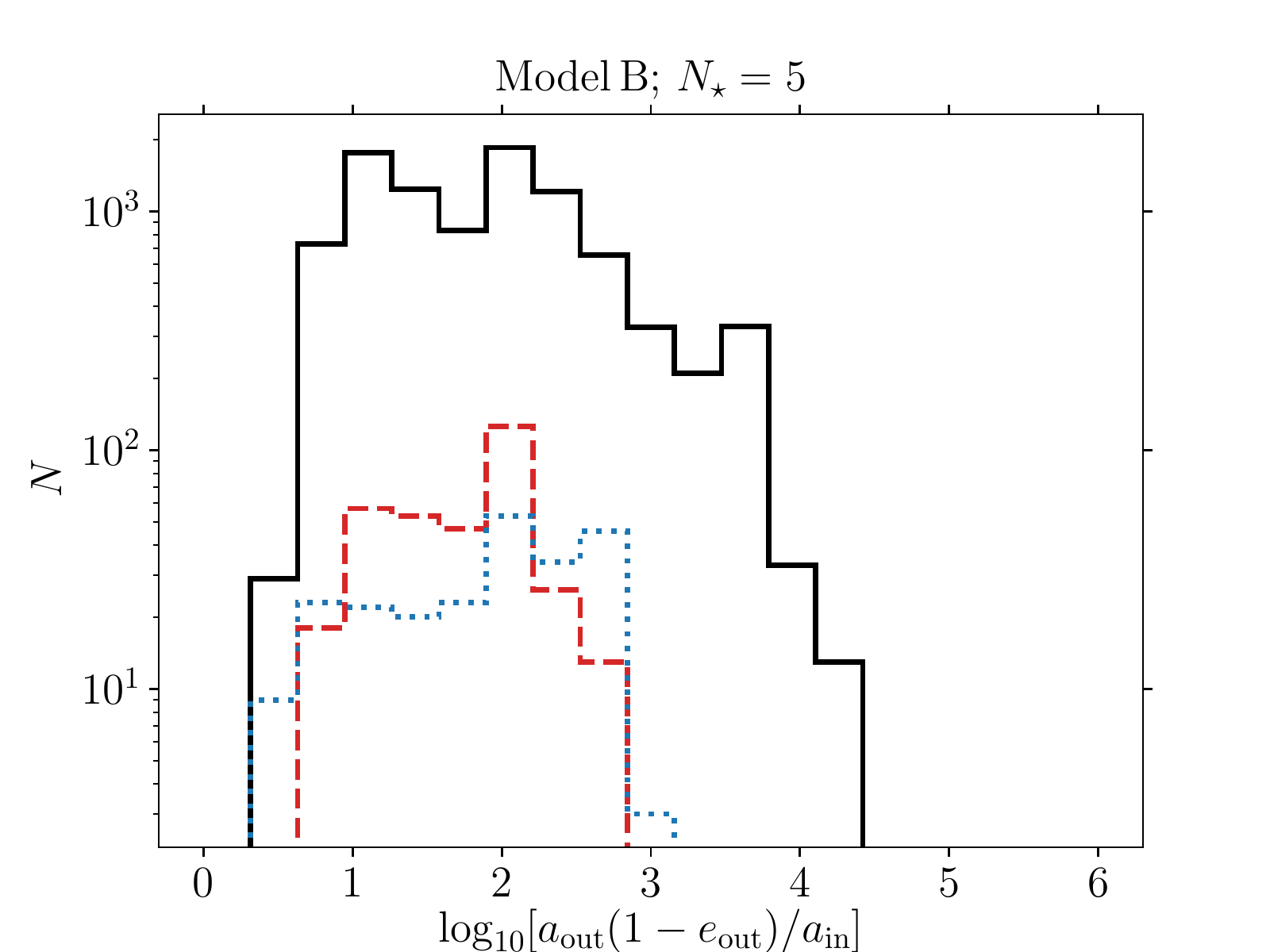}
\includegraphics[scale = \figsize, trim = 0mm 0mm 0mm 0mm]{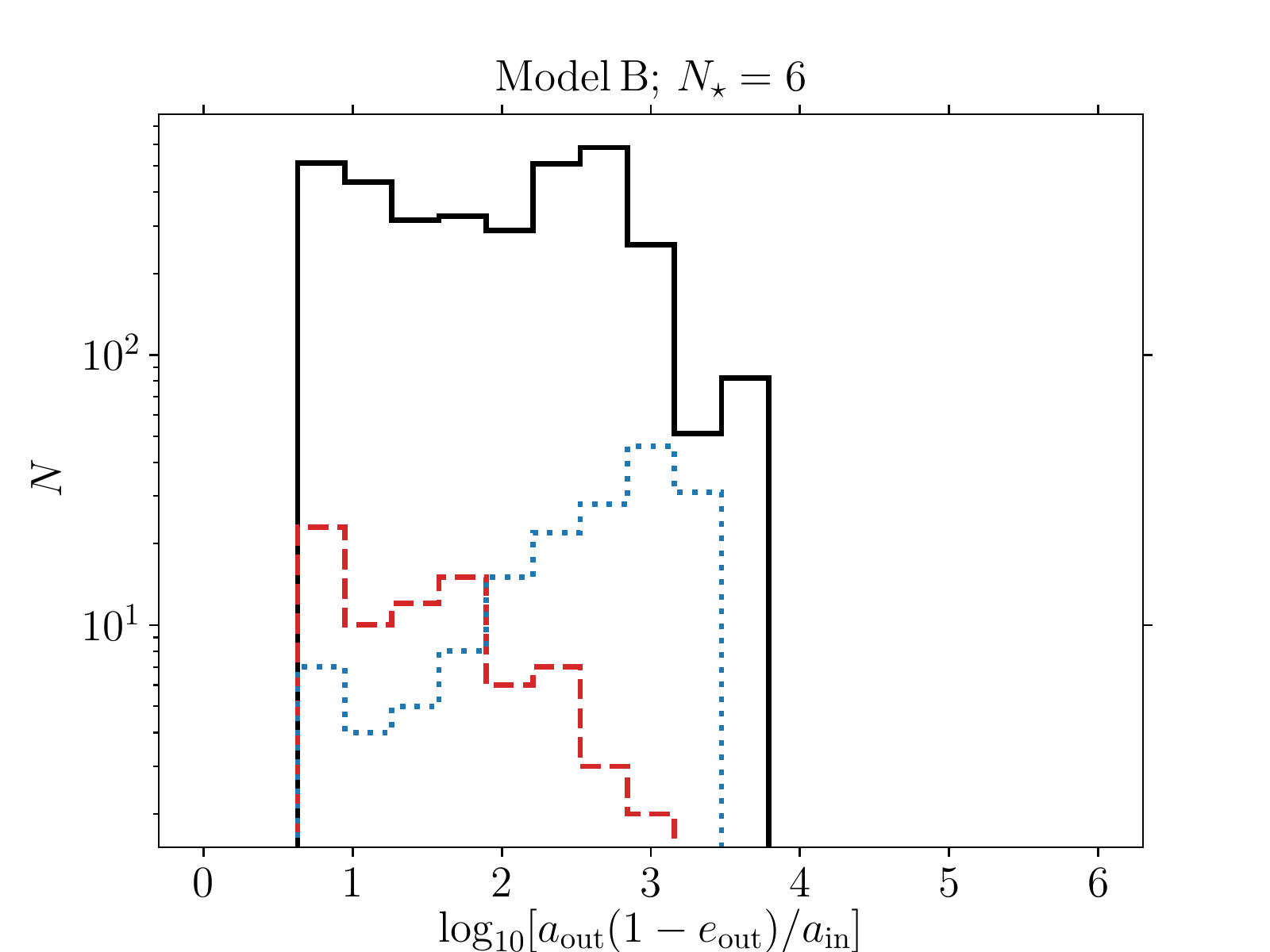}
\caption{ Distributions of the ratio of initial outer orbit periapsis distance to inner orbit semimajor axis, $a_\mathrm{out}(1-e_\mathrm{out})/a_\mathrm{in}$. Black solid lines correspond to noninteracting systems, red dashed lines to strongly interacting systems, and blue dotted lines to dynamically unstable systems. Each panel corresponds to a specific model and number of stars, indicated at the top. }
\label{fig:smas}
\end{figure*}

\subsection{Orbital distributions}
\label{sect:results:orb}
Here, we focus on the orbital parameter space associated with the different outcomes in our simulations. In \F~\ref{fig:smas}, we show the distributions of the initial ratio of outer orbit periapsis distance to inner orbit semimajor axis, $a_\mathrm{out}(1-e_\mathrm{out})/a_\mathrm{in}$. Black solid lines correspond to noninteracting systems, red dashed lines to strongly interacting systems, and blue dotted lines to dynamically unstable systems. For noninteracting systems, we determine $a_\mathrm{out}(1-e_\mathrm{out})/a_\mathrm{in}$ for {\it all} orbital pairs in the system, i.e., there can be multiple values per system depending on $\ns$. For the other outcomes, we determine a single value $a_\mathrm{out}(1-e_\mathrm{out})/a_\mathrm{in}$ for each relevant system based on the orbit for which the strong interaction condition (\eq~\ref{eq:rpmin}) or dynamical instability criterion occurred (relative to the `outer' or `parent' orbit). 

Each panel in \F~\ref{fig:smas} corresponds to a specific model and number of stars, indicated at the top. As discussed in \S~\ref{sect:results:frac}, the qualitative differences between models A and C are minor, as are those between models B and D. We therefore show only results for models A and B. Furthermore, when comparing panels with different numbers of stars, one should bear in mind the relative number of available systems in the MSC (see \S~\ref{sect:meth:IC}), which affects the statistical quality and significance. In particular, the number of triples is much larger than the number of all higher-order systems combined. 

\F~\ref{fig:smas} shows that the strongly interacting systems (red dashed lines) tend to have similar orbital distributions across all types of systems. This is different for the dynamical instability systems; in triples, only highly compact systems can become dynamically unstable ($a_\mathrm{out}[1-e_\mathrm{out}]/a_\mathrm{in} \lesssim 10$), whereas dynamical instability can be triggered in higher-order systems ($\ns>3$) for a much wider range of $a_\mathrm{out}(1-e_\mathrm{out})/a_\mathrm{in}$. This can be attributed to secular evolution: the latter can significantly reduce the ratio $a_\mathrm{out}(1-e_\mathrm{out})/a_\mathrm{in}$ in systems with $\ns>3$, whereas this is not the case for triples. 

A comparison of models A and B in \F~\ref{fig:smas} reveals no major qualitative differences between the two models, although there appears to be a slight preference for smaller $a_\mathrm{out}(1-e_\mathrm{out})/a_\mathrm{in}$ in the strongly interacting systems in triples. This is likely because Model B on average has lower inclinations compared to Model A, so smaller ratios $a_\mathrm{out}(1-e_\mathrm{out})/a_\mathrm{in}$ are required in order to drive strong interactions. 

\begin{figure*}
\center
\includegraphics[scale = \figsize, trim = 0mm 0mm 0mm 0mm]{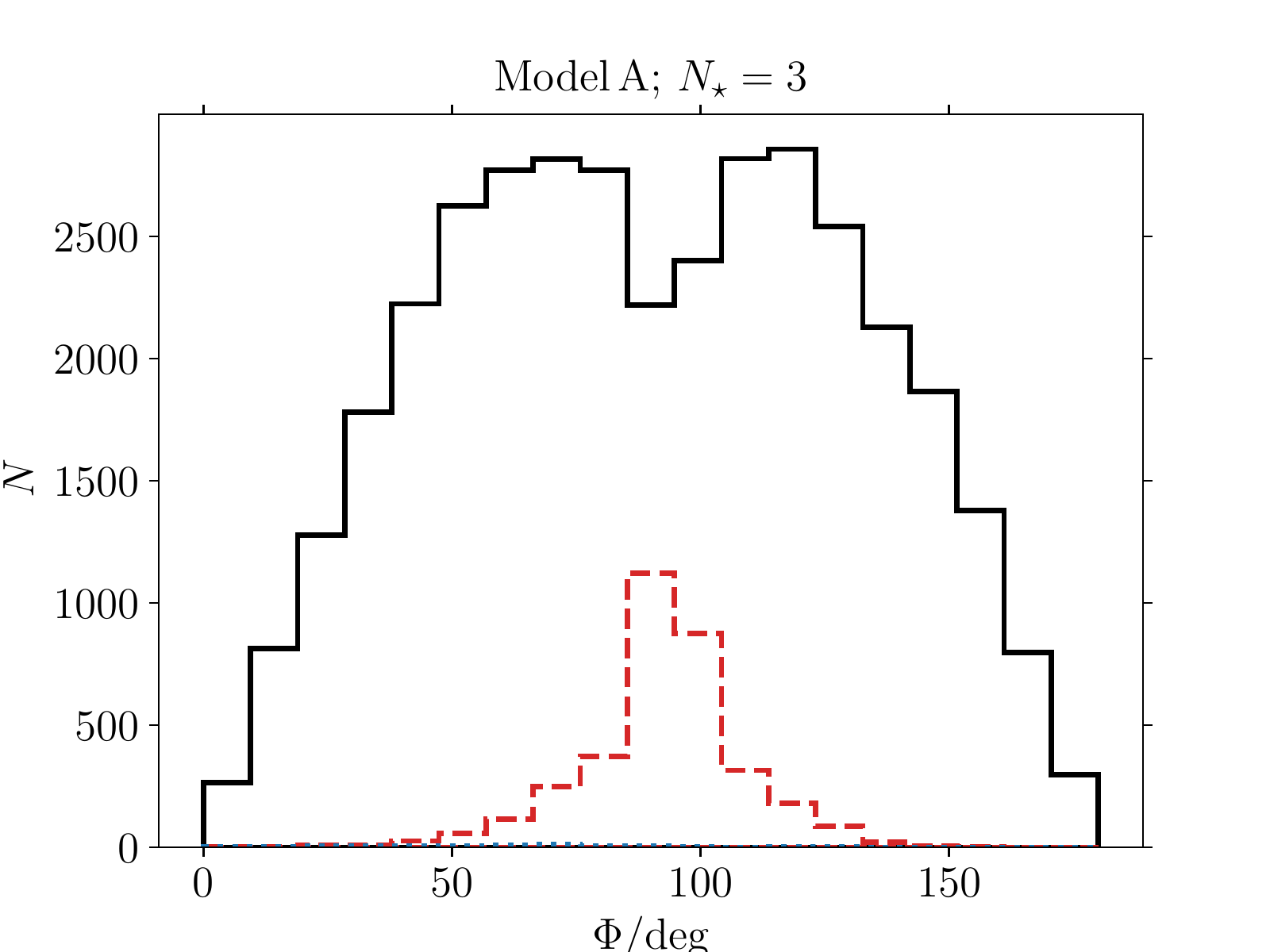}
\includegraphics[scale = \figsize, trim = 0mm 0mm 0mm 0mm]{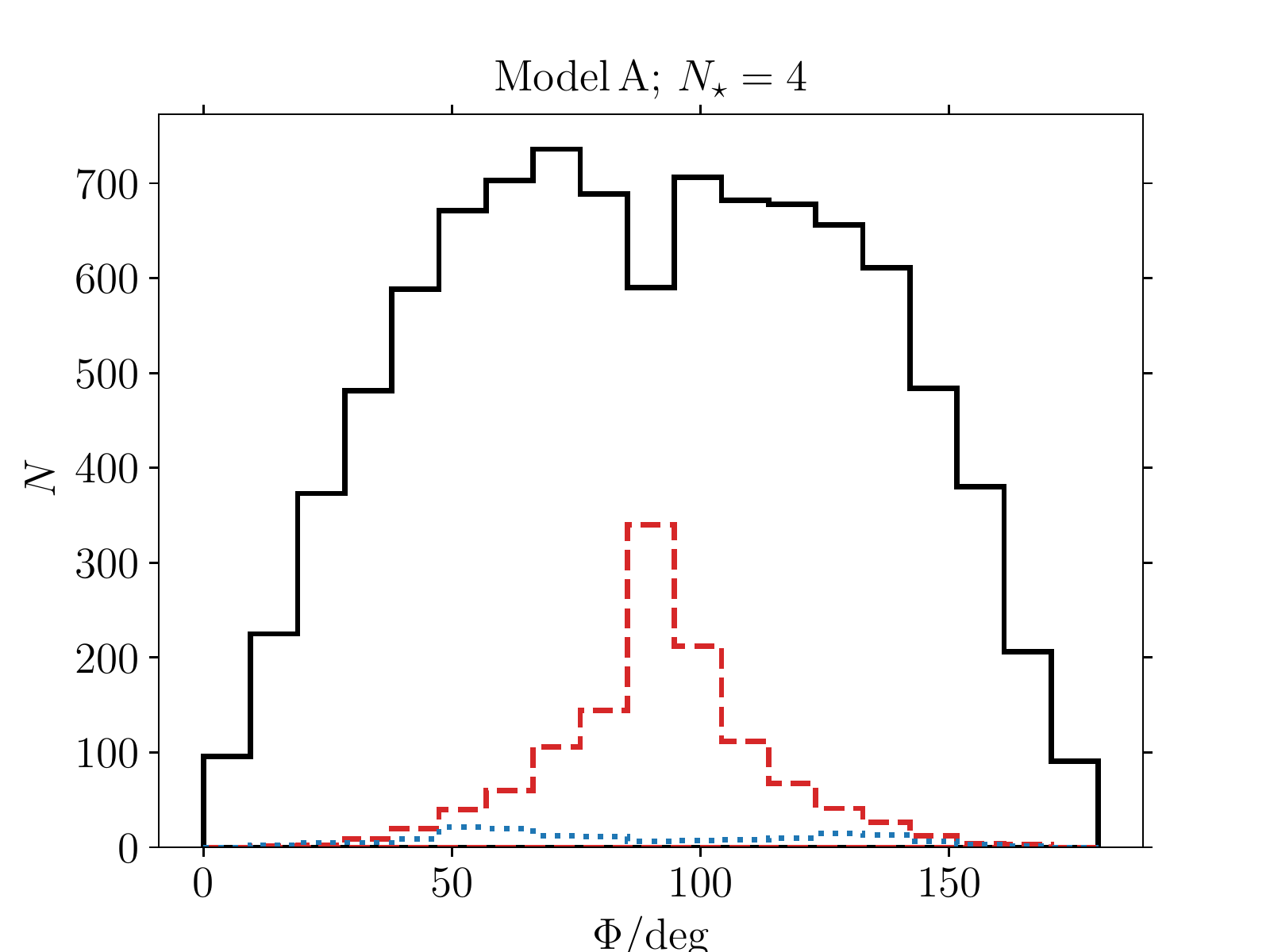}
\includegraphics[scale = \figsize, trim = 0mm 0mm 0mm 0mm]{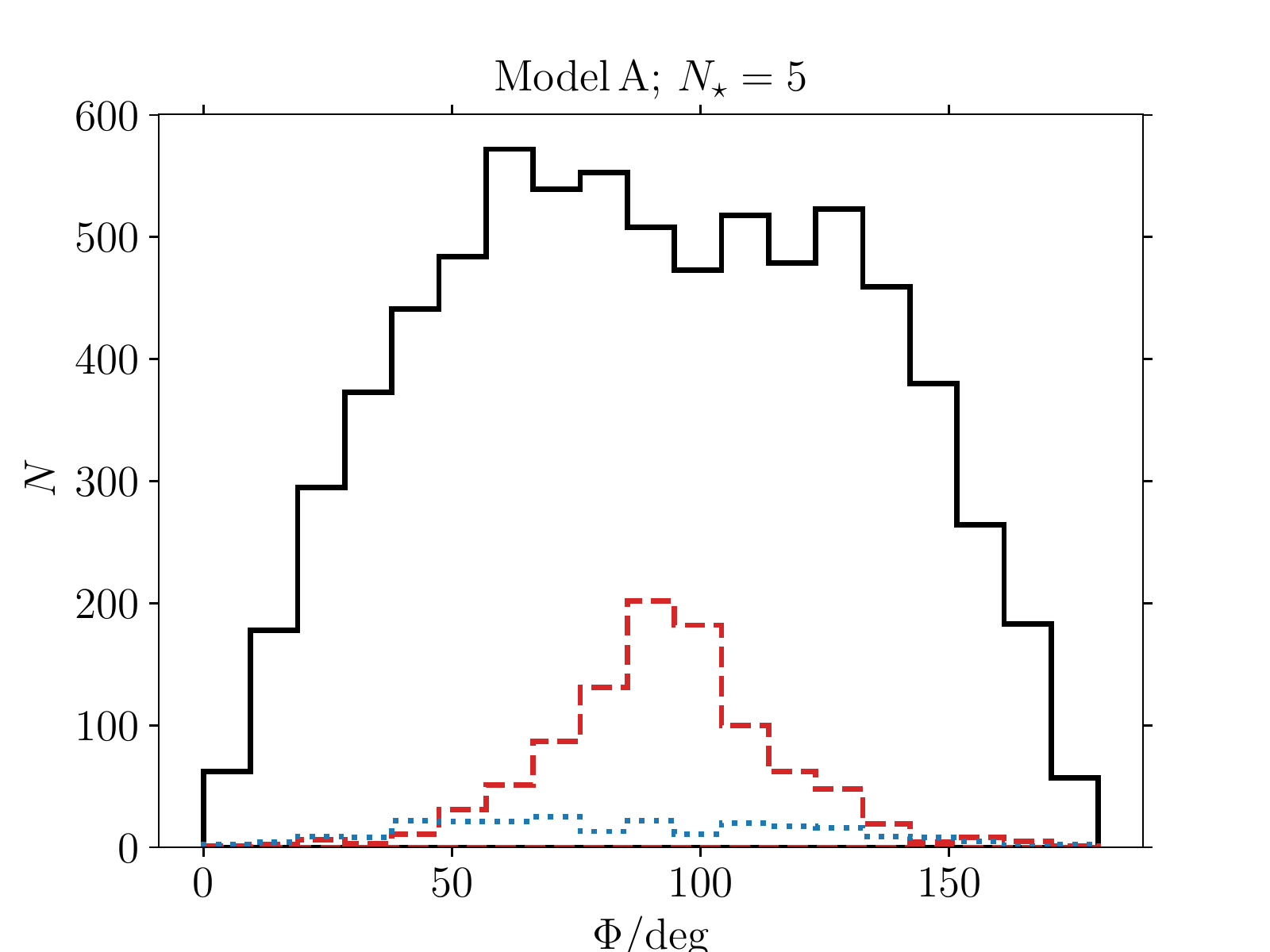}
\includegraphics[scale = \figsize, trim = 0mm 0mm 0mm 0mm]{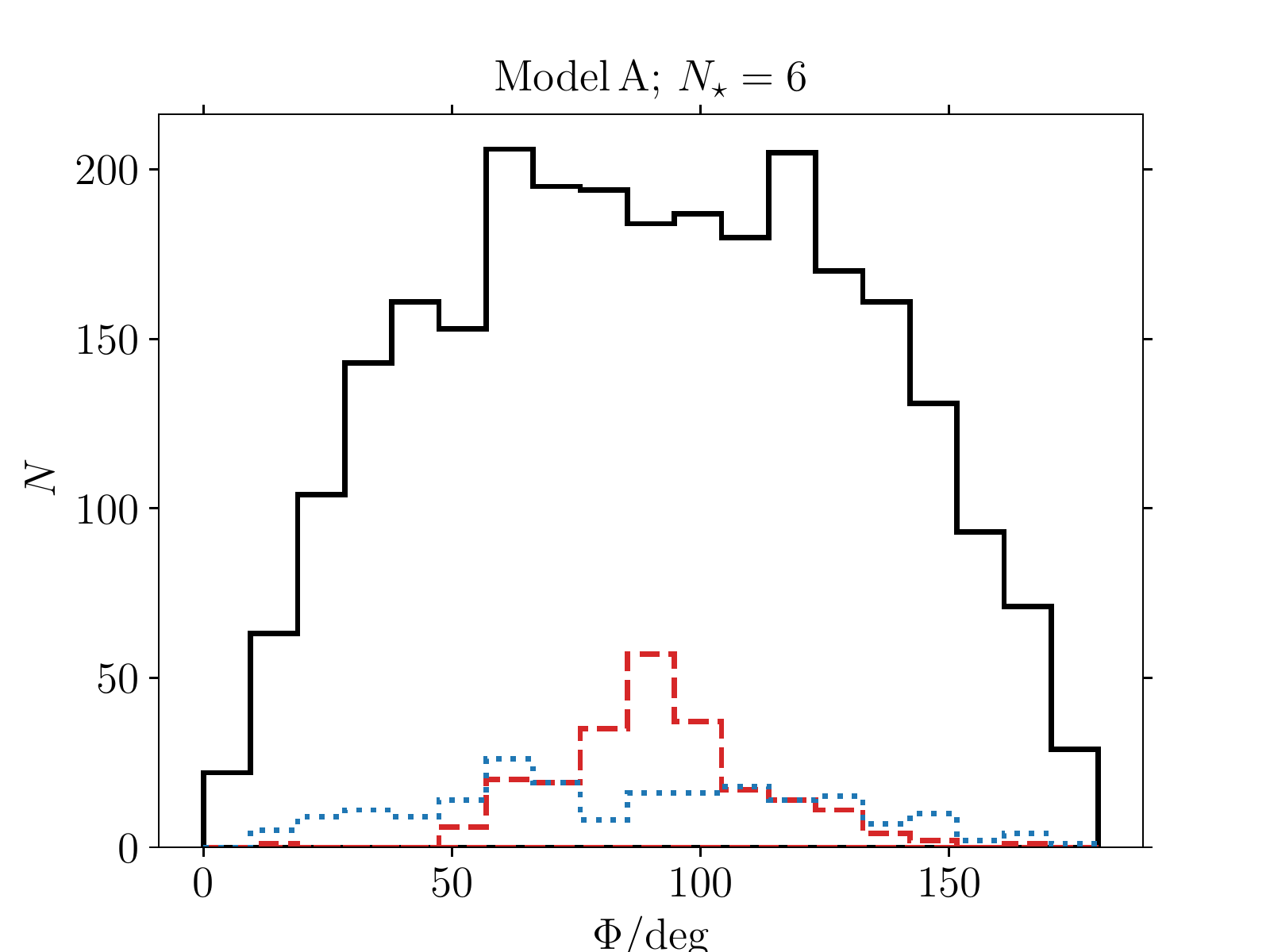}
\includegraphics[scale = \figsize, trim = 0mm 0mm 0mm 0mm]{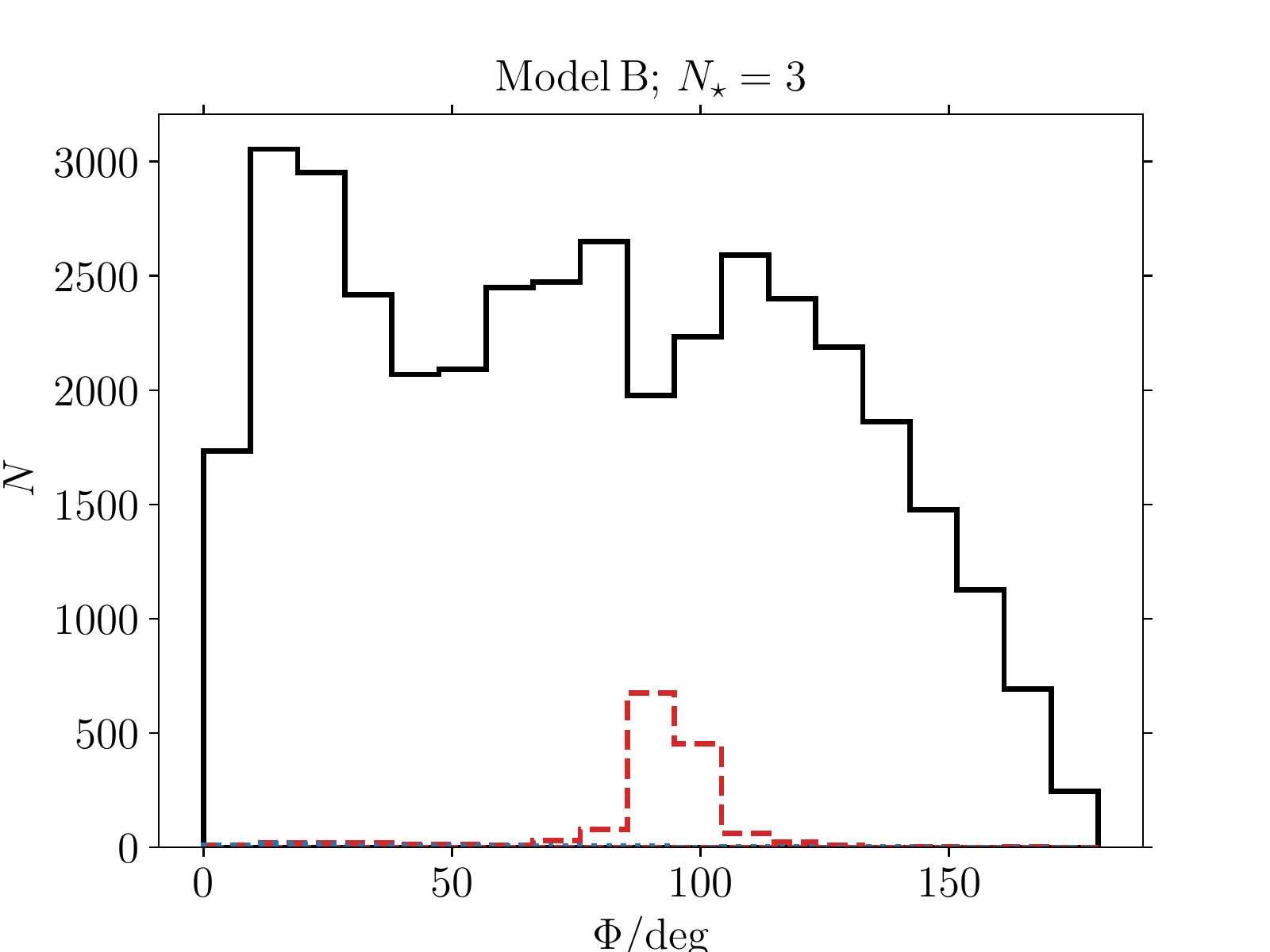}
\includegraphics[scale = \figsize, trim = 0mm 0mm 0mm 0mm]{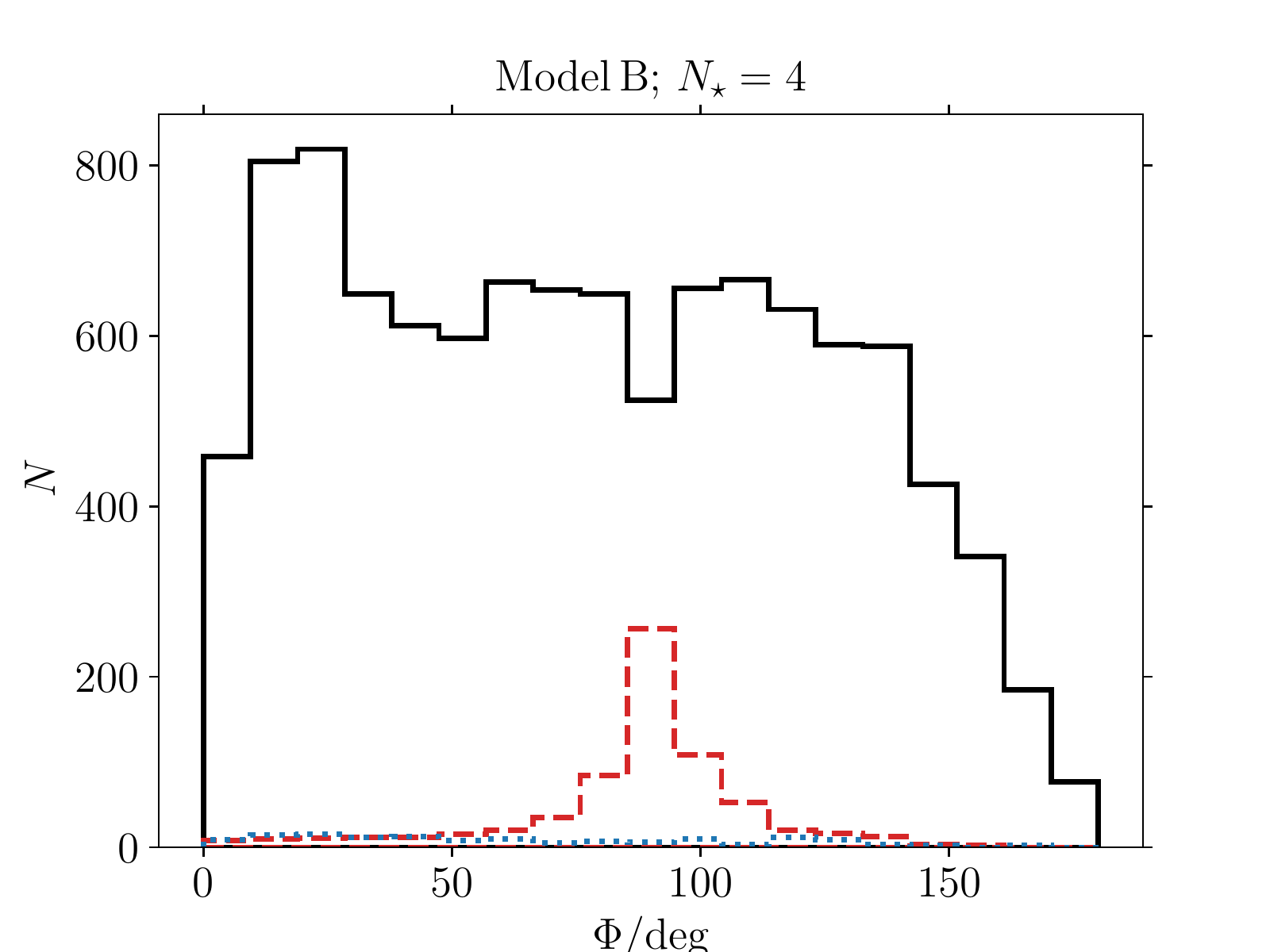}
\includegraphics[scale = \figsize, trim = 0mm 0mm 0mm 0mm]{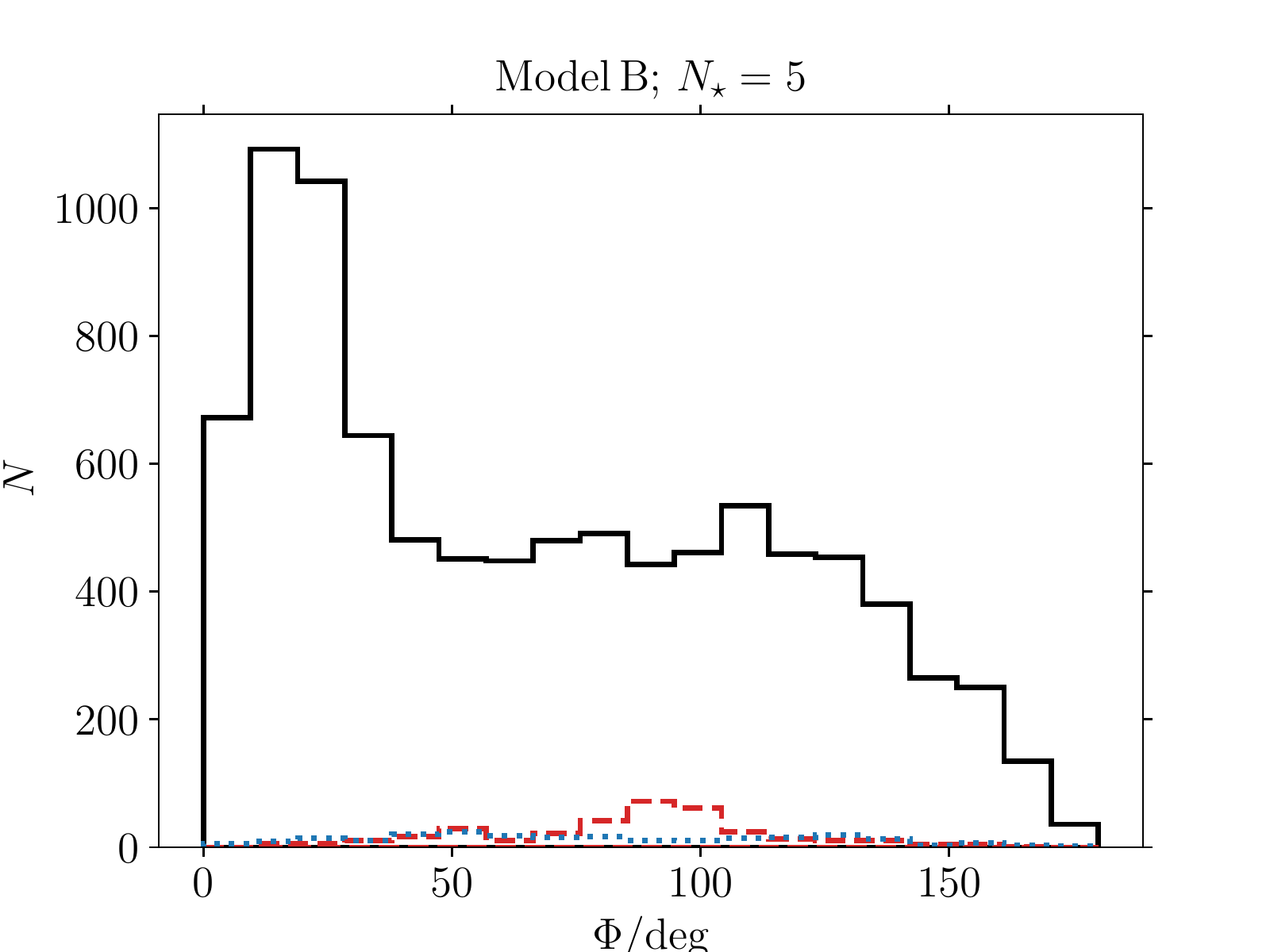}
\includegraphics[scale = \figsize, trim = 0mm 0mm 0mm 0mm]{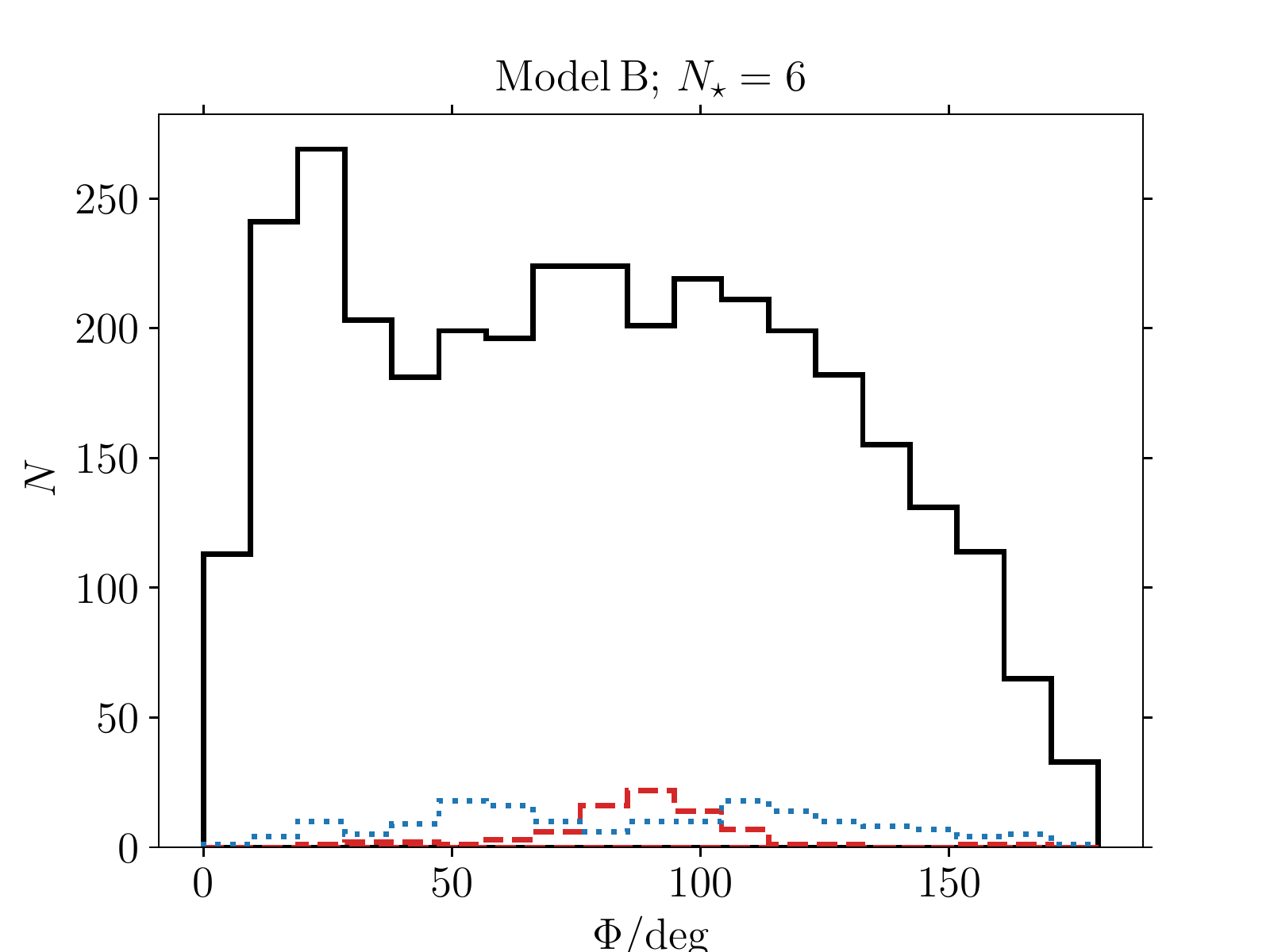}
\caption{Distributions of the (initial) mutual inclinations $\Phi$ for the different outcomes (black solid: no interaction; red dashed: strong interaction; blue dotted; dynamical instability). }
\label{fig:INCLs}
\end{figure*}

\F~\ref{fig:INCLs}, which has a similar format to \F~\ref{fig:smas}, shows the distributions of the (initial) mutual inclinations $\Phi$ for the different outcomes. Analogously to \F~\ref{fig:smas}, we include all mutual inclinations for a given system for the noninteracting outcome, whereas in the other cases, we include for a given system only the mutual inclination associated with the strongly interacting orbit, or the orbit associated with dynamical instability. 

Generally, the mutual inclination distribution for strongly interacting systems is highly peaked around $\Phi=90^\circ$. This is expected since, to lowest order, the maximum eccentricity in LK cycles is given by the canonical expression
\begin{align}
e_\mathrm{max} = \sqrt{1-\frac{5}{3} \cos^2\Phi},
\end{align}
where $\Phi$ is the initial inclination. When comparing systems with different $\ns$ in more detail, however, it becomes apparent that the $\Phi$ distributions become slightly wider. For triples, very few systems interact strongly if $\Phi$ lies outside the canonical LK window $40^\circ \lesssim \Phi \lesssim 130^\circ$ (the ones that do interact strongly are affected by the higher-order octupole term). This window increases in size for $\ns>3$, and the distributions in $\Phi$ become broader. This illustrates that strong secular evolution can be driven in systems with $\ns>3$ for a larger range of inclinations, as has been shown in detail for quadruples (e.g., \citealt{2013MNRAS.435..943P,2015MNRAS.449.4221H,2017MNRAS.470.1657H,2018MNRAS.474.3547G}). The difference in the $\Phi$ distributions between $\ns=3$ and $\ns>3$ become particularly apparent in Model B, in which fewer systems are initially highly inclined.  

Dynamical instability systems (which are unlikely in triples, as discussed above) tend to have broad distributions for $\ns>3$. For dynamical instability to occur in a system with $\ns>3$, the parent orbit typically becomes highly eccentric. The latter is determined by the mutual inclination of the {\it parent} orbit with respect to its own parent, which is not $\Phi$ in this case. Therefore, dynamical instability is approximately independent of $\Phi$.

\begin{figure*}
\center
\includegraphics[scale = \figsize, trim = 0mm 0mm 0mm 0mm]{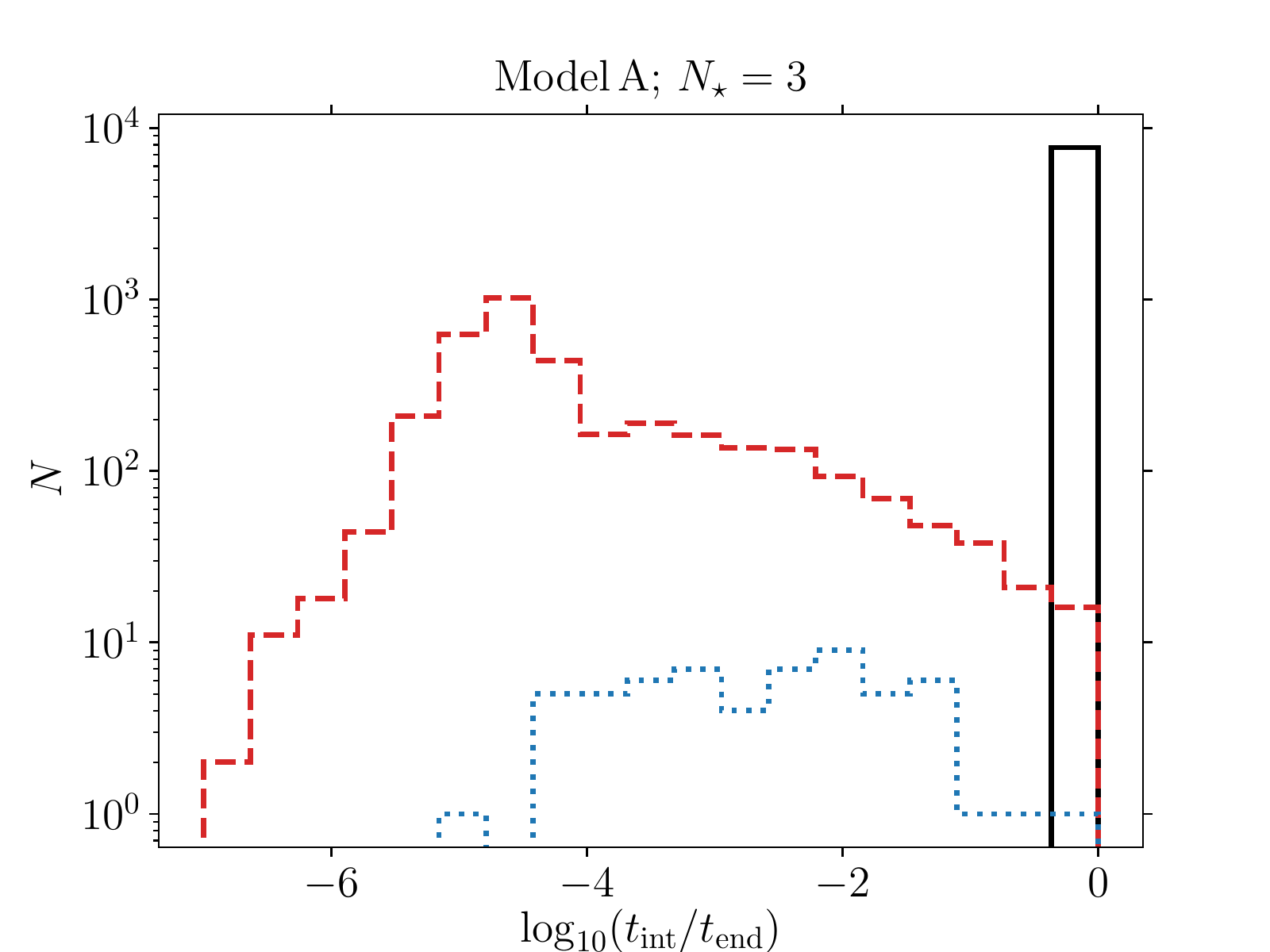}
\includegraphics[scale = \figsize, trim = 0mm 0mm 0mm 0mm]{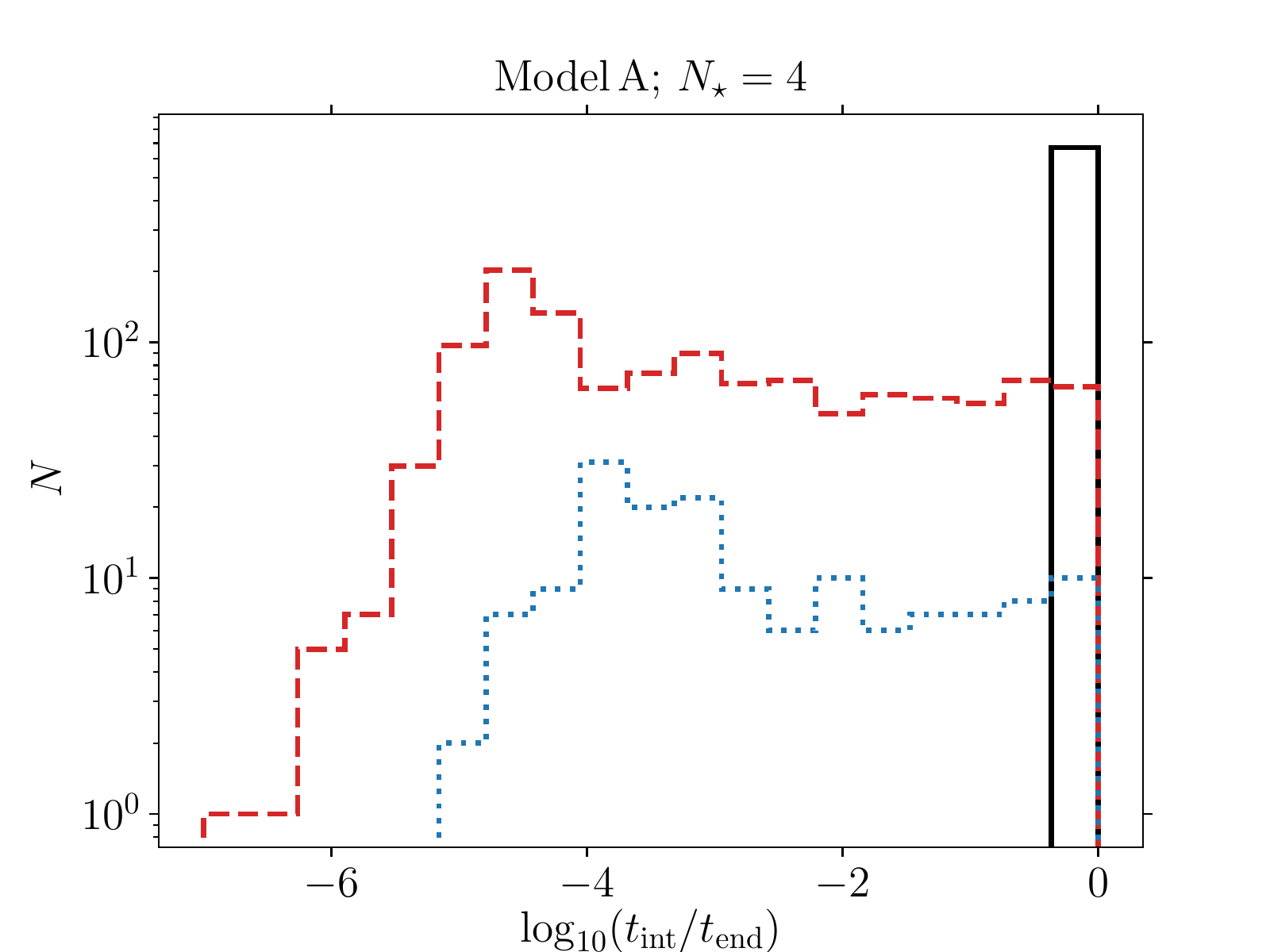}
\includegraphics[scale = \figsize, trim = 0mm 0mm 0mm 0mm]{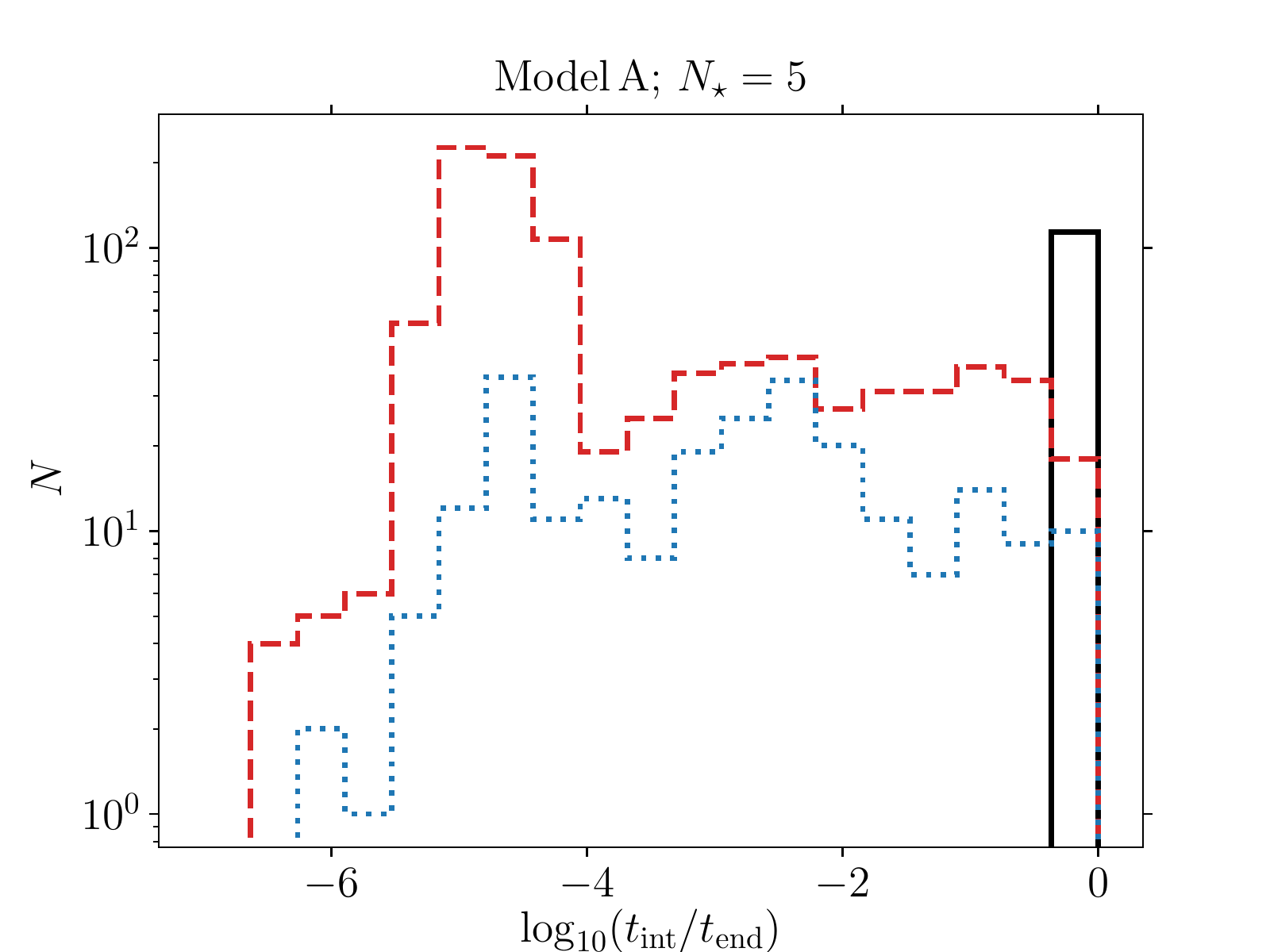}
\includegraphics[scale = \figsize, trim = 0mm 0mm 0mm 0mm]{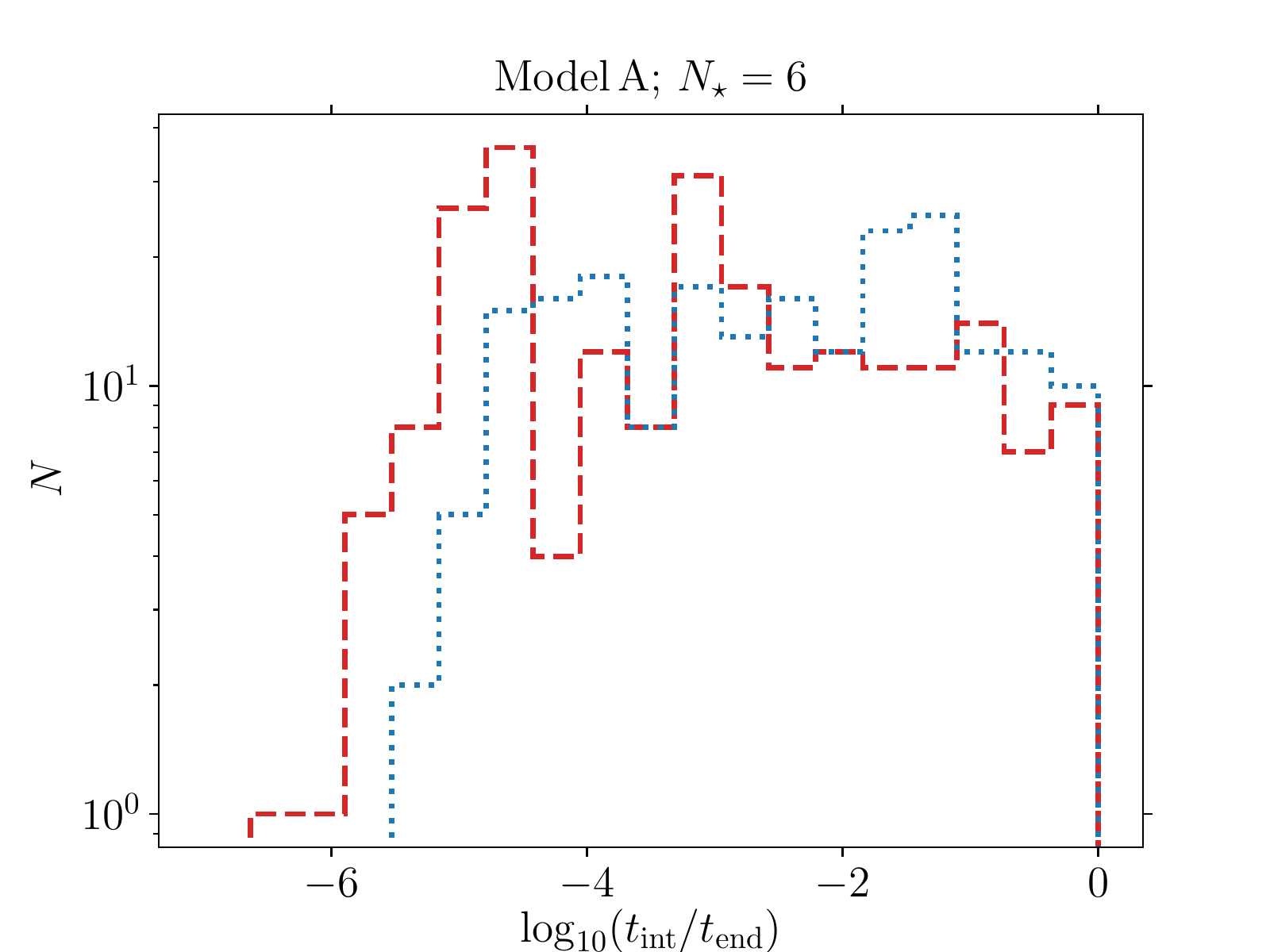}
\includegraphics[scale = \figsize, trim = 0mm 0mm 0mm 0mm]{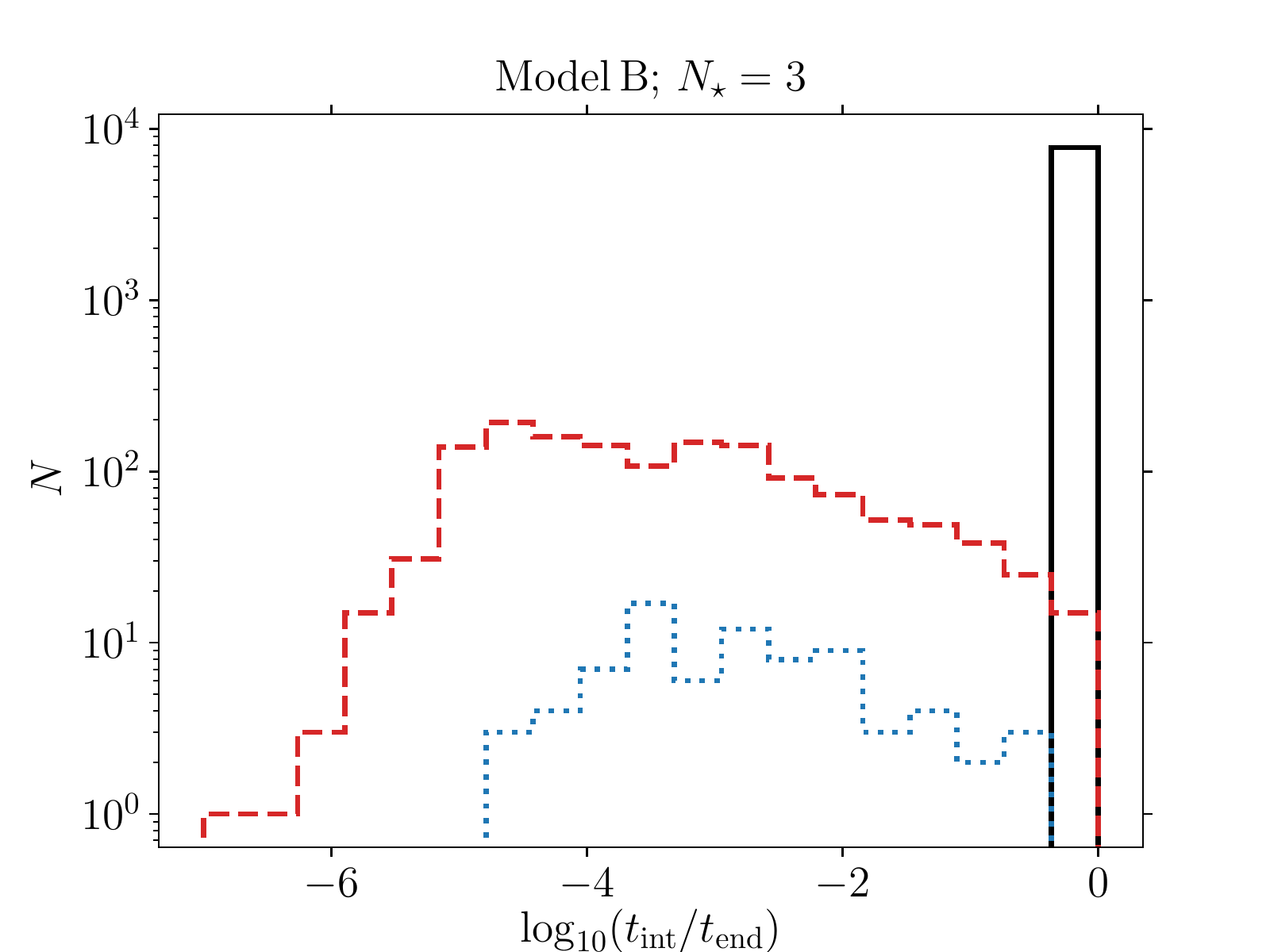}
\includegraphics[scale = \figsize, trim = 0mm 0mm 0mm 0mm]{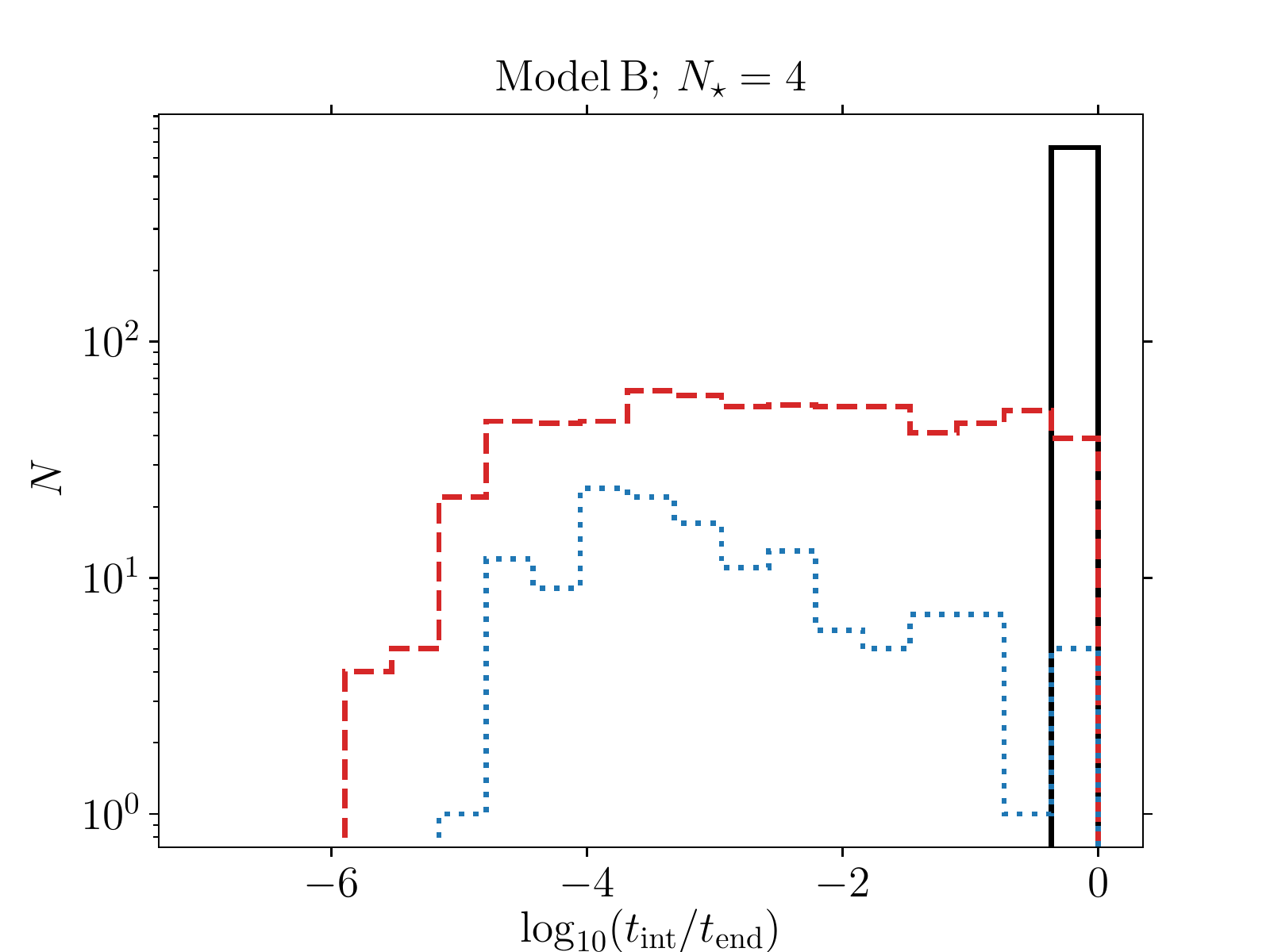}
\includegraphics[scale = \figsize, trim = 0mm 0mm 0mm 0mm]{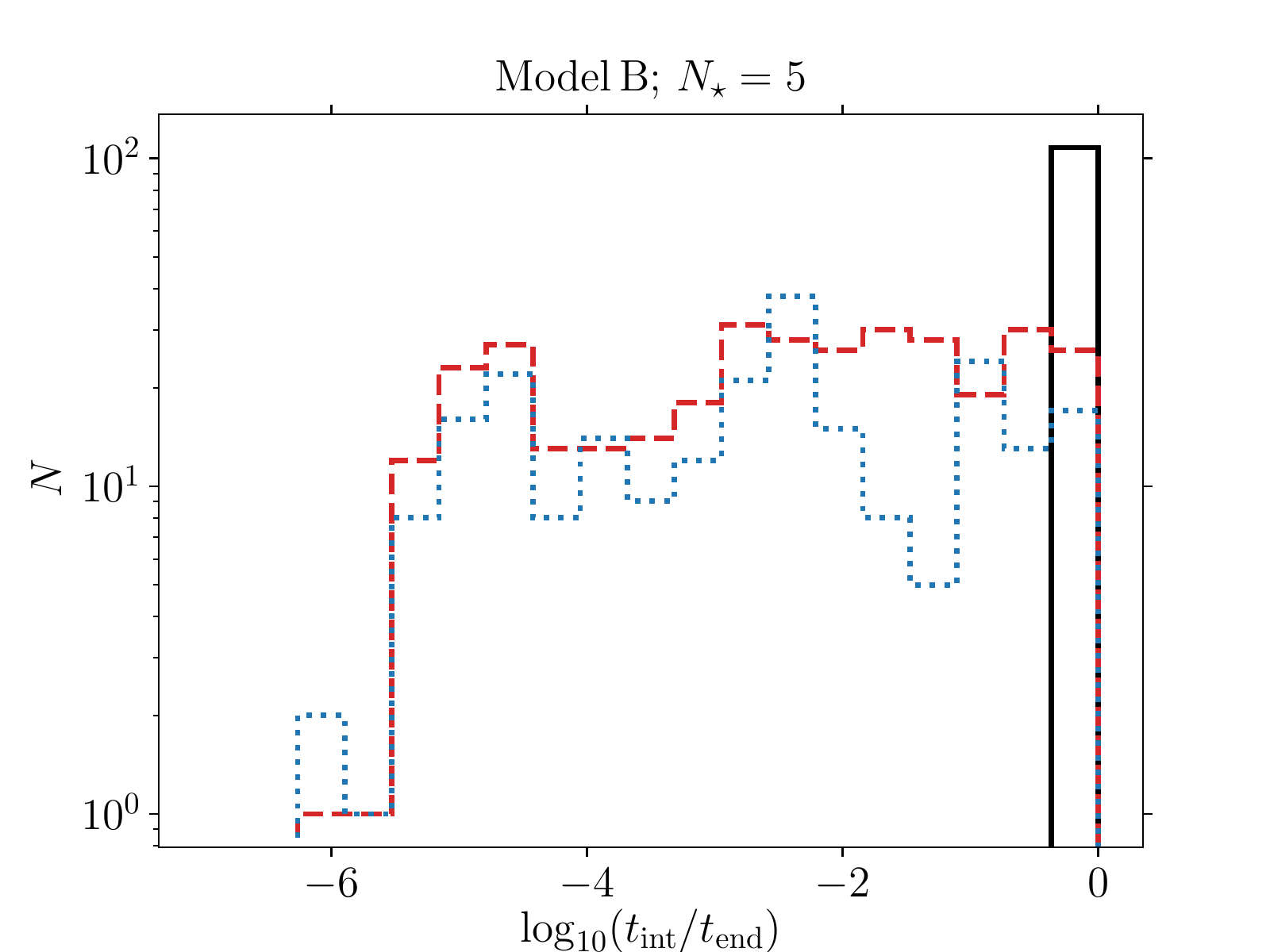}
\includegraphics[scale = \figsize, trim = 0mm 0mm 0mm 0mm]{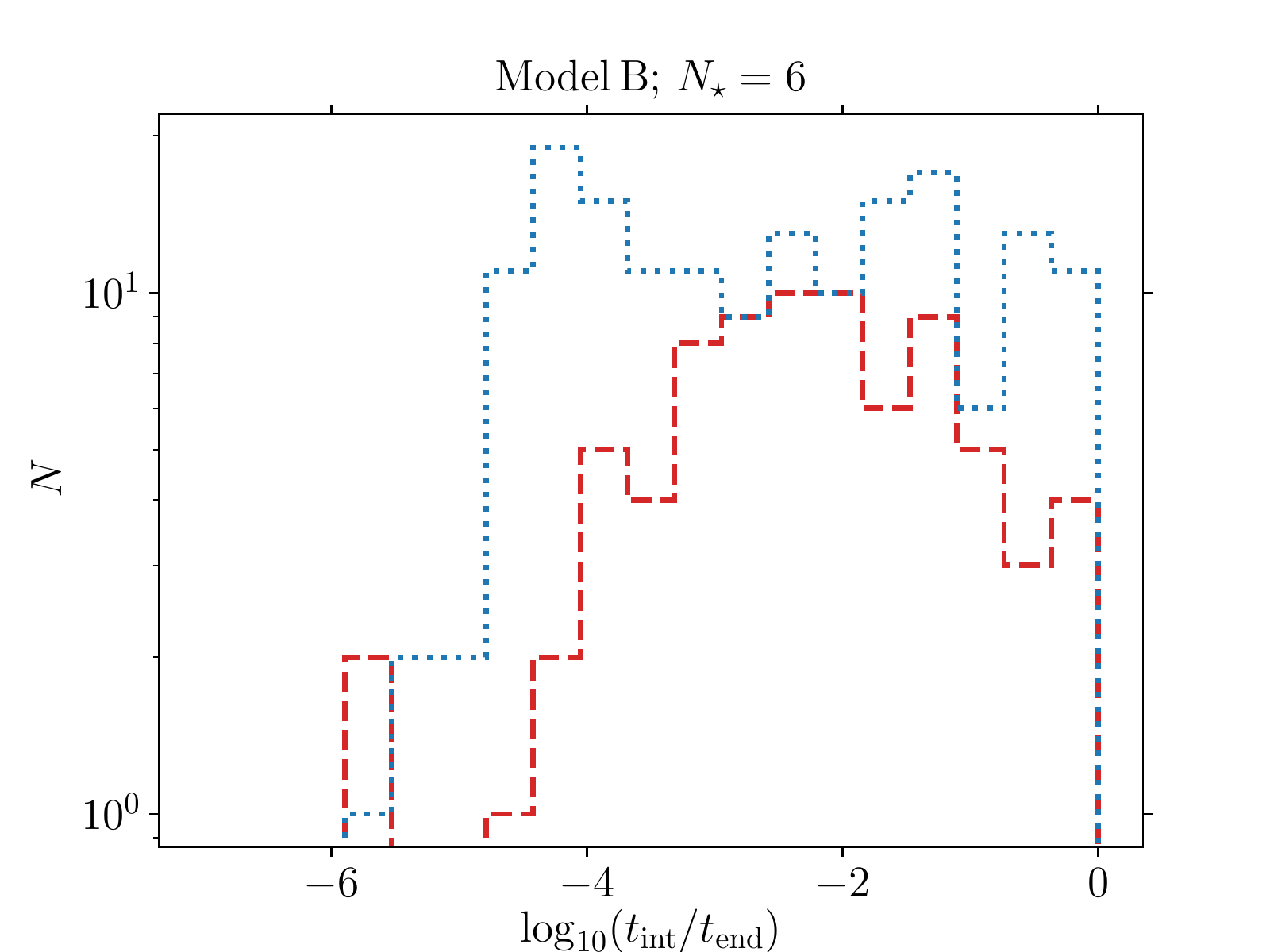}
\caption{Distributions of the interaction times for the strong interaction (red dashed lines) and dynamical instability systems (blue dotted lines), normalised to the integration time, $t_\mathrm{end}$ (cf. \eq~\ref{eq:tend}). Noninteracting systems (black dashed lines) have $t_\mathrm{int}=t_\mathrm{end}$ by design. }
\label{fig:tend}
\end{figure*}

\subsection{Interaction times}
\label{sect:results:time}
Lastly, we show in \F~\ref{fig:tend} the distributions of the interaction times $t_\mathrm{int}$, i.e., for `strong interactions', $t_\mathrm{int}$ is the time when condition \eq~(\ref{eq:rpmin}) occurred; for `dynamical instability', $t_\mathrm{int}$ is the time when the dynamical instability condition was met. For noninteracting systems, $t_\mathrm{int}=t_\mathrm{end}$ by definition. Since the integration time $t_\mathrm{end}$ of each system depends on several factors (most importantly, the MS lifetime, see \S~\ref{sect:meth:num}), we normalise the interaction times in \F~\ref{fig:tend} to $t_\mathrm{end}$. 

Strongly-interacting systems tend to form at early times; many systems have $t_\mathrm{int} \sim 10^{-5} \, t_\mathrm{end}$. Dynamical instability tends to occur somewhat later. There is a tail in the interaction time distribution for strong interactions and dynamical instability systems, up to $t_\mathrm{int}=t_\mathrm{end}$ (suggesting that some systems would interact strongly or become dynamically unstable after $t_\mathrm{end}$, which in many cases is set by the shortest MS lifetime). Interestingly, the distributions of $t_\mathrm{int}$ tend to fall off relatively quickly for triples, whereas they remain much flatter for systems with $\ns>3$. This illustrates that, due to the potentially more chaotic nature of the secular evolution, systems with $\ns>3$ can experience high eccentricities at later times in their evolution compared to triples. 

A comparison between models A and B in \F~\ref{fig:tend} reveals no qualitative differences in terms of the interaction time distributions. Similarly, the other models also show qualitatively similar results.

\section{Discussion}
\label{sect:discussion}
\subsection{Interaction times}
\label{sect:discussion:times}
As shown in \S~\ref{sect:results:time}, interactions in our simulations (both strong interactions, and cases of dynamical instability) can occur early compared to $t_\mathrm{end}$, which, in most cases, is equal to the shortest MS lifetime in the system (cf. equation~\ref{eq:tend}). This is a consequence of initial conditions and secular dynamics: an initially mutually highly inclined system (with mutual inclinations close to $90^\circ$) is likely to yield high eccentricities (and therefore trigger interaction) during the first secular oscillation; depending on the system, the secular timescale can be very short compared to the MS lifetime. In reality, such a system would likely not have large mutual inclination(s); otherwise, it would likely not be observed in its current state. 

Here, we chose an agnostic approach, and included early-interacting systems in most of our analysis. This should be taken into account when interpreting the interaction fractions (Table~\ref{table:fractions}), and for this reason, we also included a table in which interacting systems with $t_\mathrm{int}<10^{-2}\,t_\mathrm{end}$ were removed (Table~\ref{table:fractions_long}). As expected based on the interaction time distributions, the interaction fractions are markedly lower when removing systems with $t_\mathrm{int}<10^{-2}\,t_\mathrm{end}$. Nevertheless, the decreased interaction fractions in this case are still significant. Moreover, our conclusion that the interaction fraction increases strongly with $\ns$ remains robust.

\subsection{Observational biases in the MSC}
\label{sect:discussion:bias}
When interpreting our results, it is also important to bear in mind that the MSC is not based on a volume-limited sample of stars, and is therefore distorted by observational selection effects. However, the MSC does contain the most up-to-date information on the statistics of high-multiplicity systems, which motivated our choice for using this database. Our results should be interpreted with the observational biases of the MSC in mind. Also, in order to maximise the number of systems in the MSC, we integrated systems for a duration based on the shortest MS lifetime of the stars in the system, whereas some systems actually contain giant stars. This implies that we overestimated the true remaining lifetime of the components in some systems. We also ignored stellar evolution, which becomes important in those systems. Nevertheless, in our view, these complications do not affect our result of a higher interaction probability with increasing $\ns$.

It should also be taken into account that we considered interactions during the relatively long-term MS lifetime of the stars. However, strong interactions and dynamical instability can also already occur during the formation of the system (e.g., \citealt{2018ApJ...854...44M,2020MNRAS.491.5158T}). In that sense, our interaction and dynamical stability fractions are lower limits on the `true' fractions if the initial formation phase of the stars is also taken into consideration. 

\begin{figure}
\center
\includegraphics[scale = 0.42, trim = 0mm 0mm 0mm 10mm]{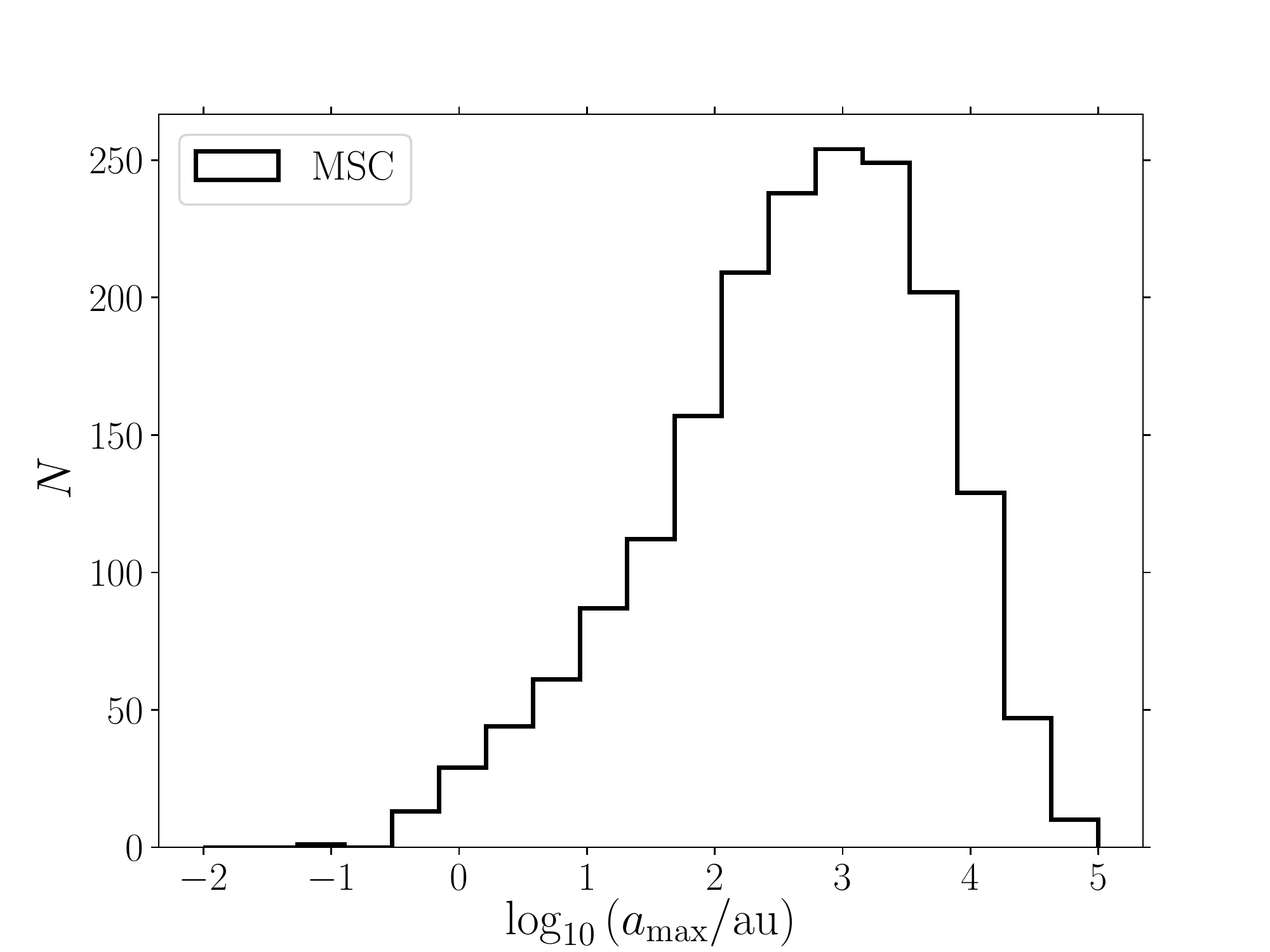}
\caption{ Distribution of the largest semimajor axes, $a_\mathrm{max}$, of systems in the MSC satisfying our selection criteria (see \S~\ref{sect:meth:IC}).}
\label{fig:MSC_max_smas}
\end{figure}

\subsection{Flybys and Galactic tides}
\label{sect:discussion:flygal}
We did not include the effects of flybys and Galactic tides in our simulations. Flybys tend to become important for orbits with separations on the order of $10^4\,\au$ and wider, and Galactic tides for separations on the order of $10^5\,\au$ and wider (e.g., \citealt{1986Icar...65...13H}). In \F~\ref{fig:MSC_max_smas}, we show the distribution of the largest semimajor axes, $a_\mathrm{max}$, of systems in the MSC satisfying our selection criteria (see \S~\ref{sect:meth:IC}). A relatively small fraction of systems has a largest separation $\gtrsim 10^4\,\au$, indicating that flybys and Galactic tides are likely not very important for our sampled systems. We remark that the MSC is likely biased to shorter orbital periods because of selection effects, since very wide orbits are generally hard to detect. 

\subsection{Future directions}
\label{sect:discussion:fut}
Our integrations were based on the orbit-averaged equations of motion. The orbit-averaging approximation can break down in cases when the timescale for secular changes is comparable to some of the orbital periods in the system. For triples, correction terms have been derived that take into account (some of) the suborbital effects that can change the long-term secular evolution (e.g., \citealt{2016MNRAS.458.3060L,2018MNRAS.475.5215B,2018MNRAS.481.4602L,2019MNRAS.490.4756L}). It is left for future work to investigate the impact of orbit-averaging corrections in higher-order systems. 

Other aspects that should be addressed in future work include continuing the phase of strong interactions taking into account secular evolution, and post-MS evolution. These aspects can produce interesting behaviour in triples (e.g., \citealt{2013MNRAS.430.2262H,2016ComAC...3....6T,2016MNRAS.460.3494S,2019ApJ...882...24H}) and quadruples (e.g., \citealt{2018MNRAS.478..620H,2019MNRAS.482.2262H}). In addition, it is of interest to investigate the outcomes of systems that are marked as becoming dynamically unstable using direct $N$-body integrations. Such an endeavour would also reveal quantitatively the true extent to which the stability criterion of \citet{2001MNRAS.321..398M} can be generalised to higher-order systems.

\section{Conclusions}
\label{sect:conclusions}
We studied the long-term evolution of systems in the MSC and estimated the probability of strong interactions (leading to tidal evolution and possibly mass transfer), and dynamical instability. These quantities give insight into the importance and efficiency of the decay of $\ns\geq3$ systems after their formation, and during their MS lifetime. Our main conclusions are listed below. 

\medskip \noindent 1. We extracted data of multiple systems from the MSC, limiting to systems with up to and including six stars ($\ns=6$), and with all component masses and orbital periods known or estimated. We adopted four different models to sample unknown orbital orientations and eccentricities. For the eccentricities, we either assumed flat distributions in all orbits subject to dynamical instability (models A and B), or period-dependent distributions (models C and D) that have a sine shape at orbital periods $P<100\,\yr$, and are thermal for $P\geq 100\,\yr$, both subject to dynamical stability \citep{2016MNRAS.456.2070T}. For the mutual inclinations, we either assumed isotropic orientations (models A and C), or we assumed that more compact systems tend to be more coplanar (models B and D; \citealt{2017ApJ...844..103T}). 

\medskip \noindent 2. In our simulations of approximately $2\times 10^5$ systems and without imposing a cut on the interaction time (see Table~\ref{table:fractions}), we found that the fraction of noninteracting systems, $\fnon$, is largest for triples ($\fnon\sim 0.9$), and decreases to $\fnon \sim 0.6$ ($\fnon \sim 0.8$) for sextuples and models A/C (B/D). The fraction of strong interactions increases from $\fint\sim 0.1$ ($\fint \sim 0.04$) to $\fint \sim 0.2$ ($\fint \sim 0.1$) from triples to sextuples in models A/C (B/D), and the fraction of dynamically unstable systems increases from $\fdyn \sim 0.001$ to $\fdyn \sim 0.2$ ($\sim0.1$) in models A/C (B/D). We interpret the increases in both $\fint$ and $\fdyn$ with increasing $\ns$ from the larger available parameter space in which strong secular evolution can arise in more complex hierarchical systems. This aspect has  has been explored before for quadruples (e.g., \citealt{2013MNRAS.435..943P,2015MNRAS.449.4221H,2017MNRAS.470.1657H,2018MNRAS.474.3547G}), but has here been shown to be the case for higher-order systems as well. Our results depend somewhat on the assumed inclination distribution, whereas there are only minor differences between the models with different eccentricity distributions. However, the absolute fractions decrease strongly when early-interacting systems are removed from the analysis (see Table~\ref{table:fractions_long}). Nevertheless, in this case, the interaction probabilities are still significant, and still increase strongly with increasing $\ns$. 

\medskip \noindent 3. The distributions of the initial ratio of outer orbit periapsis distance to inner orbit semimajor axis, $a_\mathrm{out}(1-e_\mathrm{out})/a_\mathrm{in}$, do not depend strongly on $\ns$ for the strongly interacting systems. However, only few triples become dynamically unstable during the MS, and the ones that do tend to have small values of $a_\mathrm{out}(1-e_\mathrm{out})/a_\mathrm{in}$, i.e., $a_\mathrm{out}(1-e_\mathrm{out})/a_\mathrm{in}\lesssim 10$. In contrast, in systems with $\ns>3$, the initial distributions of $a_\mathrm{out}(1-e_\mathrm{out})/a_\mathrm{in}$ for systems that become dynamically unstable are much broader, with values of $a_\mathrm{out}(1-e_\mathrm{out})/a_\mathrm{in}$ of up to $\sim 10^4$. This can be attributed to secular evolution: the latter can significantly reduce the ratio $a_\mathrm{out}(1-e_\mathrm{out})/a_\mathrm{in}$ in systems with $\ns>3$ (due to increased $e_\mathrm{out}$), whereas this is not the case for triples. 

\medskip \noindent 4. The mutual inclination distributions for strongly interacting systems tend to be broader for systems with $\ns>3$ compared to triples. Strong secular evolution can be driven in systems with $\ns>3$ for a larger range in inclinations, as has been shown before for quadruples, and has been demonstrated here to apply also to higher-order systems.

\section*{Acknowledgements}
I thank Andrei Tokovinin for carefully reading and providing feedback on an earlier version of the manuscript, and the anonymous referee for a helpful report. 

\bibliographystyle{mnras}
\bibliography{literature}

%\appendix

%\onecolumn

\label{lastpage}

\end{document}